\documentclass[sts]{imsart-arxiv}

\RequirePackage{amsthm,amsmath,amsfonts,amssymb}
\RequirePackage[authoryear]{natbib}
\RequirePackage[colorlinks,citecolor=blue,urlcolor=blue]{hyperref}
\RequirePackage{graphicx}
\RequirePackage{enumitem}
\usepackage[ruled,linesnumbered]{algorithm2e}
\SetKwInput{KwInput}{Input} 
\SetKwInput{KwOutput}{Output} 
\SetKwInput{KwNotation}{Notation}
\RequirePackage{microtype} 

\startlocaldefs

\def\E{\mathbb{E}}
\def\P{\mathbb{P}}
\def\T{\mathsf{T}}

\def\hbeta{\hat{\beta}}

\def\cL{\mathcal{L}}

\newcommand{\argmin}{\mathop{\mathrm{argmin}}}

\newcommand{\minimize}{\mathop{\mathrm{minimize}}}
\newcommand{\subjectto}{\mathop{\mathrm{subject\,\,to}}}
\def\hx{\hat{x}}
\def\hp{\hat{p}}
\def\hF{\hat{F}}
\def\hS{\hat{S}}
\def\bx{\bar{x}}
\def\bp{\bar{p}}

\def\bS{\bar{S}}
\def\hP{\hat{P}}
\def\tP{\tilde{P}}
\def\tD{\tilde{D}}
\def\D{\mathcal{D}}

\def\T{\mathcal{T}}
\def\th{^\mathrm{th}}
\def\st{^\mathrm{st}}
\def\rd{^\mathrm{rd}}

\theoremstyle{plain}

\theoremstyle{definition}

\theoremstyle{remark}

\newtheorem{assumption}{Assumption}
\endlocaldefs

\begin{document}
\begin{frontmatter}
\title{Real-Time Estimation of COVID-19 Infections: Deconvolution and Sensor
  Fusion}  
\runtitle{Real-Time Estimation of COVID-19 Infections}

\begin{aug}
\author[A]{\fnms{Maria} \snm{Jahja}\ead[label=e1]{maria@stat.cmu.edu}},
\author[B]{\fnms{Andrew} \snm{Chin}\ead[label=e2]{achin23@jhu.edu}},
\and
\author[C]{\fnms{Ryan J.} \snm{Tibshirani}\ead[label=e3]{ryantibs@cmu.edu}}
\address[A]{Maria Jahja is Ph.D. Candidate, 
  Department of Statistics \& Data Science, 
  Machine Learning Department,  
  Carnegie Mellon University, 
  Pittsburgh, PA \printead{e1}.} 
\address[B]{Andrew Chin is Statistical Developer, 
  Machine Learning Department,
  Carnegie Mellon University, 
  Pittsburgh, PA \printead{e2}.} 
\address[C]{Ryan J. Tibshirani is Professor, 
  Department of Statistics \& Data Science, 
  Machine Learning Department, Carnegie Mellon University,
  Pittsburgh, PA \printead{e3}.} 
\end{aug}

\begin{abstract}
We propose, implement, and evaluate a method to estimate the daily number of new
symptomatic COVID-19 infections, at the level of individual U.S.\ counties, by
deconvolving daily reported COVID-19 case counts using an estimated
symptom-onset-to-case-report delay distribution. Importantly, we focus on
estimating infections in real-time (rather than retrospectively), which poses
numerous challenges. To address these, we develop new methodology for both the 
distribution estimation and deconvolution steps, and we employ a sensor fusion 
layer (which fuses together predictions from models that are trained to track
infections based on auxiliary surveillance streams) in order to improve accuracy
and stability. 
\end{abstract}


\end{frontmatter}

\section{Introduction}
\label{sec:intro}

Accurate, real-time estimates of incident infections play a critical role in
informing the public health response to the spread of a disease through a
population. However, official metrics on disease activity published by
traditional public health surveillance systems in the United States do not in
fact reflect activity in real-time, as they suffer from some degree of latency
due to the way their reporting pipelines are set up and implemented.

With addressing the latency in traditional public health reporting a part of the
motivation, the last decade has seen a rise in the development of \emph{digital
  surveillance} streams in public health. Search and social media trends have
constituted much of the focus \citep[e.g.,][]{Brownstein:2009, Ginsberg:2009,
Salathe:2012, Kass-Hout:2013, Paul:2017}. More broadly, \emph{auxiliary
surveillance} streams that operate outside of traditional public health
surveillance, like online surveys, medical device logs, or electronic medical 
records, have also received significant attention
\citep[e.g.,][]{Kass-Hout:2011, Carlson:2013, Viboud:2014, Smolinski:2015,
Santillana:2016, Charu:2017, Yang:2019, Ackley:2020, Leuba:2020, Radin:2020}.

Auxiliary surveillance can improve not only on the timeliness but also on the  
accuracy and robustness of traditional public health reporting. 
Auxiliary data streams have therefore become an integral part of modern systems for
disease \emph{nowcasting} \citep[e.g.,][]{McIver:2014, Santillana:2015,
  Yang:2015, Farrow:2016, Jahja:2019, Brooks:2020}, which, put broadly, are used
to estimate the contemporaneous value of a signal that will only be fully
observed at a later date, using partial or noisy data.    

\subsection{Surveillance During the Pandemic}

During the COVID-19 pandemic, public health surveillance has produced, on one
hand, some of the most detailed public health data that the U.S.\ has ever seen,
such as daily, county-level data on reported COVID-19 cases and deaths. It has
also, on the other hand, painted an imperfect picture of situational awareness,
which created a number of downstream challenges for the public health 
response. See, e.g., \citet{Rosenfeld:2021} and references therein for an
overview of the issues. In this paper, we identify a few issues surrounding
COVID-19 case reporting in particular, propose methodology to address 
them, and implement and evaluate this proposal over eight months of pandemic
data.   

To give some background, in the early days of the pandemic, a handful of
non-gonvermental groups such as JHU CSSE \citep{Dong:2020} (and also the COVID
Tracking Project, the New York Times, and USAFacts) became known as the most 
trustworthy sources for aggregate public health reporting data on COVID-19 in
the U.S. They were founded around the idea of scraping COVID-19 data published
daily on dashboards that are run by local public health authorities (such as
state and county departments of public health), which, at the time, provided
more accurate and timely data than federal health authorities (probably due to
unrecoverable failures at one or more points along the reporting pipeline). In
fact, not only in the early days of the pandemic, but throughout, the data
published by these groups has been invaluable for decision-makers, modelers,
journalists, and the general public; for example, data from JHU CSSE remains the
gold standard for COVID-19 case and death forecast evaluation in the COVID-19
Forecast Hub \citep{ForecastHub}, a community-driven repository of forecasts
that serves as the official source for forecasting communications by the U.S.\
CDC.

Turning our focus now to case reporting, JHU scrapes cumulative case numbers
that are published daily on local health authority dashboards, and subsequently
derives a notion of case incidence based on day-to-day differences in cumulative
counts. Note that, by construction, this definition of incidence reflects the
number of new COVID-19 cases that are \emph{reported} (to the public) on any
given day. Of course, this is not the same as the number of new cases by date
tested, specimen collection date, or symptom onset date. Any of the latter
options would be more informative (increasingly so) as a definition of
incidence; revamping our surveillance systems so that they can directly provide
these and other aggregates of interest to the public health response is a
critical task for future public health crises.

 The reality of the current pandemic: alignment by report date is the only
option available, given the data published broadly on local health authority
dashboards, hence collected and aggregated by data scrapers. JHU publishes the
number of new COVID-19 case reports per U.S.\ county, daily, at a 1-day
lag. However, since report dates can lag behind symptom onset dates by many days
(a typical lag is around 5-10, but lags can be up to 30 days or more; see
Figure~\ref{fig:line_list_time}), this is actually giving us a glimpse into
COVID activity in the recent past, rather than the present.

Importantly, the CDC publishes a de-identified patient-level data set (``line
list'') on COVID-19 infections \citep{cdc_public}, which provides a symptom 
onset date column. In principle, this should allow us to construct a notion of
case incidence that is aligned by symptom onset date, but this is not possible
in practice, due to two barriers. First, the CDC only publishes updates to the 
line list monthly (due to the complexity of managing this data set). Second, 
and more problematically, this line list is fraught with missingness,
extending well beyond missingness in the symptom onset column: the \emph{total}
number of COVID-19 cases according to this line list (whether the symptom onset
date is observed or not) is far less than the total number of cases from JHU
(e.g., in early September 2021, the CDC line list reports about 30 million total
versus about 40 million from JHU), and some states (such as Texas) appear to 
missing nearly all of their cases in the line list altogether (see
Figure~\ref{fig:line_list_state}). 

\subsection{Nowcasting by Deconvolution}

\begin{figure*}[tb]
\centering
\includegraphics[width=0.95\linewidth]{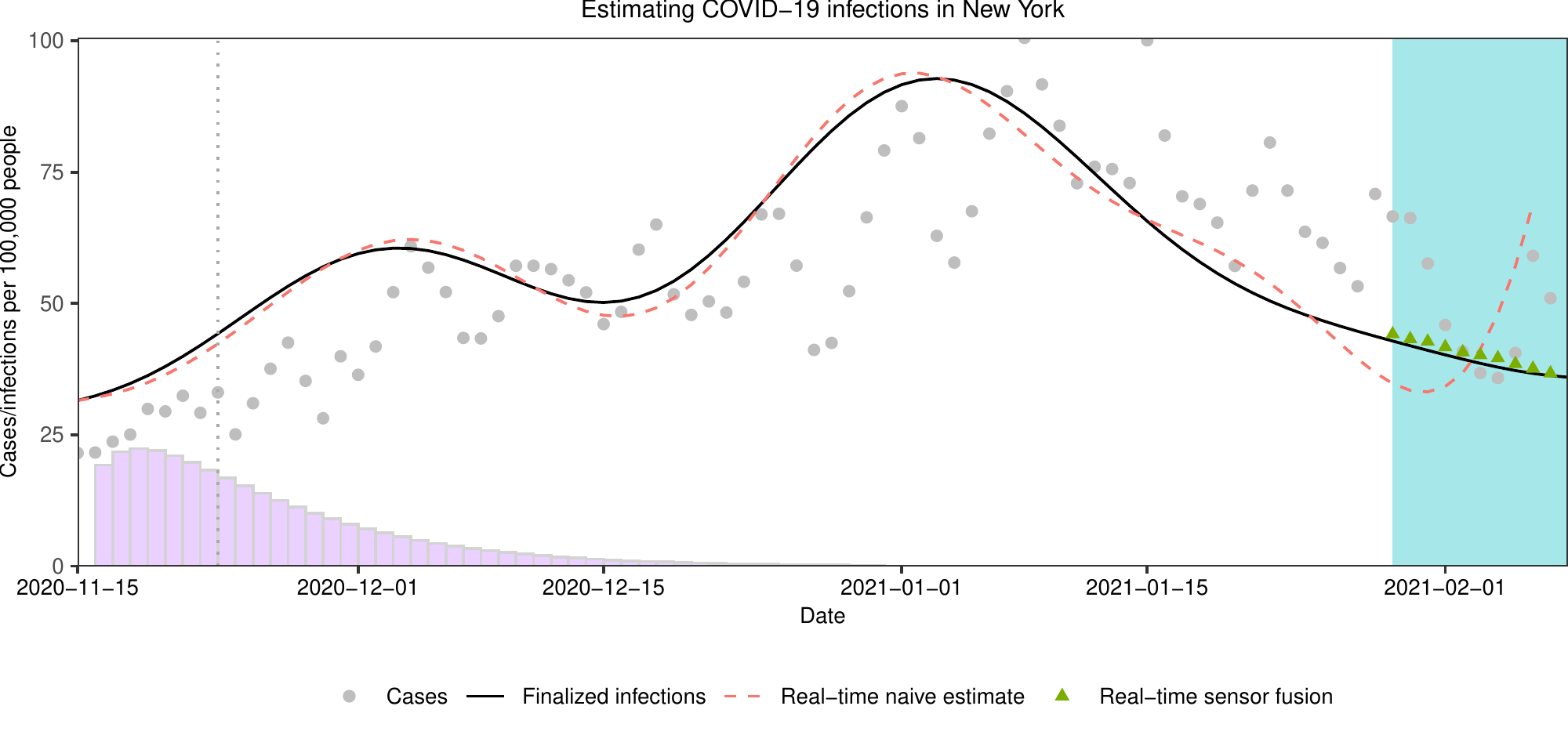}
\caption{Illustration of estimating latent infections from reported cases. The 
  dashed red line displays infection rates estimated ``naively'' in real-time,
  by directly deconvolving case data up through early February 2021, while
  the solid black line display infection rates estimated using finalized data
  from roughly four months afterwards. The blue region on the
  right-hand side highlights a period in which the real-time estimate deviates 
  substantially from the finalized one, due to the fact that we are lacking 
  sufficient (future) case observations needed to perform a ``full''
  deconvolution. The green triangles represent real-time nowcasts made by sensor
  fusion, which reduces the volatility of the real-time estimate and tracks the
  finalized estimate nicely. Lastly, the (scaled) reporting delay distribution
  estimated at the midpoint of November 2020 is drawn in purple, with the median
  reporting delay (8 days) marked as a dotted gray line.} 
\label{fig:deconv_demo}
\end{figure*}

In what follows in this paper, we use the CDC line list to estimate a delay
distribution between symptom onset and report dates, and then use this delay
distribution to deconvolve daily numbers of new case reports published by JHU
CSSE to estimate daily numbers of new symptomatic infections. Moreover, we train
models that track historical trends between past infection estimates and
auxiliary signals of COVID-19 activity from Delphi's COVIDcast project
\citep{Reinhart:2021}, and we fuse together predictions from these models in
order to improve the accuracy and robustness of our estimates of new infections
for the most recent 10 days (where deconvolution is particularly
challenging). An illustration is given in Figure~\ref{fig:deconv_demo}.

We focus on estimating new infections in \emph{real-time}, laying out a
framework for an operational nowcasting system that is forced to cope with all
of the challenges of disease tracking using provisional data. At any given
nowcast date $t$, to estimate the number of symptomatic infections at day $t-k$
(for small values of $k$, such as $k=1,2,\ldots$), we make sure to use data that
would have actually been available at $t$. This not only affects the way we
carry out all of our experiments (model training and evaluation), it also leads
us to develop some new interesting methodology to deal with the issue of
\emph{right truncation} (highlighted in Figure~\ref{fig:deconv_demo} by the blue
region). For example, in order to estimate the delay distribution in real-time,
we develop a Kaplan-Meier-like procedure to deal with a kind of right censoring
that occurs in the line list. We also develop specialized regularization techniques
to control the volatility of estimates around the nowcast date in an
optimization problem that we solve for real-time deconvolution.    

An outline for this paper is as follows. In Section~\ref{sec:preliminaries}, we
cover various preliminary details about the problem setup. Retrospective
construction of the delay distribution and deconvolution are described in
Section~\ref{sec:deconv_retro}, whereas real-time estimation is the focus in
Section~\ref{sec:deconv_realtime}. Sensor fusion is covered in
Section~\ref{sec:leverage_aux}, and extensive evaluations---comparing nowcasts
made in real-time to those made retrospectively (using ``finalized'' data that
would have only been available much later), are performed in
Section~\ref{sec:evaluation}. In Section~\ref{sec:discussion}, we conclude with
a discussion and describe a few directions for future work. 

R and Python code
for reproducing all figures and results in this paper can be found at
\url{https://github.com/cmu-delphi/stat-sci-nowcast}. 

\subsection{Related Work}
\label{sec:related_work}

In the computational epidemiology literature, the term ``nowcasting'' has been
applied to a variety of related but distinct estimation problems. Broadly
speaking, what these problems have in common is that they are about real-time
estimation of some quantity, based on partial or noisy data. They differ in
\emph{what} is being estimated, and whether this quantity will eventually be
fully observed (after enough time has passed) or whether it is latent. Examples
in the former non-latent setting, which span applications in influenza, dengue,
and COVID-19, include \citet{Yang:2015, Farrow:2016, Jahja:2019, Brooks:2020,
  McGough:2020, Hawryluk:2021}. 

The latent setting exhibits another degree of diversity within itself. In our
work, we target symptomatic COVID-19 infections, which, to be perfectly clear,
is a latent time series. Another example along similar lines is
\citet{Goldstein:2009}, who estimate influenza infection incidence via Bayesian
deconvolution of mortality data. Meanwhile, other authors might view inferring 
latent infections as just a stepping stone toward ultimately estimating the 
instantaneous reproductive number $R_t$, a key epidemic parameter. Important 
contributions to the methodology on real-time estimation of $R_t$ include: 
\citet{Bettencourt:2008}, who use a local approximation to the SIR model, and 
\citet{Cori:2013, Thompson:2019}, who use a discretization of the renewal
equation within a Bayesian framework. For a thorough review and comparison of
these methods, see \citet{Gostic:2020}. The latter paper also discusses in some
detail the importance of properly modeling the delay between infection onset and
case report, and the issue of right truncation, which, as we will see, are
central issues in our paper as well. 

The aforementioned methods have been applied and extended to build systems for
real-time $R_t$ nowcasting during the COVID-19 pandemic by \citet{Abbott:2020,
  rtlive, Chitwood:2021}. A key difference between these approaches and ours is 
that they infer infections through forward-filling: loosely speaking, they
convolve forward a candidate estimate of infections, obtain feedback by
comparing the result to measured cases, and iterate to refine estimates. This
can be effective given accurate prior knowledge, but of course it can be hard to
judge the accuracy of prior knowledge in practice. We take a more flexible
approach and estimate infections via direct deconvolution. Our approach is
nonparametric, but is still fairly simple and computationally efficient. We also
focus on fusing in auxiliary sources of information in order to improve
real-time accuracy and robustness. We remark that, if estimates of $R_t$ were
desired, then these could certainly be inferred as a by-product of our infection
nowcasts.

Finally, deconvolution has been extensively studied for many years in many
fields, notably signal and image processing, where deconvolution is sometimes
called deblurring. As an inverse problem, deconvolution is ill-posed in settings
in which the convolution operator is not known exactly or observations are made
with noise \citep{Oppenheim:2017}. Approaches to overcome this traditionally
involve regularization, as in the classical Wiener deconvolution
\citep{Wiener:1964}, which stabilizes the inversion using an estimated
signal-to-noise ratio. Alternative approaches employ familiar regularization
techniques such as $\ell_1$ and $\ell_2$ penalities \citep{Taylor:1979,
  Debeye:1990}. Most related to our paper is deconvolution using total variation 
regularization, first proposed by \citet{Rudin:1994}, and now a central tool in
signal and image processing.

\section{Preliminaries}
\label{sec:preliminaries}

In the remainder of this paper, we develop a framework for estimating the daily
symptomatic COVID-19 infection rate (where by ``rate'' we mean a count per
100,000 people, the standard units in epidemiology), concentrating on infections
that will eventually result in a reported COVID-19 case. To be clear on
nomenclature: for convenience, we will often abbreviate ``symptomatic
infection'' by ``infection'' (and so, terms like ``infection onset'' and
``infection rate'' should be implicitly interpreted as symptomatic). To
estimate infection rates, we deconvolve reported case rates with an estimated
symptom-onset-to-case-report delay distribution. To reiterate, we use case data
from JHU CSSE \citep{Dong:2020}, and to infer the delay distribution, we use a
de-identified line list on patient-level infections from the CDC
\citep{cdc_public}. 

\smallskip
\paragraph*{Auxiliary Indicators.}

After deconvolution, we improve our infection rate estimates by incorporating a
number of contemporaneous signals that track COVID activity---we will also 
refer to these as \emph{indicators}---which are publicly available through
Delphi's COVIDcast API \citep{Reinhart:2021}. The five indicators that we
consider, described below, provide auxiliary information on COVID-19 outside of 
traditional public heath reporting. Here and throughout, we abbreviate
COVID-like illness by CLI. 

\begin{enumerate}
\item Change Healthcare COVID (CHNG-COVID): The percentage of outpatient
  visits that have confirmed COVID-19 diagnostic codes, based on de-identified 
  Change Healthcare medical claims data. 
\item Change Healthcare CLI (CHNG-CLI): The percentage of outpatient visits that
  have COVID-like diagnostic codes, based on the same data.
\item Doctor Visits CLI (DV-CLI): The same definition as CHNG-CLI, but applied
  to de-identified medical claims data from other health systems partners. 
\item COVID Trends and Impact Survey CLI in the community
  (CTIS-CLIIC): The estimated percentage of people reporting  
  illness in their household or local community, based on Delphi's COVID
  Trends and Impact Survey (CTIS), in partnership with Facebook. 
\item Google searches for anosmia and ageusia (Google-AA): A measure of volume
  for Google queries related to anosmia or ageusia (loss of smell or taste),
  from Google's COVID-19 Search Trends data set. 
\end{enumerate}

Roughly speaking, we study these particular indicators (ordered roughly from
``late'' to ``early'') because conceptually they reflect data measurements that
would be made at some period of time in between infection onset and case report
to a public health authority, and therefore would be relevant in inferring
latent infection rates. More information on these indicators and their
underlying data sources is given in \citet{Reinhart:2021}. For more information
on CTIS in particular, see \citet{Salomon:2021}; and for a study of how these
and similar indicators can improve COVID-19 forecasting, see
\citet{McDonald:2021}.

\smallskip
\paragraph*{Sensor Fusion.}  

For each of the auxiliary indicators described above, we train a model to
estimate latent infection rates from indicator values, using historical data
(described in Section~\ref{sec:sensor_models}). At each nowcast date, we then  
use such a model to estimate the latent infection rate from the current
indicator value, which gives a total of five estimates (one from each of the
five models), along with a sixth estimate coming from an autoregressive model 
trained on historical estimated infection rates. We will refer these six 
contemporaneous estimates as \emph{sensors}.

In this paper, we consider (as described in Section~\ref{sec:sensor_fusion})
various methods for combining these estimates into a single estimate of the  
infection rate, which we will call \emph{sensor fusion} methods. Broadly
speaking, sensor fusion is a form of ensembling, which is ubiquitous in in
predictive modeling in statistics and machine learning, as it can often help
improve both accuracy and robustness. In our particular application, the sensors
themselves are constructed from data streams operating outside of traditional
public health reporting, which itself contributes an additional important angle
in terms of robustness. 

\subsection{Problem Setup}
\label{sec:problem_setup}

\paragraph*{Estimation Period.}

For every day $t$ in between October 1, 2020 and June 1, 2021 inclusive (243
days in total), we estimate the symptomatic infection rate at day $t-k$, using  
only data that would have been as of time $t$, which in this context we call the  
\emph{nowcast date}. Estimation of the latent infection rate at time $t-k$ (for
positive $k$) is technically a backcast, though we will not be careful to
distinguish this notationally from nowcasting, and will generally refer to this
as nowcasting at lag $k$. We produce estimates for each $k=1,\ldots,10$, a total  
of 10 targets per nowcast date $t$. 

When we say above that nowcasts are made using data that would have been
available \emph{as of} a given nowcast date $t$, we mean that we adhere not to
only the real-time availability (latency) of signals at $t$, but also the
\emph{version} of the data published at $t$---simply put, imagine that we
``rewind'' the clock to time $t$ and query the API to receive the data that
would have been returned then. This is possible becausse the COVIDcast API
records and provides access to all historical versions of data, as described in
\citet{Reinhart:2021}. As epidemic data is often subject to revision, if we
train and evaluate models on ``finalized'' data (that would have been available
only at a much later time point) then this can lead to inaccurate conclusions
about real-time model performance; see, e.g., \citet{McDonald:2021}.

Further, it is worth noting that reported case data from JHU is available at a
1-day lag, and we assume that there is at least another 1-day lag between
symptom onset and case report (explained in
Section~\ref{sec:delay_distribution}). Hence through real-time deconvolution
alone we would be able to make nowcasts at a 2-day lag at the earliest. Making
nowcasts at a 1-day lag is possible with sensor fusion, using auxiliary signals 
with 1-day latency (explained in Section~\ref{sec:leverage_aux}). In this sense,
sensor fusion is able to improve not only accuracy, but also latency, and buys
us 1 extra day. 

\smallskip
\paragraph*{Geographic Scope.}

We produce nowcasts at the county resolution, but for computational purposes, we
restrict our attention to the 200 U.S.\ counties with the highest
population. We additionally produce estimates for each of the 50 U.S.\ states. 
(Some of the methodology that we use for sensor fusion requires a geographical
hierarchy, thus using the remaining $\approx$ 3000 U.S.\ counties we aggregate
these within each state to create ``rest-of-state'' jurisdictions, and make
estimates for these as well, for the purposes or maintaining such a hierachy.) 

\smallskip
\paragraph*{Evaluation.}

We evaluate all nowcasts made in between October 1, 2020 and June 1, 2021
inclusive (243 days in total) and at each of the 250 locations in consideration 
(50 states and the 200 largest counties) against latent infection rate estimates 
obtained by deconvolving the case rate data available as of August 30, 2021. 
We will refer to the latter as \emph{finalized} infection rate estimates (as 
opposed to real-time ones); details are given in Section~\ref{sec:ground_truth}. 

\subsection{Confounding}
\label{sec:confounding}

Estimates of COVID infections obtained by deconvolving reported cases will
generally underestimate the true number of infections, because many infections
are undetected or untested, and as such, do not appear later on in case
reports. If we wanted to estimate the true number of symptomatic infections
from case reports, then we would need to have some sense of the fraction of 
symptomatic infections that go untested. Of course, this only gets more
complicated if we extend our consideration to both symptomatic and asymptomatic
infections.  

Other authors, e.g., \citet{Chitwood:2021}, have taken the ambitious step of
proposing and implementing frameworks with parameters that account for such
confounding. However, adjustments for case ascertainment and asymptomatic
infections generally rely, at least to some nontrivial extent, on model
assumptions (typically, mechanistic ones) that are difficult to substantiate.

We take a different perspective and pose the problem as one of real-time 
deconvolution only. We seek to answer the question:   
\begin{quote}
\it
Can we estimate---in real-time---the number of new symptomatic COVID-19
infections that will eventually appear in case reports?  
\end{quote}
Hence, by construction, confounding is not a problem that we even attempt   
to reconcile (because the target we track, infections that eventually show up
in case reports, simply inherits any confounding that would be present in the
case reporting stream in the first place). 

Our approach can be seen as one that runs in parallel (rather than in
contradiction) to an approach that explicitly models and removes the effects of
confounding in case reporting. We focus on addressing the deconvolution problem
as carefully as possible, with a concern for real-time estimation, and an eye
toward using auxiliary signals to improve accuracy and robustness. Estimates of
parameters that account for confounding (that comes from other work focused on  
these aspects) could certainly be applied to our deconvolution estimates post
hoc in order to adjust them appropriately; we revisit this idea in the
discussion. 

Lastly, under an assumption that the confounding acts as a multiplicative bias
that changes slowly over time, our real-time infection rate
estimates---themselves subject to confounding, as explained above---can be
post-processed to derive real-time \emph{approximately unconfounded} estimates
of $R_t$. This is also described in the discussion.

\section{Retrospective Deconvolution}
\label{sec:deconv_retro}

In this section, we study and fit a convolutional model between infections and  
reported cases. We adopt a \emph{retrospective} angle here and do not concern
ourselves with data availability or versioning issues; this is covered in the
next section.   

\subsection{Convolutional Model}
\label{sec:conv_model}

For simplicity, we introduce the convolutional model in just a single
location. We denote by $y_t$ the number of new cases that are reported at time
$t$, and by $x_t$ the number of new infections that have onset at time $t$. Our
jumping-off point is the following model:
\begin{equation}
\label{eq:conv_model1}
\E[y_t \,|\, x_s, s \leq t] = \sum_{s=1}^t \pi_t(s) \, x_s, 
\end{equation}
where for each $s \leq t$,
\begin{equation}
\label{eq:delay_prob1}
\pi_t(s) = \P\big( \text{case report at $t$} \,|\, \text{infection onset at   
  $s$} \big). 
\end{equation}
We refer to the probabilities above as \emph{delay probabilities} at time $t$,
and the entire sequence $(\pi_t(s) : s \leq t)$, as the \emph{delay
  distribution} at time $t$. 

The justification for \eqref{eq:conv_model1}, \eqref{eq:delay_prob1} is
elementary: to count $y_t$, we enumerate all infections that ever occurred in
the past:  
\[
y_t = \sum_{s=1}^t \sum_{i=1}^{x_s} 1\{\text{the $i\th$ \hspace{-3pt} infection   
  at $s$ gets reported at $t$}\}.
\]
Taking a conditional expectation on both sides above, and using linearity,
delivers \eqref{eq:conv_model1}, \eqref{eq:delay_prob1}.

In the next subsections, we will describe how to estimate the probabilities
$\pi_t(s)$ in \eqref{eq:delay_prob1}, and how to use this alongside the observed
case reports $y_t$ in order to estimate the latent infections in
\eqref{eq:conv_model1}. 

\subsection{Estimating the Delay Distribution} 
\label{sec:delay_distribution}

At the outset, we place the following assumptions on the delay distribution in 
order to make its estimation (using the CDC line list data, to be described
shortly) more tractable.   

\begin{assumption}
\label{asm:delay_support} 
Infections are always reported within $d=45$ days; that is, $\pi_t(s) = 0$
whenever $s < t-d$. 
\end{assumption}

\begin{assumption}
\label{asm:zero_at_zero}
The probability of zero delay is zero; that is, $\pi_t(t) = 0$.
\end{assumption}

\begin{assumption}
\label{asm:geo_invar}
The delay distribution is geographically invariant (it is the same for any
location). 
\end{assumption}

Assumption~\ref{asm:delay_support} is innocuous. The vast majority of pairs of
recorded infection dates and report dates in the CDC line list data fall within
$d=45$ days of one another. Assumption~\ref{asm:zero_at_zero} is perhaps less
innocuous but still fairly minor, and it is a consequence of the fact that a
delay of zero (infection date equal to report date) has been used inconsistently
in the CDC line list: this could mean a true delay of zero, or it could be a
code for missingness.

\begin{figure*}[tb]
\centering
\includegraphics[width=0.95\linewidth]{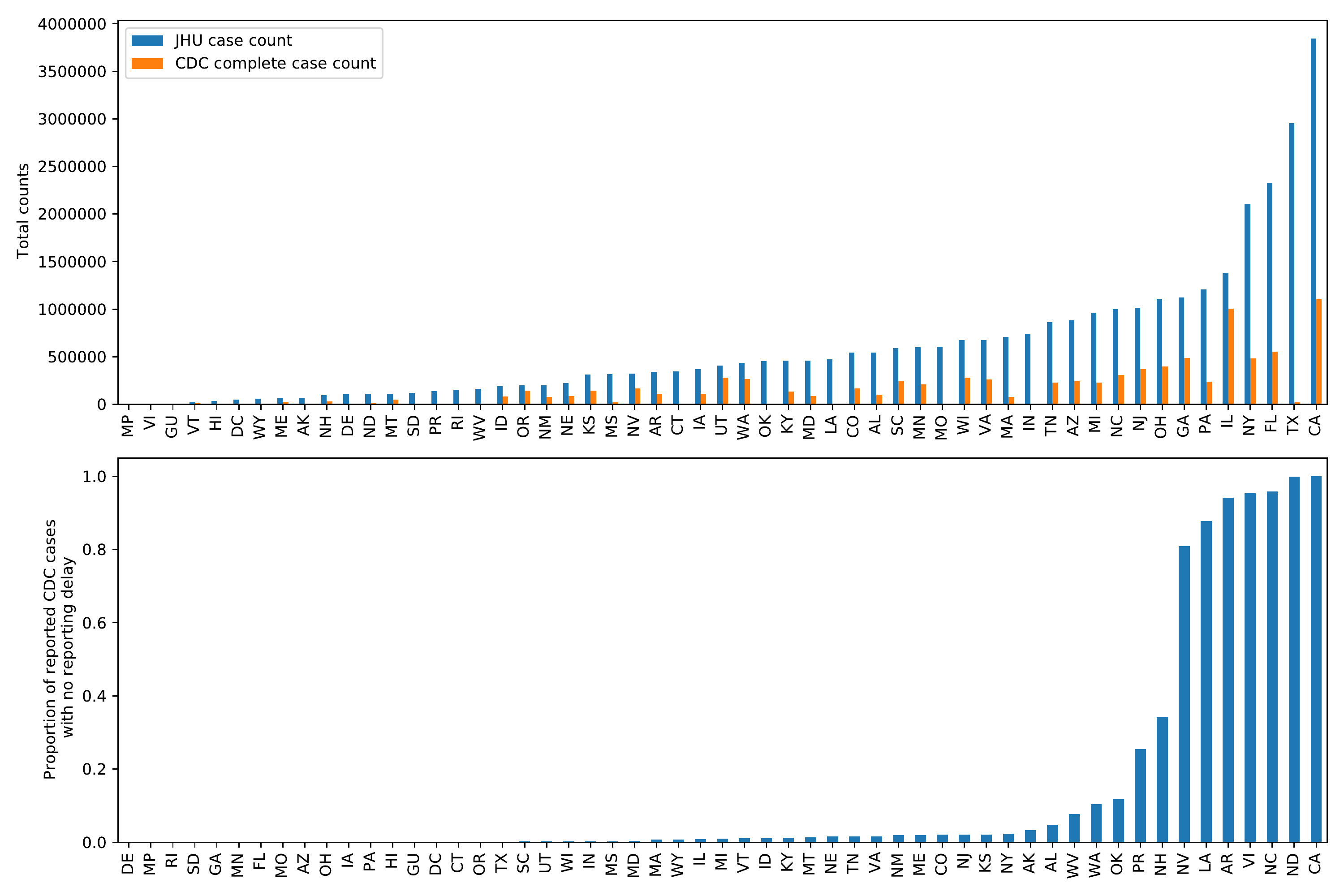}
\caption{Top: cumulative case count per state on June 1, 2021, as reported by
  JHU CSSE, compared to the complete case count (where both onset date and
  report date are observed) per state up through the same date, in the CDC
  restricted line list. Most states have less than 50\% of the cases appear in
  complete form in the line list, and some (e.g., Texas) have almost none at
  all. Bottom: proportion of complete cases with zero delay per state in the
  same line list data. There is very wide variation between these proportions.} 
\label{fig:line_list_state}
\end{figure*}

\begin{figure}[tb]
\centering
\includegraphics[width=0.9\linewidth]{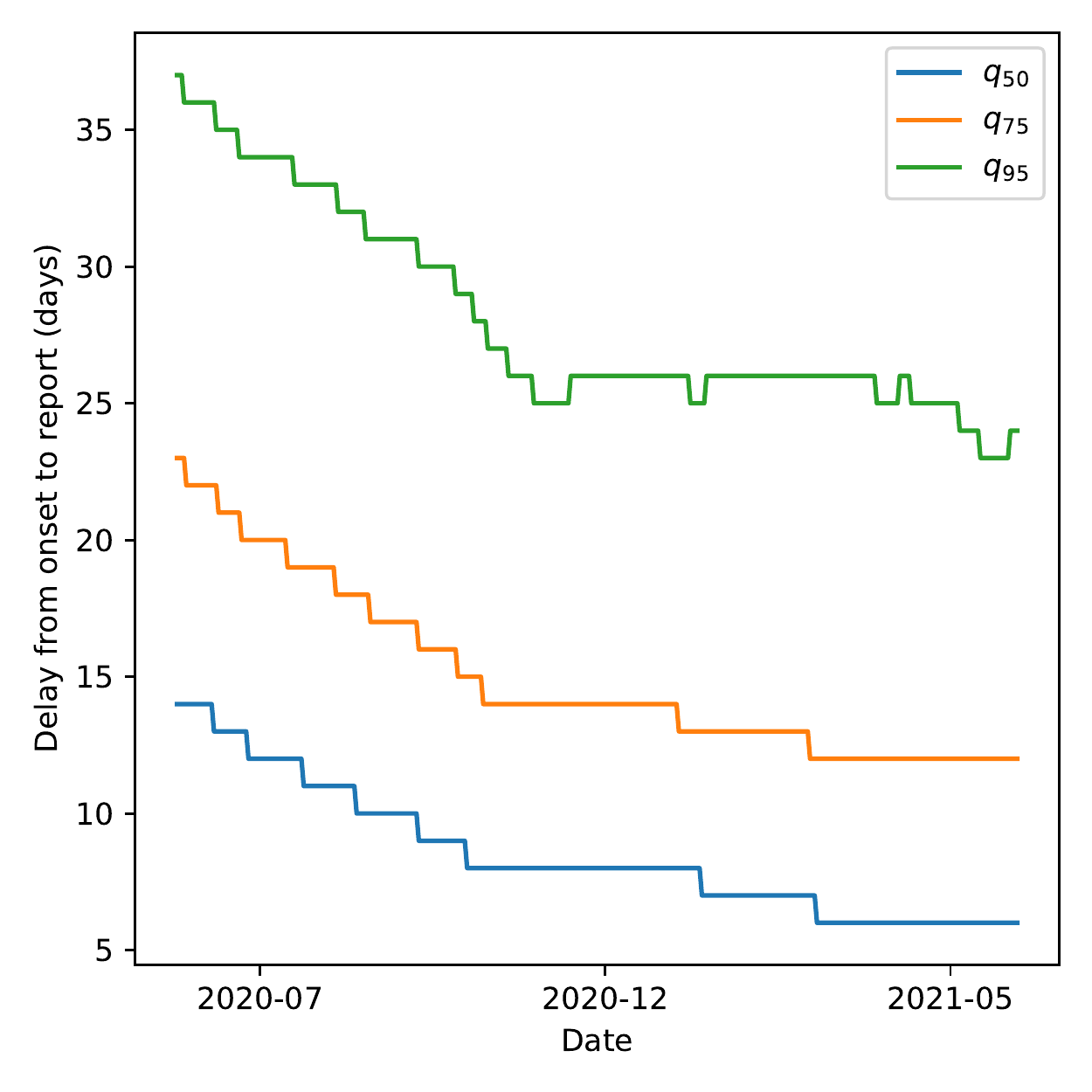}
\includegraphics[width=0.9\linewidth]{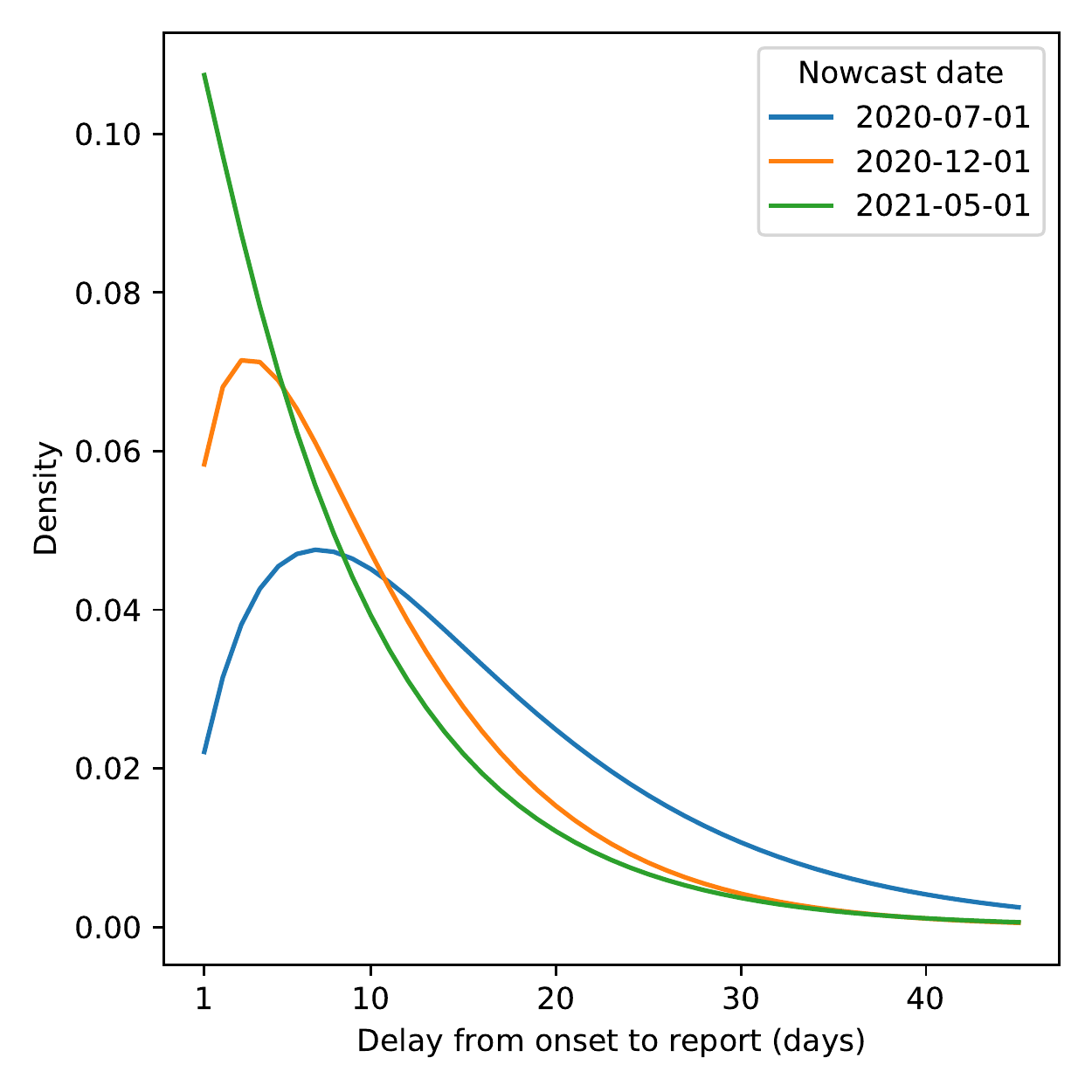}
\caption{Top: quantiles of the estimated delay distribution returned by 
  Algorithm~\ref{alg:delay_dist_retro} at the levels 50\%, 75\%, and 95\%, as 
  $t$ varies from June 1, 2020 to June 1, 2021. Bottom: estimated delay
  distributions overlaid for three nowcast dates within the same time
  interval.}    
	\label{fig:line_list_time}
\end{figure}

Assumption~\ref{asm:geo_invar} is the most noteworthy and troublesome. We do
\emph{not} believe it to be true that different locations actually have
identical patterns of delay between infections and case reports; conversely, we
expect there to be a considerable amount of variability between locations in
this regard. While we do allow the delay distribution to change over time (see
Figure~\ref{fig:line_list_time} for evidence for the importance of this), we
consider Assumption~\ref{asm:geo_invar} to be a weakness of our work. However,
the \emph{data is simply not there} in the CDC line list to warrant
location-specific estimation of the delay distribution (see
Figure~\ref{fig:line_list_state}), thus we resort to estimating a nation-wide
delay distribution.

Meanwhile, it is worth pointing out that better (location-specific) estimates of
the delay distribution could be simply plugged into our deconvolution
methodology (detailed in Section~\ref{sec:ground_truth}) to yield better
estimates of latent infections. This would carry over to all of the real-time  
methodology for deconvolution and sensor fusion (in
Section~\ref{sec:deconv_realtime}) as well. In other words, a strength of our
methodology is that it can treat the delay distribution as an input, and a user
(say, a local health official) can replace the default nation-wide delay
distribution with a more-informed local one in order to get more-informed local 
estimates. 

In light of Assumptions~\ref{asm:delay_support} and \ref{asm:zero_at_zero}, we
change our notation henceforth, and rewrite \eqref{eq:conv_model1},
\eqref{eq:delay_prob1} as: 
\begin{equation}
\label{eq:conv_model2}
\E[y_t \,|\, x_s, s \leq t] = \sum_{k=1}^d p_t(k) \, x_{t-k},  
\end{equation}
where for $k=1,\ldots,d$,
\begin{equation}
\label{eq:delay_prob2}
p_t(k) = \P\big( \text{case report at $t$} \,|\, \text{onset at $t-k$} \big). 
\end{equation}

\smallskip
\paragraph*{CDC Line List.}

The CDC provides de-identified patient-level surveillance data on COVID-19 in
both public and restricted forms \citep{cdc_public, cdc_restricted}. The
restricted one is made available under a data use agreement. The public line
list contains the same patient-level records as the restricted one, but it has
geographic details withheld. (There is another publicly available that contains
geographic details, but withholds temporal details). We use the public data
set\footnote{The CDC does not take responsibility for the scientific validity or
  accuracy of methodology, results, statistical analyses, or conclusions
  presented.} 
in this paper for estimating the delay distribution, since missingness compels
us to make nation-wide (rather than location-specific) estimates.

It is worth noting that the line list is itself provisional and subject to
revision. Furthermore, the CDC only publishes updates to the line list
monthly. In this paper, for simplicity, we use a single version of the CDC line
list---released on September 9, 2021---to construct all delay
distributions. Nonetheless, in our real-time nowcasting experiments, we 
restrict our access to data in this line list that would have been available at
each nowcast date $t$ (rows whose report date to the CDC is at most $t$) to
construct delay distribution estimates at $t$. This is highly nontrivial, due to
bias induced by truncation of data after $t$ (see
Section~\ref{sec:delay_adjust}). 

\smallskip
\paragraph*{Missing Values.}

The CDC line list (both public and restricted data sets) is subject to a high
degree of missingness. Such missingness manifests itself in a variety of
ways. For the public line list published on September 9, 2021: 
\begin{itemize}
\item it has 29,851,450 rows, compared to 39,365,080 cumulative cases reported  
  by JHU CSSE on September 9, 2021;
\item 8.64\% of rows are missing the case report date (the
  \texttt{cdc\_report\_dt} column);    
\item 53.6\% of rows are missing the symptom onset date (the \texttt{onset\_dt}
  column);   
\item of all rows in which symptom onset date is present, the case report date
  is also present, but when a report date is missing in practice it sometimes 
  gets filled in with the onset date, clouding the interpretation of a zero 
  delay.\footnote{Confirmed by personal communication with the CDC.}     
\end{itemize}
Due to the last point, we exclude zero in the construction of all delay
distribution estimates, in what follows. 

The restricted line list is no better with respect to such missingness, 
exhibiting nearly exactly the same patterns as those described above. It
does additionally provide geographic details, which allows us to examine how  
missingness is dispersed across different locations. 
Figure~\ref{fig:line_list_state} displays results to this end, using  
the restricted line list released on October 12, 2021. The top panel shows that
there is a high degree of missingness in complete case counts (those with
both onset date and report date observed) in most states, often well over 50\%,
and moreover, missingness is far from uniform at random: e.g., Texas has barely
any of its cases present in the line list. The latter observation is why we
resort to estimating nation-wide delay distributions, in what follows. 

The bottom panel in the figure shows that there is also a high degree of
heterogeneity in the fraction of complete cases with zero delay (between onset
date and report date) across states. Some states (e.g., California) have zero
delays for nearly all of their complete cases, while others (e.g., Delaware)
have zero delays for none of their complete cases, suggesting that the practice
of setting a missing report date equal to the associated onset date is highly
inconsistent between states. This only further corroborates the decision to
exclude zero delays from the data set when estimating the delay distribution. 

\smallskip
\paragraph*{Delay Distribution Estimation.}

From the public line list, we estimate the delay distribution at each time $t$, 
namely the probabilities in \eqref{eq:delay_prob2} for $k=1,\ldots,d$, using the  
empirical distribution of all lags, excluding zero, between complete onset and
report dates, for all onset dates falling in $[t-2d+1, t]$. Then, we fit a gamma
density to the empirical distribution by the method of moments, and discretize
the resulting density over the support $\{1,\ldots,d\}$. For concreteness,
this procedure is described in Algorithm~\ref{alg:delay_dist_retro}.   

\begin{algorithm}[tb]
\caption{Delay distribution estimation, retrospective}
\label{alg:delay_dist_retro}
\DontPrintSemicolon
\KwInput{Time $t$, support size $d$, window size $w=2d$, line list $\D$ 
  with onset dates $a_i$ and report dates $b_i$.} 
\KwOutput{Estimated delay probabilities \smash{$\hp_t(1), \ldots, \hp_t(d)$}.}  

Find all pairs in $\D$ with onset dates within a recent time window: $I_t = \{i 
: a_i \in (t-w, t]\}$. 

Compute the empirical distribution of lags $1,\ldots,d$ among these pairs: 
$$
\bp_t(k) = \frac{|\{i \in I_t: b_i - a_i = k\}|}
{\sum_{\ell=1}^d |\{i \in I_t: b_i - a_i = \ell\}|}, \;\; k=1,\ldots,d. 
$$

Fit a gamma density to \smash{$\bp_t(1), \ldots, \bp_t(d)$} using the method of
moments (matching the mean and variance). 

Discretize this gamma density to the support set $\{1,\ldots,d\}$, call the 
result \smash{$\hp_t(1), \ldots, \hp_t(d)$}, and return these probabilities.
\end{algorithm}

We use only ``recent'' pairs of onset and report dates at time $t$ (whose onset
date lies in $[t-2d+1, t]$) in order to adapt to the nonstationarity in
reporting delays over time. The top panel in Figure~\ref{fig:line_list_time}
plots quantiles of the estimated delay distribution from
Algorithm~\ref{alg:delay_dist_retro}, as $t$ ranges from June 1, 2020 to June 1,
2021. We see sharp drops in all quantiles during the first half of this period,
and then a more gradual decline over time. The bottom panel in the figure gives
a qualitative sense of how the delay distribution estimates change in shape over
time.

\subsection{Defining Ground Truth}
\label{sec:ground_truth}

Given the estimated delay distributions over time from the previous subsection, 
we now describe how to estimate latent infections in the model
\eqref{eq:conv_model2}. In short, we will solve one large optimization problem
to perform deconvolution. To define the best possible retrospective estimates of
latent infections over the period October 1, 2020 to June 1, 2021, which we will
treat as \emph{ground truth} in what follows (in the sense that they will be the
point of comparison for all of our real-time estimates), we will perform
deconvolution over a wider time period than the previously specified one in
order to avoid any bias issues at the boundaries (where there is insufficient
data for accurate deconvolution; more details are provided in the next
section): our retrospective deconvolution runs from May 1, 2020 to August 28,
2021, a period we denote by $\T$, and uses case data published on August 30,
2021. 

For location $\ell$, denote by $y_{\ell,t}$ and $x_{\ell,t}$ the number of new
cases reported and number of new infections that onset at time $t$,
respectively, per 100,000 people. Note that $y_{\ell,t},x_{\ell,t}$ obey
\eqref{eq:conv_model2}, \eqref{eq:delay_prob2}, because we have just rescaled
the underlying counts here by a constant (in order to put them on the scale of
rates), and recall, we assume that all locations have the same delay
distribution (Assumption~\ref{asm:geo_invar}).

Given the delay distribution estimates from
Algorithm~\ref{alg:delay_dist_retro}, \smash{$\hp_t = (\hp_t(1), \ldots, 
  \hp_t(d))$} for $t \in \T$, we estimate the full vector \smash{$x_\ell =
  (x_{\ell,t})_{t \in \T}$} of latent infection rates across time, separately
for each location $\ell$, by solving the problem:   
\begin{multline}
\label{eq:tf_retro}
\minimize_{x_\ell} \; \sum_{t \in \T} \bigg( y_{\ell,t} - \sum_{k=1}^d \hp_t(k)
\, x_{\ell, t-k} \bigg)^2 +{} \\ \lambda \|D^{(4)} x_\ell\|_1,   
\end{multline}
where $D^{(4)}$ is a matrix such that $D^{(4)} v$ gives all 4$\th$-order
differences of a vector $v$, and $\|\cdot\|_1$ is the $\ell_1$ norm. Problem 
\eqref{eq:tf_retro} could be called a trend-filtering-regularized least squares 
deconvolution problem. We solve it (as well as all related optimization problems
in this paper) numerically with an adaption of the ADMM algorithm of 
\citet{ramdas2016fast}, detailed in Appendix~\ref{app:tf_admm}. 

The solution \smash{$\hx_\ell$} in problem \eqref{eq:tf_retro} takes the form of
a cubic piecewise polynomial (discrete spline) with adaptively chosen
knots \citep{tibshirani2014adaptive, tibshirani2020divided}. The tuning
parameter $\lambda \geq 0$ controls its complexity, and we choose it using  
3-fold cross-validation: we hold out every third value from training, and impute
it by the average of the neighboring trained estimates; to compute the
validation error, we reconvolve the full vector of imputed infections and
measure against observed cases. 
 
\section{Real-Time Deconvolution}
\label{sec:deconv_realtime}

Real-time deconvolution refers to the the task of deconvolving case reports
observed up until time $t$ to estimate latent infections up until $t$,
repeatedly, as $t$ marches over the period of interest. We are particularly
focused on estimating recent latent infections---nowcasting at a $k$-day lag,
which means estimating at $t$ the latent infection rate at time $t-k$.

Compared to retrospective deconvolution, real-time deconvolution differs in two
important ways. The first is that we are forced to work with provisional case
data, subject to revision at times in the future, as discussed earlier in
Section~\ref{sec:problem_setup}. All of our experiments in what follows use
properly-versioned data that would have been available as of the nowcast date.
We use the notation \smash{$y^{(t)}_{\ell,s}$} to reflect the reported case rate
in location $\ell$ at time $s$ as of time $t$. Reported case data from JHU is
available at a 1-day lag and therefore, as of time $t$, we only observe
\smash{$y^{(t)}_{\ell,s}$} up through $s=t-1$ (we use analogous superscript
notation for all auxiliary signals and estimates). This means we can only
produce deconvolution estimates \smash{$\hx^{(t)}_{\ell,s}$} up through $s=t-2$ 
(recall we exclude zero delays, in Assumption~\ref{asm:zero_at_zero}). 

\begin{figure}[tb]
\centering	
\includegraphics[width=0.95\linewidth]{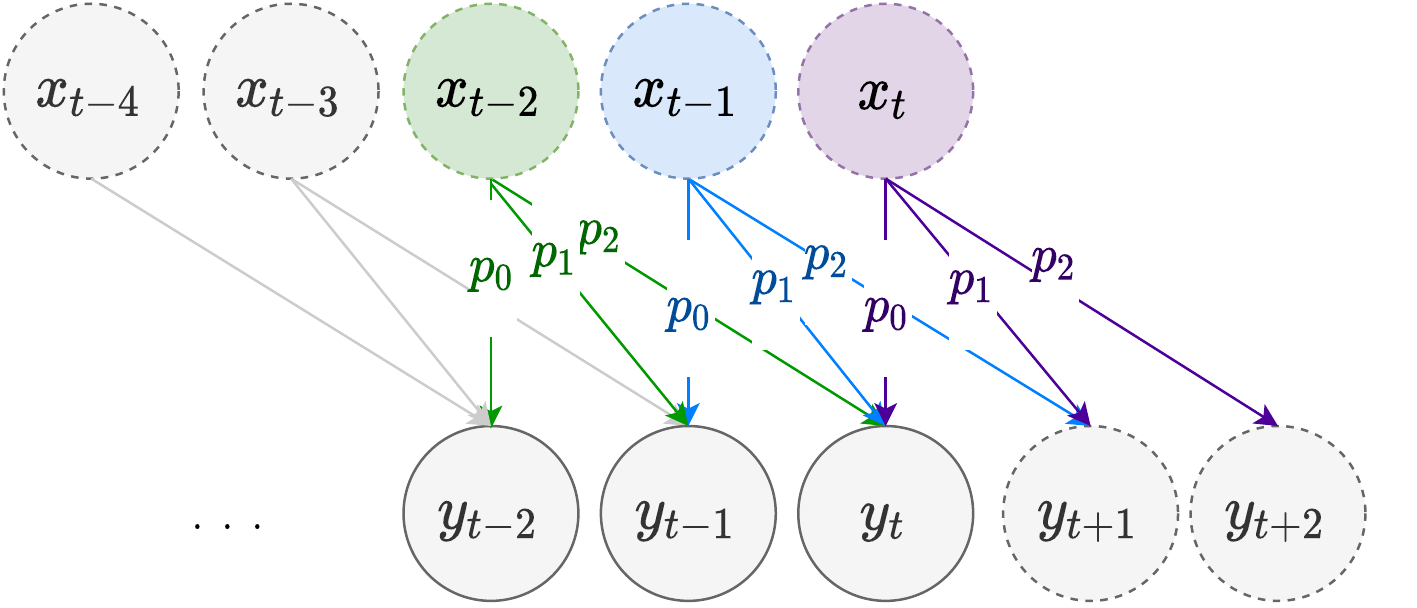}
\caption{Illustration of right truncation with a delay distribution of length 3
  (which is taken to be stationary for simplicity). At the nowcast time $t$,
  some ``part'' of the latent signal $x_t$ will appear in $y_{t+1},y_{t+2}$;
  likewise, some ``part'' of $x_{t-1}$ will appear in $y_{t+1}$.}   
\label{fig:right_truncation}
\end{figure}

The second issue of note, in real-time deconvolution, is \emph{right
  truncation}: in nowcasting at lag $k$, where $k$ is small (compared to $d$),
we are only able to carry out a ``partial'' deconvolution, as much of the needed  
information would come from case reports occurring in the future, past time the
nowcast date $t$. Figure~\ref{fig:right_truncation} gives an illustration. Thus,
if we simply performed real-time deconvolution by solving the problem analogous
to \eqref{eq:tf_retro}, using data that would have been available at time $t$,   
\begin{multline}
\label{eq:tf_realtime1}
\minimize_{x^{(t)}_\ell} \; \sum_{s < t} \bigg( y^{(t)}_{\ell,s} -
\sum_{k=1}^d \hp^{(t)}_s(k) \, x^{(t)}_{\ell, s-k} \bigg)^2 +{} \\  
\lambda \big\|D^{(4)} x^{(t)}_\ell\big\|_1,
\end{multline}
then we would find that the solution \smash{$\hx^{(t)}_\ell = (\hx^{(t)}_{\ell,s}
  : s < t)$} has highly volatile components for $s$ close to $t$.

The problem does not stop there; the truncation of data after the nowcast time
$t$ also affects estimation of the delay distribution itself. Most rows in the
line list with an onset date of $s=t-k$, for small $k$, will only have a report
date (and thus not appear in the line list) until after time $t$. This means
that the estimate \smash{$\hp^{(t)}_s$} of $p_s$ given by the empirical
distribution of all available line list data, with report date less than $t$,
will be biased toward smaller lag values (i.e., it will place too little weight
on larger lag values).    

In the next two subsections, we work through each of these truncation issues in
turn, by incorporating extra regularization around the right boundary into the  
criterion in \eqref{eq:tf_realtime1}, and estimating the delay distribution
from truncated data using a Kaplan-Meier-like approach. 

\subsection{Incorporating Extra Regularization}
\label{sec:extra_regularization}

We consider two forms of extra regularization to dampen the variability of trend
filtering estimates toward the right boundary. 

\smallskip
\paragraph*{Natural Trend Filtering.}

A natural cubic spline places additional regularity on top of the cubic spline,
by maintaining that the function be linear beyond the left and right boundary
points of the underlying domain. Natural trend filtering proceeds in a similar
vein, but operating in the space of discrete splines; see
\citet{tibshirani2020divided}. Transporting this idea over to our real-time  
deconvolution problem \eqref{eq:tf_realtime1}, and applying it to the right
boundary only, gives:
\begin{equation}
\label{eq:tf_realtime2}
\begin{alignedat}{2}
&\minimize_{x^{(t)}_\ell} \; && \sum_{s < t} \bigg( y^{(t)}_{\ell,s} -
\sum_{k=1}^d \hp^{(t)}_s(k) \, x^{(t)}_{\ell, s-k} \bigg)^2 +{} \\  
& && \hspace{100pt} \hfill
\lambda \big\|D^{(4)}  x^{(t)}_\ell \big\|_1 \\
&\subjectto && \;\; x^{(\ell)}_t - 2x^{(\ell)}_{t-1} + x^{(\ell)}_{t-2} = 0.
\end{alignedat}
\end{equation}

\begin{figure*}[tb]
\centering
\includegraphics[width=.315\linewidth]{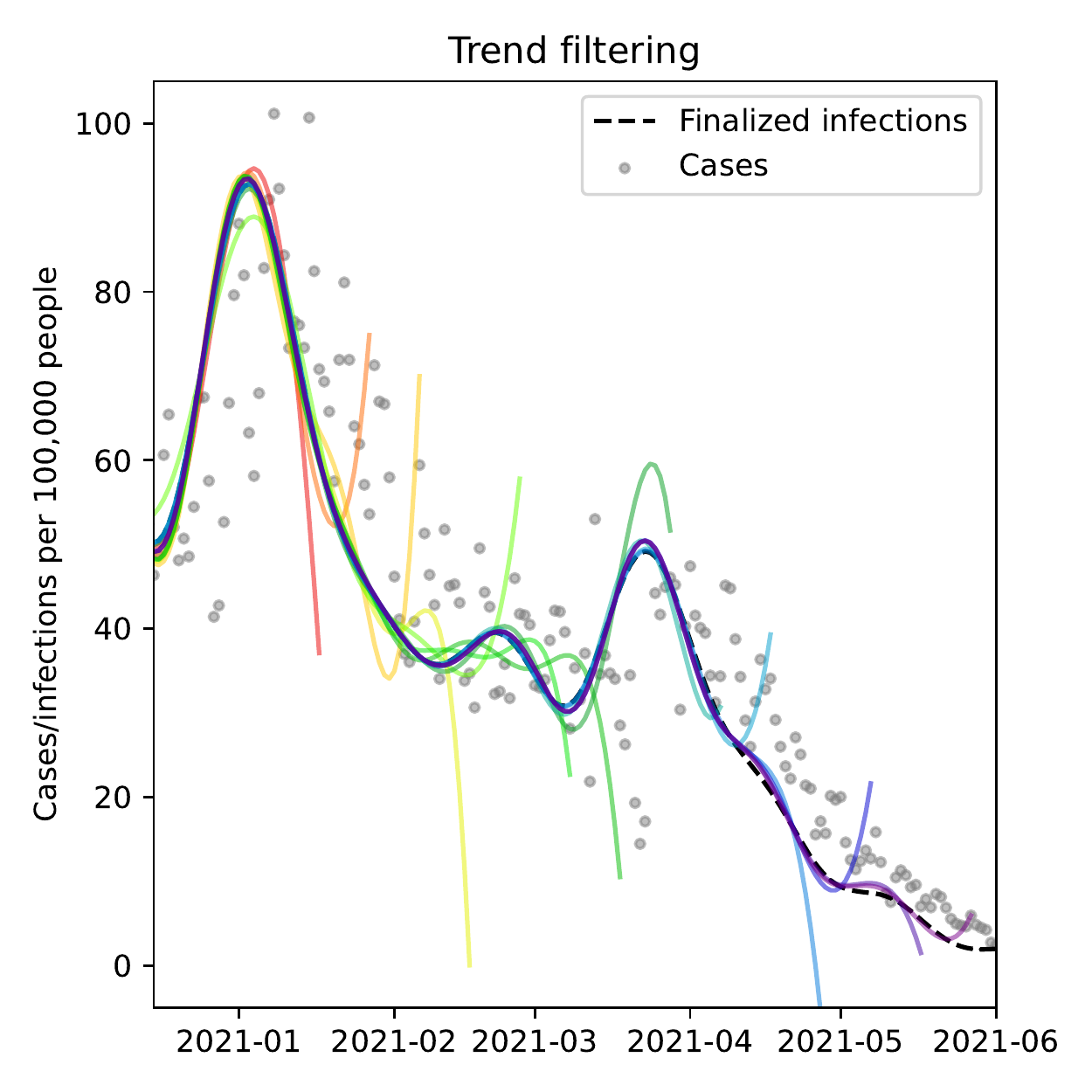}
\includegraphics[width=.315\linewidth]{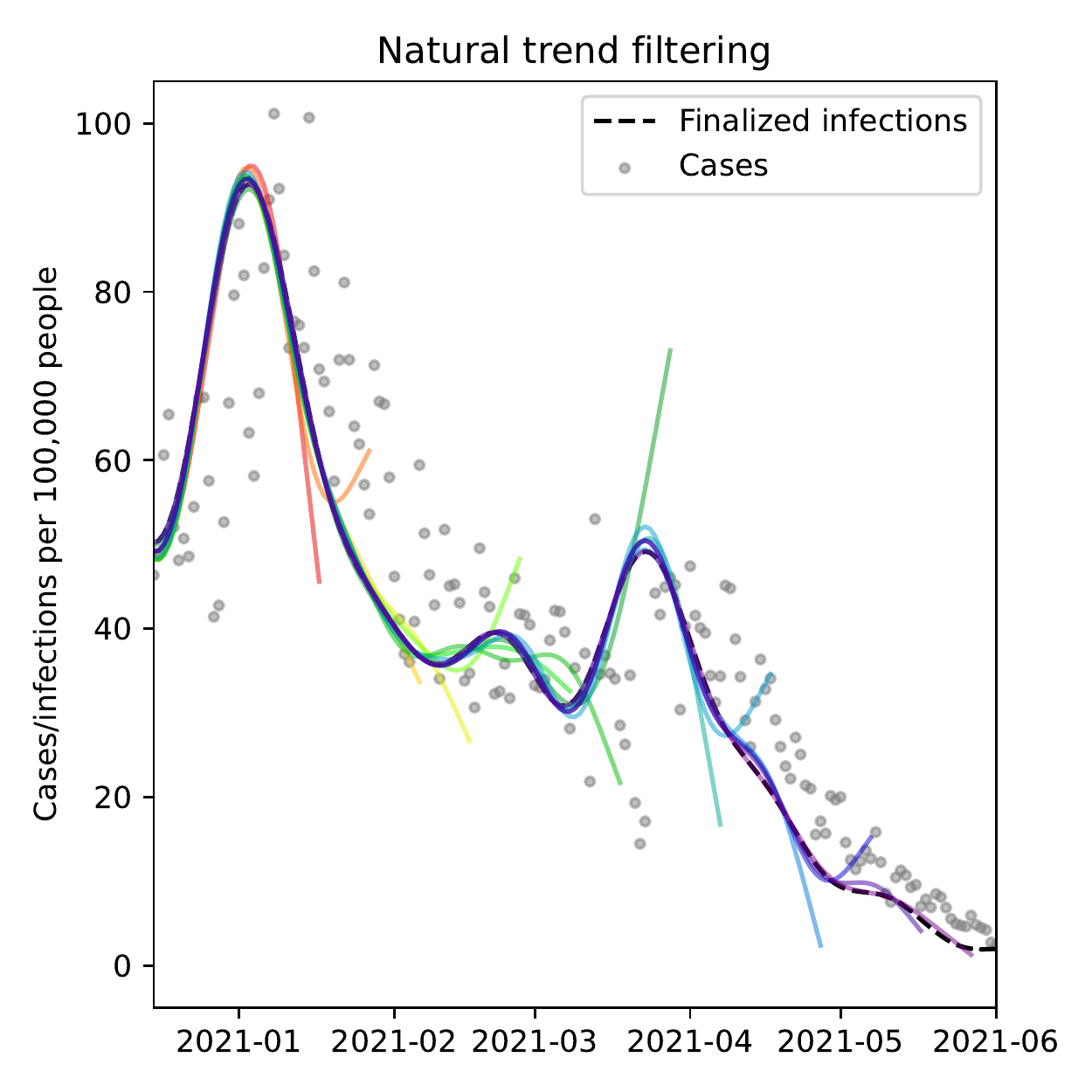}
\includegraphics[width=.315\linewidth]{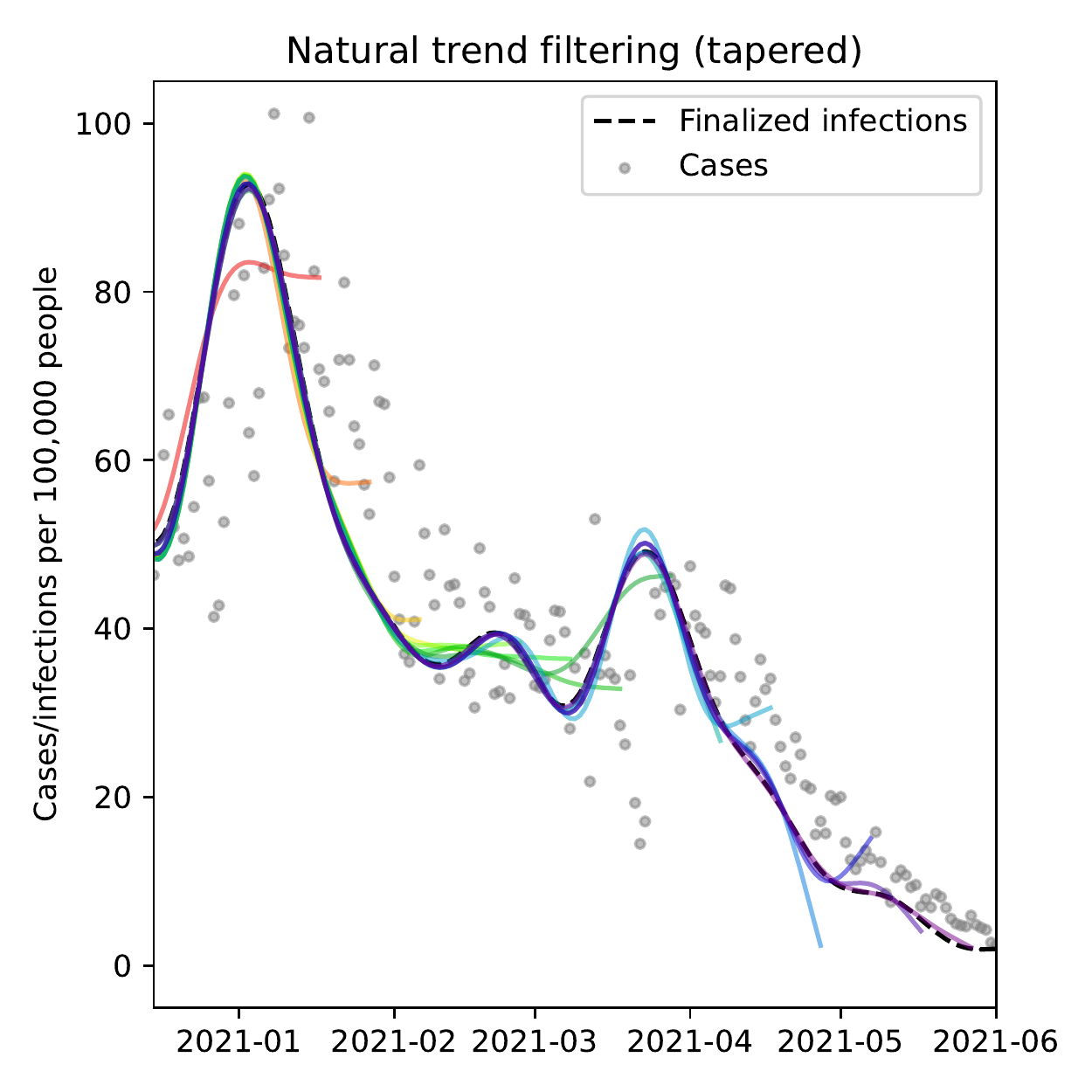}
\caption{Comparison of boundary behavior for real-time deconvolution in New 
  York, displayed for a sample of different nowcast dates (where each colored
  curve traces out the deconvolution estimates for a different nowcast
  date). The black dashed line indicates finalized infections, estimated roughly
  three months after June 1, 2021.}  
\label{fig:boundary_comparison}
\end{figure*}

The left and middle panels of Figure~\ref{fig:boundary_comparison} demonstrate
the improvement that the additional constraints in \eqref{eq:tf_realtime2} can
have on the boundary estimates, particularly during periods of dynamic change in
the underlying case trajectories.

\smallskip
\paragraph*{Tapered Smoothing.}

The right truncation phenomenon is not a binary one and there is increasingly
less and less information available for deconvolution as we move the time index
$s$ up toward the nowcast date $t$. Therefore, we design a second penalty to add
to the criterion in \eqref{eq:tf_realtime2} to gradually increase the amount of
regularization accordingly:  
\begin{equation}
\label{eq:tf_realtime3}
\begin{alignedat}{2}
&\minimize_{x^{(t)}_\ell} \; && \sum_{s < t} \bigg( y^{(t)}_{\ell,s} -
\sum_{k=1}^d \hp^{(t)}_s(k) \, x^{(t)}_{\ell, s-k} \bigg)^2 +{} \\  
& && \hspace{25pt} \hfill
\lambda \big\|D^{(4)} x^{(t)}_\ell \big\|_1 +
\gamma \big\|W^{(t)} D^{(1)} x^{(t)}_\ell \big\|_2^2 \\ 
&\subjectto && \;\; x^{(\ell)}_t - 2x^{(\ell)}_{t-1} + x^{(\ell)}_{t-2} = 0, 
\end{alignedat}
\end{equation}
where $D^{(1)} v$ gives the first-order differences of a vector $v$, and 
$W^{(t)}$ is a diagonal matrix that is supported on the last $d$ diagonal
entries, these being (in reverse order, starting with the last entry): 
$$
\frac{1}{\sqrt{\hF^{(t)}_{t-1}(k)}}, \;\; k=1,\dots,d, 
$$
where \smash{$\hF^{(t)}_{t-1}$} is the cumulative distribution function (CDF) 
corresponding to the estimated delay distribution \smash{$\hp^{(t)}_{t-1}$} at
the most recent time $t-1$. The parameter $\gamma \geq 0$ controls the
strength of the additional ``tapered'' penalty in \eqref{eq:tf_realtime3}, and
we tune $\lambda,\gamma$ with a two-stage cross-validation procedure: 
\begin{enumerate}
\item fix $\gamma =0$, and tune $\lambda$ using 3-fold cross-validation,
  as before;    
\item fix $\lambda$ at the value in Step 1, and tune $\gamma$ using 
  7-fold forward-validation: for $s= t-2, \ldots, t-8$, we solve the
  deconvolution problem with a working nowcast date of $s$, linearly extrapolate
  to impute an estimate at $s+1$, and then we reconvolve the solution vector
  along with this imputed point and measure error against observed cases at time
  $s+1$; the validation error is obtained by averaging these errors over the
  iterations $s =  t-2, \ldots, t-8$. 
\end{enumerate}

\begin{figure*}[tb]
\centering
\includegraphics[width=0.95\linewidth]{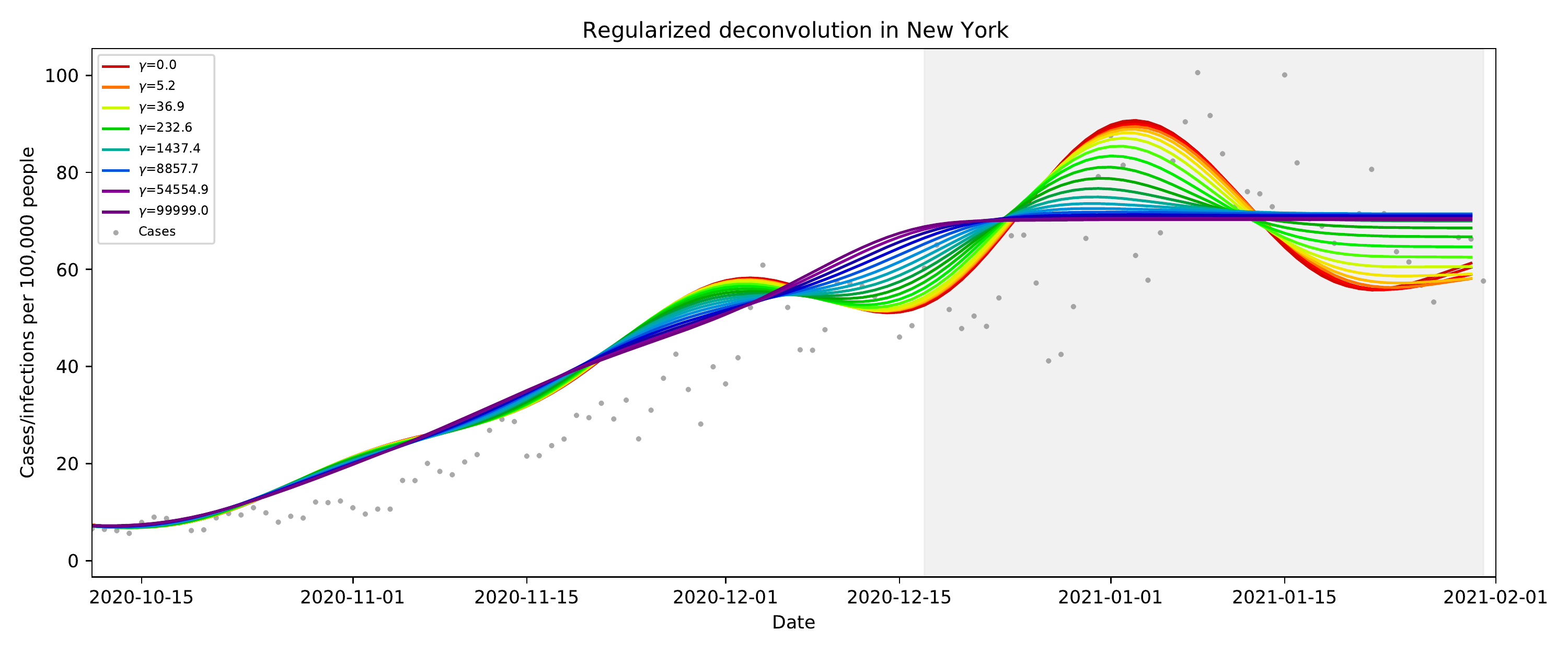}
\caption{Effect of the tapered smoothing penalty, as we vary the 
corresponding tuning parameter $\gamma$, for a single real-time 
deconvolution example with on nowcast date February 1, 2021. The gray 
region highlights the  components on which the tapered smoothing penalty 
acts.}
\label{fig:tapered_smooth}
\end{figure*}

Figure~\ref{fig:tapered_smooth} displays the effect of varying $\gamma$ on the
solution in \eqref{eq:tf_realtime3}, for a particular deconvolution example, to
give a qualitative sense of the role of the tapered penalty. Furthermore, the
right panel in Figure~\ref{fig:boundary_comparison} demonstrates the benefit
this penalty can provide in nowcasting.

\begin{figure}[tb]
\centering
\includegraphics[width=0.85\linewidth]{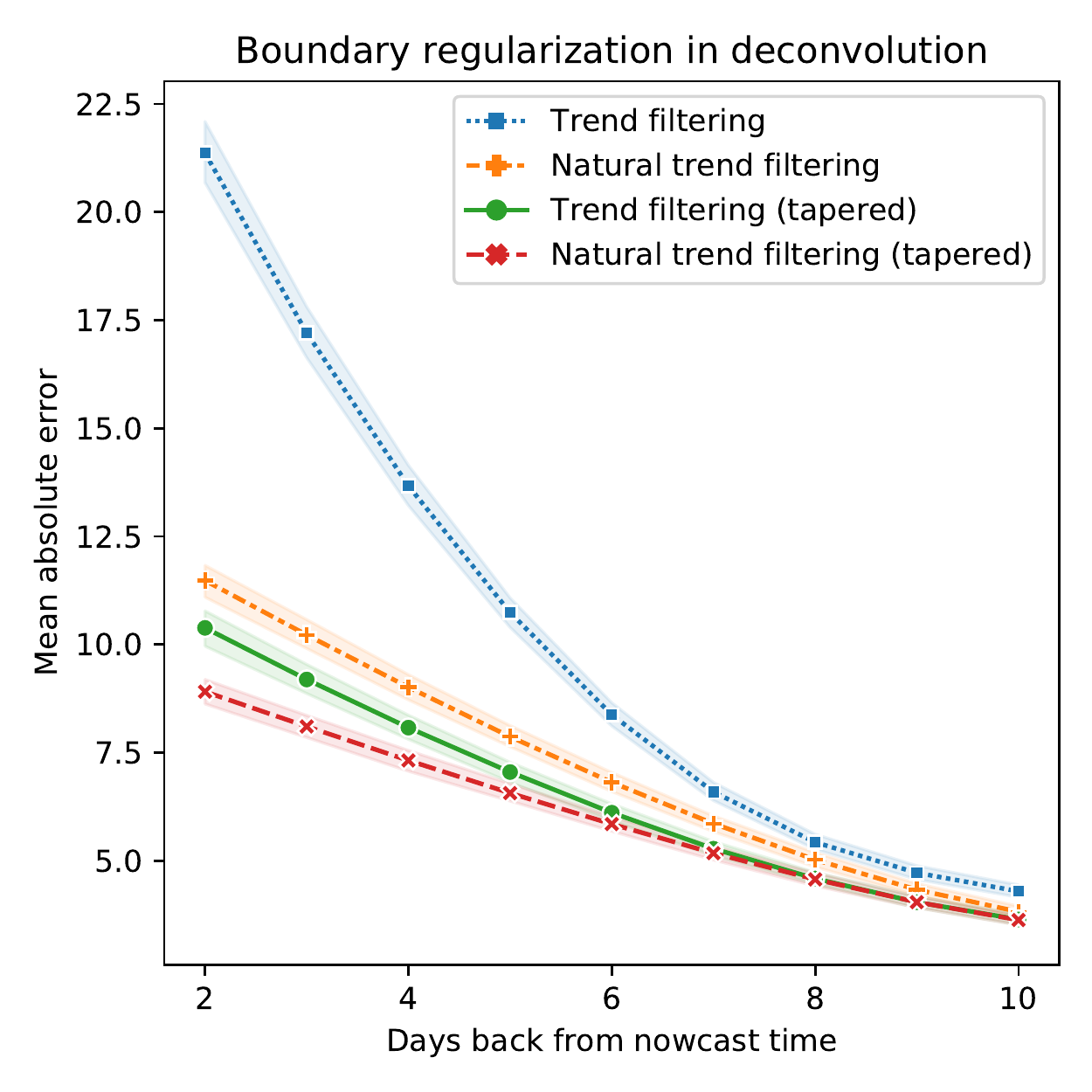}
\caption{Comparing regularization approaches by MAE for nowcasting (the shaded
  bands here and henceforth, in all MAE figures, correspond to 95\% bootstrap
  confidence intervals.) Both approaches for additional regularization give a
  huge improvement on trend filtering. The biggest improvement comes from
  combining the two approaches.}
	\label{fig:extra_reg}
\end{figure}

Lastly, and importantly, Figure~\ref{fig:extra_reg} quantifies the improvement
offered by the additional regularization mechanisms, in terms of mean absolute
error (MAE) measured against finalized infections in nowcasting at a $k$-day
lag, for each $k=2,\ldots,10$. This is averaged over all locations and every
10th nowcast date in the evaluation set. We see a considerable improvement in 
both the natural trend filtering and tapered smoothing modifications, with the
biggest improvement occurring when the two are combined as in
\eqref{eq:tf_realtime3}, and hence we stick with this framework in what follows. 

\subsection{Adjusting the Delay Distribution for Truncation}   
\label{sec:delay_adjust}

Now we propose an iterative adjustment to the empirical distribution of
truncated line list data in order to overcome the truncation bias. To develop
intuition, we first describe the problem using a simple abstraction, formulate a
general solution, and then we translate this back over to our particular
setting.   

\smallskip
\paragraph*{KM-Adjustment Under Truncation.}

Suppose $p$ is a distribution that is supported on $\{1,\ldots,d\}$, and we observe
independent random draws that we can partition into two sets: $\D_1$ and $\D_2$,
where $\D_2$ contains draws from $p$ and $\D_1$ contains draws from $p$
conditional on the random variable lying in $[1,z_1]$, for a fixed $z_1 \in
\{1,\ldots,d\}$. Denote by \smash{$\hp_\D$} the empirical distribution based on
a data set $\D$. Clearly \smash{$\hp_{\D_2}$} is unbiased for $p$, but
\smash{$\hp_{\D_1}$} is generally biased (it always places zero mass above
$z_1$), and thus the pooled estimate \smash{$\hp_{\D_1 \cup \D_2}$} would be
biased as well.

To build a more informed estimate based on the pooled sample, the intuition 
is as follows. First, observe that the only way we can estimate $p(k)$ for $k > 
z_1$ is by using $\D_2$. Then, this gives an estimate of \smash{$S(z_1) =
  \sum_{k >  z_1} p(k)$}, the survival function of $p$ at $z_1$, and we can
estimate $p(k)$ for $k \leq z_1$, denoting $Z \sim p$, by observing that  
$$
p(k) = \P(Z = k \,|\, Z \leq z_1) (1-S(z_1)).
$$
where we estimate $\P(Z = k | Z \leq z_1)$ using the empirical distribution     
over the set $\D_1 \cup \D_2 \cap [1, z_1]$. In other words, we construct our   
distribution estimate \smash{$\bp$} using two steps:     
\begin{enumerate}
\item define $\bp(k) = \hp_{\D_2}(k)$ for $k > z_1$, and also $\bS(z_1) =
  \sum_{k >  z_1} \bp(k)$;
\item define $\bp(k) = \hp_{\D_0}(k) (1-\bS(z_1))$ for $k \leq z_1$, where we
  let $\D_0 = \D_1 \cup \D_2 \cap [1, z_1]$. 
\end{enumerate}

We can readily generalize the above to a setting in which we observe $N$ data 
sets, with varying levels of truncation:
\begin{equation}
\label{eq:data_seq_trunc}
\text{$\D_i$ contains draws $Z \sim p \,| Z \leq z_i$, $i=1,\ldots,N$},
\end{equation}
where $1 \leq z_1 < \cdots < z_N = d$, and we set $z_0=0$ for notational
simplicity. To construct an estimate of $p$ based on all the samples, we proceed 
iteratively as before: first we estimate $p(k)$ for $k > z_{N-1}$ based on the
data in $\D_N$, then we estimate $p(k)$ for $k \in (z_{N-2},z_{N-1}]$ based on
data in $\D_{N-1} \cup \D_2 \cap [1, z_{N-1}]$, and so on. 
Algorithm~\ref{alg:dist_seq_trunc} spells out the procedure in full. 

\begin{algorithm}[tb]
\caption{Distribution estimation under sequential truncation} 
\label{alg:dist_seq_trunc}
\DontPrintSemicolon
\KwInput{Data sets and truncation limits $\D_i$ and $z_i$, for $1,\ldots,N$, as 
  in \eqref{eq:data_seq_trunc}.}    
\KwOutput{Estimated probabilities \smash{$\bp(1),\ldots,\bp(d)$}.}

Initialize $\bS(d) = 0$. 

\For{$i = N, \ldots, 1$} {
  Set $\D_0 = \bigcup_{j=i}^N D_j \cap [1, z_i]$.
  
  Compute $\bp(k)$, for $k \in (z_{i-1}, z_i]$ based on the empirical
  distribution of data in $\D_0$ and an estimate of the survival function at
  $z_i$:     
  $$
  \bp(k) = \hp_{\D_0}(k) (1-\bS(z_i)), \;\; k \in (z_{i-1}, z_i].
  $$
  
  Compute an estimate of the survival function at $z_{i-1}$: 
  $$
  \bS(z_{i-1}) = \bS(z_i) + \sum_{k \in (z_{i-1}, z_i]} \bp(k). 
  $$
}

Return \smash{$\bp(1),\ldots,\bp(d)$}. 
\end{algorithm}

The algorithm just derived may be seen as Kaplan-Meier-like, in the sense that 
it is motivated by the decomposition
$$
p(k) = \P(Z=k \,|\, Z \leq z_i) (1-S(z_i)), \;\; k \in (z_{i-1}, z_i]. 
$$
We use an unbiased plug-in estimate for each term in the product above  
based on the appropriate data. The Kaplan-Meier estimator has a similar plug-in 
foundation \citep{kaplan1958nonparametric}, so we refer to our approach as the  
\emph{KM-adjusted estimator} of the distribution under truncation. 

\smallskip
\paragraph*{Application to CDC Line List.}

Porting the last idea over to the CDC line list, we can use it to estimate the
delay distribution at time $s$ using the line list as of time $t$. Note that if
$s < t-d$ then we can still use Algorithm~\ref{alg:delay_dist_retro}, as
there is no truncation issue whatsoever. However, if $s \geq t-d$, then we would 
need to apply the KM-adjusted estimator, because we would be using the rows in
the line list whose onset date is at or shortly before $s$, but are only able to
see those whose report date is at most $t-1$ (thus would have been available 
at time $t$). After making this adjustment to the empirical distribution, we
apply gamma smoothing as before. This is detailed in
Algorithm~\ref{alg:delay_dist_realtime}. 

\begin{algorithm}[tb]
\caption{Delay distribution estimation in real-time}
\label{alg:delay_dist_realtime}
\DontPrintSemicolon
\KwInput{Nowcast time $t$, working onset time $s$, support size $d$, window size 
  $w=2d$, truncated line list \smash{$\D^{(t)}$} with onset dates $a_i$ and report
  dates $b_i$ such that $b_i < t$.}  
\KwOutput{Estimated delay probabilities \smash{$\hp^{(t)}_s(1), \ldots,
    \hp^{(t)}_s(d)$}.}   

\If{$s < t-d$} {
  Return probability estimates from Algorithm~\ref{alg:delay_dist_retro}
  (setting $t=s$ and \smash{$\D=\D^{(t)}$} in the notation of that algorithm).} 

Set $N=d-(t-s)+2$. 

\For{$i = 1, \ldots, N-1$} {
  Define 
  \begin{align*}
  \D_i &= \{ b_i - a_i :  a_i = s-i+1 \} \\
    z_i &= t-s+i-2. 
  \end{align*}
}

Define $\D_N = \{ b_i - a_i :  a_i \in (s-w, t-d) \}$ and $z_N = d$.

Use Algorithm~\ref{alg:dist_seq_trunc} (applied to $\D_i$, $z_i$,
$i=1,\ldots,N$) to compute probability estimates \smash{$\bp_t(1), \ldots, 
  \bp_t(d)$}. 

Fit a gamma density to \smash{$\bp_t(1), \ldots, \bp_t(d)$} using the method
of moments (matching the mean and variance). 

Discretize this gamma density to the support set $\{1,\ldots,d\}$, call the 
result \smash{$\hp_t(1), \ldots, \hp_t(d)$}, and return these probabilities. 
\end{algorithm}

\begin{figure}[tb]
\centering
\includegraphics[width=0.85\linewidth]{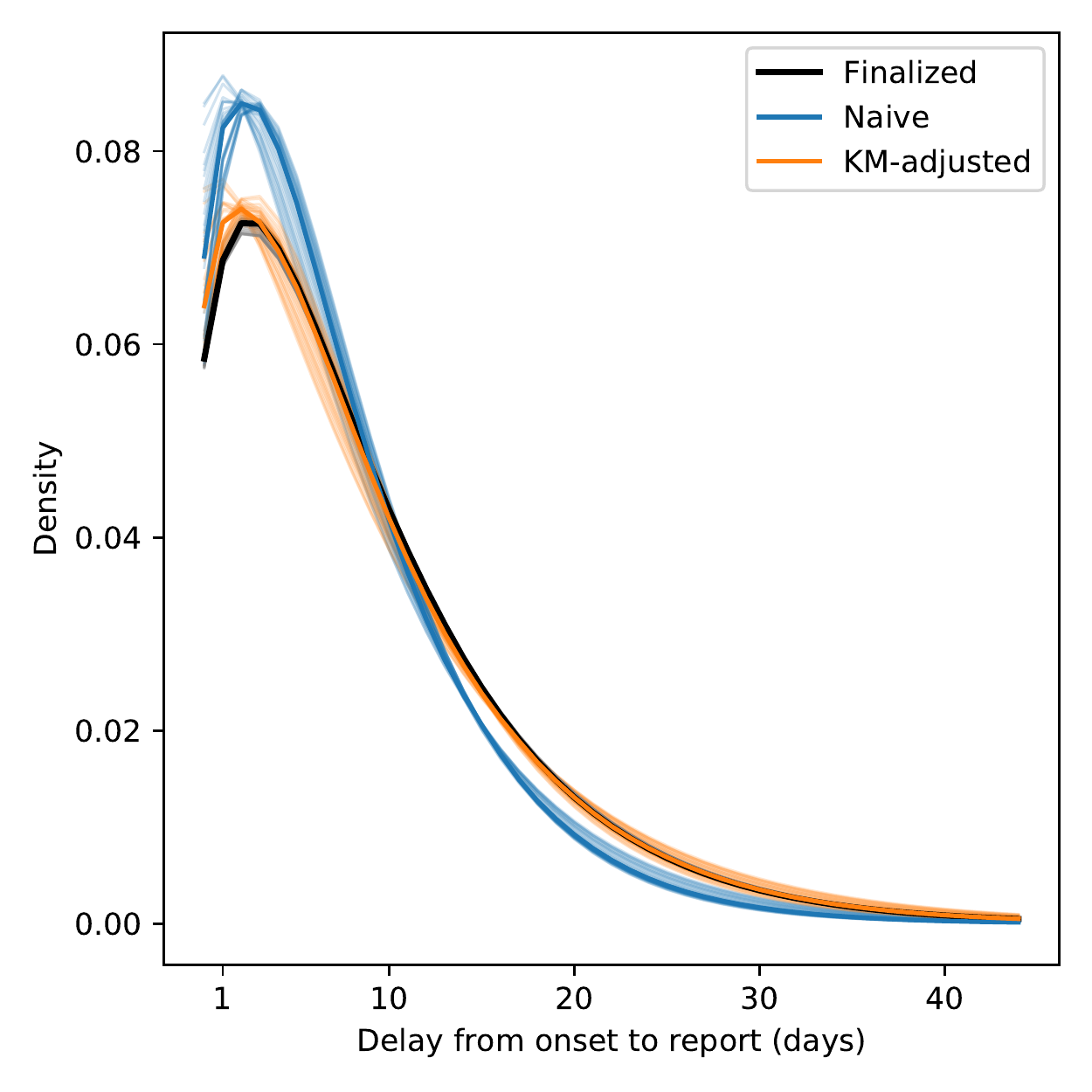}
\includegraphics[width=0.85\linewidth]{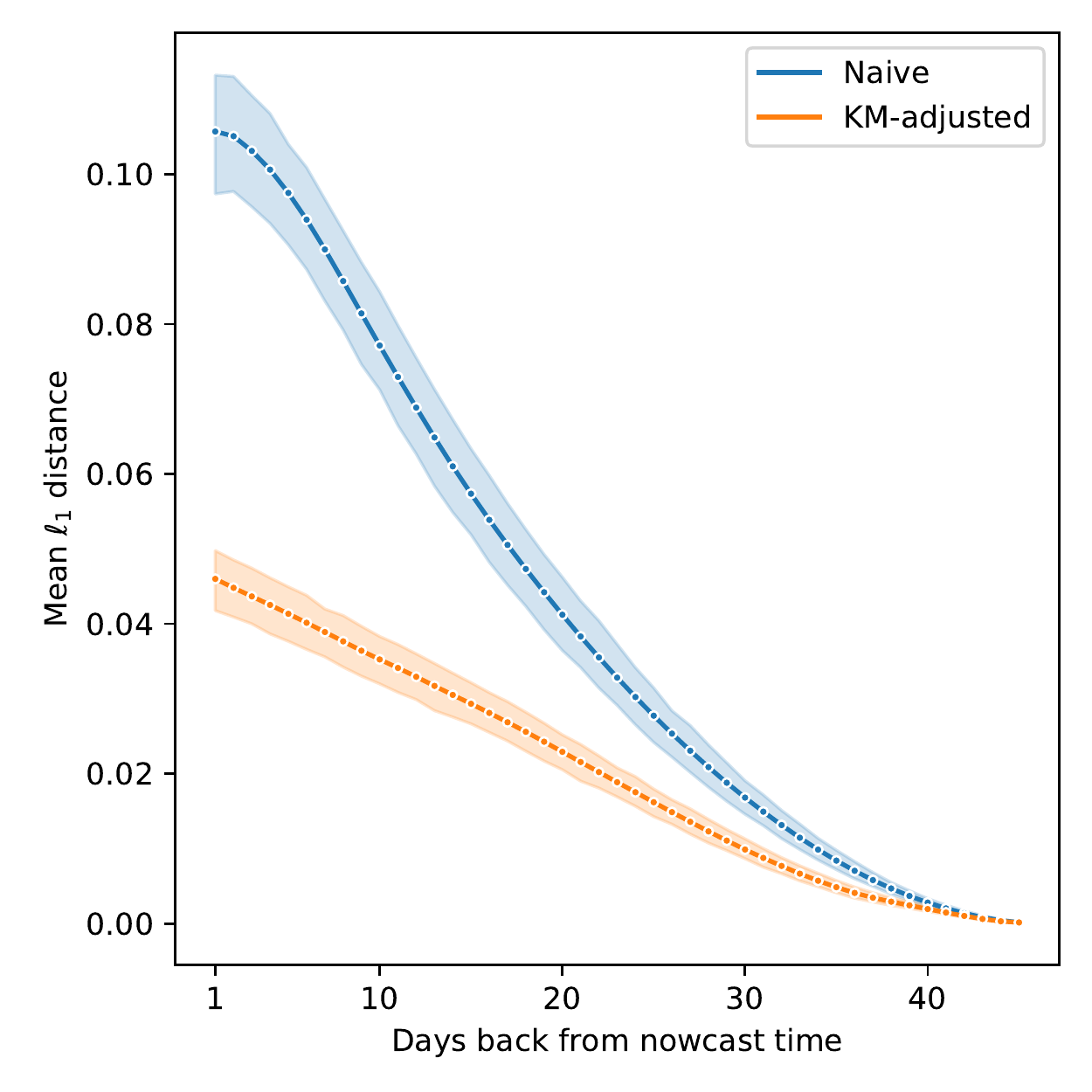}
\caption{Top: estimated delay distributions overlaid for all nowcast dates in
  the month of November 2020, when $s=t-1$ (working onset date one day 
  before the nowcast date). Bottom: mean $\ell_1$ distance to finalized
  estimate of the delay distribution, as a function of the lag $k=t-s$.}     
\label{fig:delay_dist_realtime}
\end{figure}

Figure~\ref{fig:delay_dist_realtime} compares the KM-adjusted and naive
estimates of the delay distribution, Algorithm~\ref{alg:delay_dist_realtime}
versus Algorithm~\ref{alg:delay_dist_retro} applied directly to
\smash{$\D^{(t)}$}, the line list available at each nowcast date $t$. In terms
of $\ell_1$ distance, measured to the finalized delay distribution estimate
computed retrospectively (based on the full untruncated line list), and averaged
over all  nowcast dates in the evaluation period, we see that the KM-adjustment
greatly improves the accuracy at all lags $k=2,\ldots,10$ (where $k=t-s$, the 
difference between the nowcast and working onset dates). 

\subsection{Shortening the Deconvolution Window}

Lastly, we investigate shortening the window used in the regularized
deconvolution problem \eqref{eq:tf_realtime3} so that we use only a window 
length of $w$ days before $t$:
\begin{equation}
\label{eq:tf_realtime4}
\begin{alignedat}{2}
&\minimize_{x^{(t)}_\ell} \; && \sum_{s \in [t-w, t)} \bigg( y^{(t)}_{\ell,s} - 
\sum_{k=1}^d \hp^{(t)}_s(k) \, x^{(t)}_{\ell, s-k} \bigg)^2 +{} \\  
& && \hspace{25pt} \hfill
\lambda \big\|D^{(4)} x^{(t)}_\ell \big\|_1 +
\gamma \big\|W^{(t)} D^{(1)} x^{(t)}_\ell \big\|_2^2 \\ 
&\subjectto && \;\; x^{(\ell)}_t - 2x^{(\ell)}_{t-1} + x^{(\ell)}_{t-2} = 0, 
\end{alignedat}
\end{equation}
As we are mainly interested in the components of the solution
\smash{$\hx^{(t)}_s$} for $s$ close to $t$, shortening the training window is
computationally advantageous and should not change the behavior of the solution
very much for $s$ close to $t$.

\begin{figure}[tb]
\centering
\includegraphics[width=0.85\linewidth]{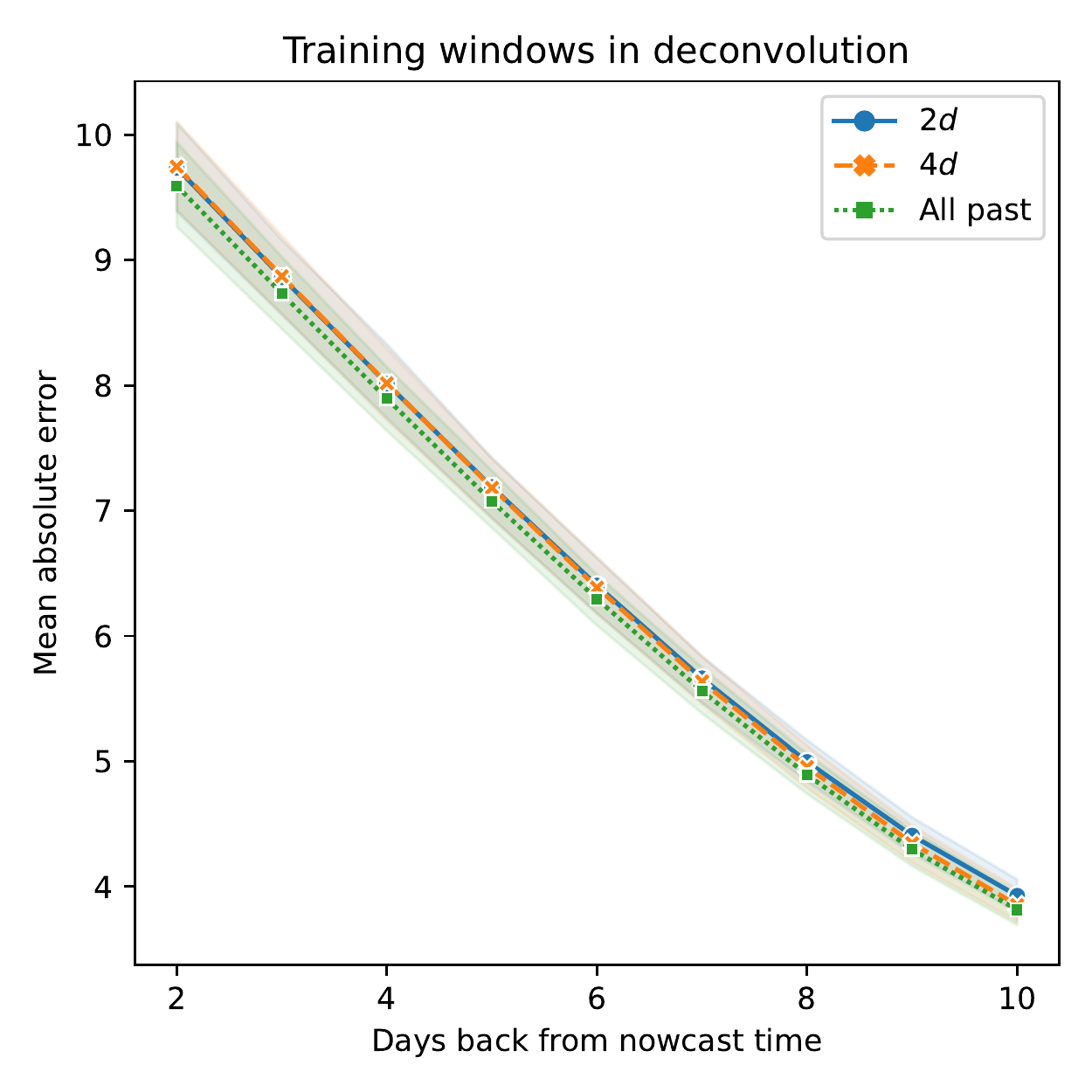}
\caption{Comparing window lengths used in regularized deconvolution by
  MAE for nowcasting. The performance is very similar throughout.}
\label{fig:deconvolution_window}
\end{figure}

Figure~\ref{fig:deconvolution_window} compares \eqref{eq:tf_realtime4} with
$w=2d$, $w=4d$, and ``all-past'', which is the original problem
\eqref{eq:tf_realtime3}, in terms of mean absolute error (MAE) measured against
finalized infections in nowcasting at a $k$-day lag, for each
$k=2,\ldots,10$. This is averaged over all locations and every 10th nowcasting
date in the evaluation set. The performance is basically identical for window
lengths $2d$ and $4d$, and though all-past may appear to have the slightest 
advantage, this does not warrant the extra computation, hence in what follows
we stick to \eqref{eq:tf_realtime4} with a window length $w=2d$ as our
real-time deconvolution estimator.    

\section{Leveraging Auxiliary Signals}
\label{sec:leverage_aux}

The indicators enumerated in Section~\ref{sec:preliminaries} have displayed
impressive correlations to reported COVID-19 cases \citep{Reinhart:2021}, and
moreover, demonstrated an ability to improve the accuracy of case forecasting
and hotspot prediction models \citep{McDonald:2021}. In this section, we
describe how to use each indicator to build a real-time \emph{sensor} that
estimates the latent infection rate, and how to fuse such estimates together
into a single nowcast.

\subsection{Sensor Models}
\label{sec:sensor_models}

At each prediction time $t$, for each location $\ell$, and for each of the five
indicators (abbreviated CHNG-COVID, CHNG-CLI, DV-CLI, CTIS-CLIIC, and
Google-AA), we will train a model to predict in real-time
latent infections from indicator values. Let \smash{$\hx^{(t)}_{\ell,s}$} denote
the solution at time $s$ in problem \eqref{eq:tf_realtime4}, which represents
our best estimate of the latent infection rate at time $s$ as of time $t$ from
deconvolution of case rates alone. 

We use \smash{$z^{i,(t)}_{\ell,s}$} to denote the value of indicator $i$ at time
$s$ and location $\ell$, as of time $t$. We fit a simple linear model to predict 
latent infections from indicator values by solving 
\begin{equation}
\label{eq:sensor_reg}
\minimize_{\beta_0, \beta_1} \; \sum_{s = t-d}^{t-\tilde{k}_i}
w^{(t)}_s \big( \hx^{(t)}_{\ell,s} - \beta_0 - \beta_1 z^{i,(t)}_{\ell,s} \big)^2,
\end{equation}
which is a weighted linear regression over the time period $[t-d,
t-\tilde{k}_i]$, where \smash{$\tilde{k}_i = \max\{k_i, 2\}$} and $k_i$ denotes
the lag at which indicator $i$ is available. This is:
\begin{itemize}
\item $k_i=1$ for CTIS-CLIIC and
  Google-AA\footnote{Our treatment of Google-AA is different from the 
  rest. Google's team did not start publishing this signal until September 2020,
  and the historical latency of this signal was sporadic, but was often longer
  than 1 week. However, unlike (say) the claims-based signals, revisions are
  never made after initial publication, and the latency of the signal is not an
  unavoidable property of the data type, and therefore we use finalized signal
  values, with a 1-day lag, in our analysis.}; and 
\item $k_i=4$ for the claims-based indicators, due to heavy revision or
  ``backfill'' over the first several days in the underlying claims data after
  an outpatient visit date \citep{Reinhart:2021}. 
\end{itemize}
Notice that, as defined, \smash{$\tilde{k}_i$} is the lag at which \emph{both}
the deconvolution estimate of infection rate and auxiliary signal $i$ are
available, which is the data we need to fit the linear sensor model (response
and covariate data, respectively). 

The observation weights in \eqref{eq:sensor_reg} are given by
$$
w^{(t)}_{t-k} = \hS^{(t)}_{t-1}(k-1), \;\; k=1,\ldots,d.
$$ 
Here \smash{$\hS^{(t)}_{t-1}$} is the survival function of
\smash{$\hp^{(t)}_{t-1}$}, the estimated delay distribution from the most recent
time point $t-1$. We define \smash{$\hS^{(t)}_{t-1}(1) = 1$}, corresponding to 
the exclusion of 0-day delays. This scheme upweights the more recent estimates
(responses in the regression) of latent infections as they contain more timely
information for nowcasting (assuming that the right-truncation bias has been
effectively mitigated in the deconvolution step).

Given the solution \smash{$\hbeta^{i,(t)}_{\ell,0}, \hbeta^{i,(t)}_{\ell,1}$} in
\eqref{eq:sensor_reg}, we then define a sensor---which is just a prediction from
the fitted linear model---based on indicator $i$, for time $s$ and location
$\ell$, as of time $t$, as:
\begin{equation}
\label{eq:sensor_def}
\bx^{i,(t)}_{\ell,s} = \hbeta^{i,(t)}_{\ell,0} + \hbeta^{i,(t)}_{\ell,1} \,
z^{i,(t)}_{\ell,s}. 
\end{equation}
This sensor is available up until $s=t-k_i$. For the CTIS-CLIIC and Google-AA
sensors, the lag is $k_i=1$, smaller than the inherent lag of 2 in the
deconvolution estimate. 

In brief, each sensor model takes a certain indicator and transforms it---using
a location-specific and time-varying mapping---to the scale of local infection
rates. While this mapping is simple (based on linear regression), it is also
highly nontrivial, as it inherently accounts for geographic biases and
nonstationarity.

Finally, in addition to defining sensors based on \eqref{eq:sensor_reg},
\eqref{eq:sensor_def} for each of the five auxiliary sensors, we also define a
sixth sensor based on a 3$\rd$ order autoregressive model trained on
\smash{$\hx^{(t)}_\ell = (\hx^{(t)}_{\ell,s} : s < t)$}. It is constructed 
exactly as in \eqref{eq:sensor_reg}, \eqref{eq:sensor_def} (same weights and
same training window). Henceforth we abbreviate it AR(3). 

\subsection{Sensor Missingness}

\begin{figure*}[tb]
\centering
\includegraphics[width=0.95\linewidth]{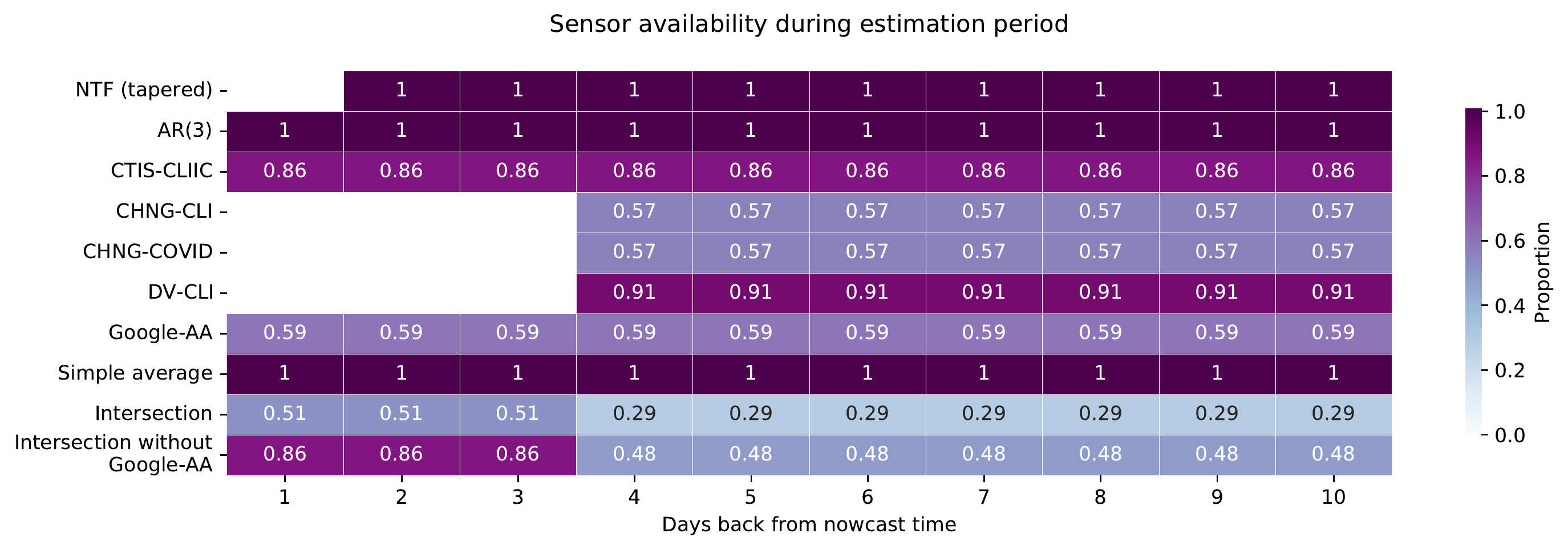}
\caption{Proportion of observed (non-missing) values over the evaluation 
  period from October 1, 2020 to June 1, 2021, and over all locations, as a 
  function of lag $k=1,\ldots,10$. (NTF refers to the real-time deconvolution
  estimator, and simple average refers to the sensor fusion method that 
  averages all available sensors.) The bottom two rows reflect the
  intersection of location-time pairs for which all data---deconvolution 
  estimates and sensors---are available for that given lag, with and
  without including the Google-AA sensor, since this sensor has a large amount
  of individual missingness. Each intersection at each given lag $k$ is restricted
  to data whose latency is not greater than $k$. For example, the bottom
  leftmost cell computes the porportions of locations and dates at which AR(3), 
  CTIS-CLIIC, and the simple average are concurrently available.} 
\label{fig:sensor_avail}
\end{figure*}

To be clear \eqref{eq:sensor_reg}, \eqref{eq:sensor_def} are to be implicitly
understood as performed over observed (non-missing) indicator values. If an
indicator value is missing at a particular location and time, then we drop it
from the training set in \eqref{eq:sensor_reg}, and do not produce a
corresponding sensor value in \eqref{eq:sensor_def}. For a summary of
missingness in the sensors, see Figure~\ref{fig:sensor_avail}.

In general, an indicator will be missing when there is insufficient underlying
data (from surveys, medical claims, etc.) to form a reliable signal value at a
given location and time. However, the situation is different for the Google-AA
indicator: here missingness occurs because the COVID-19 search trends data set
is released after using a differential privacy layer \citep{Bavadekar:2020}, and
a missing value means that the level of noise added for privacy protection is
high compared to the search count. Therefore we impute missing Google-AA signal
values by zeros in our analysis; we do this unless the Google-AA signal was
missing for a particular location and \emph{all} times in the evaluation period,
in which case we leave it as missing for this location entirely.

\subsection{Sensor Fusion}
\label{sec:sensor_fusion}

Sensor fusion (SF), broadly speaking, refers to the process of assimilating  
data sources, each of which ideally contains complementary information, in order
to produce more accurate estimates or predictions. SF falls into the general 
class of ensemble methods, and the sensors constructed in the previous section
can be thought of as base learners, to be subsequently combined.

We consider the following five ensemble methods. In each case, we describe how
to form the estimate at time $s$ and location $\ell$ as of time $t$. Though not
explicitly stated, it is to be implicitly understood that all sensor values are
as of time $t$ as well. 

\begin{enumerate}
\item Simple average: the average of available sensors at time $s$ and location 
  $\ell$. 
\item Simple regression: the prediction from a linear regression model at time
  $s$ and location $\ell$, fit to available sensors over the training period at
  location $\ell$. 
\item Ridge: the prediction from a ridge regression model at time $s$ and
  location $\ell$, fit to available sensors over the training period and over
  locations $j$ such that $j,\ell$ lie in the same U.S.\ state (including the
  state sensor itself).   
\item Lasso: same as in the last item, but using the lasso instead of ridge
  regression. 
\item KF-SF: the Kalman-Filter-inspired method for sensor fusion from
  \citet{Farrow:2016, Jahja:2019}, with covariance shrinkage, and
  operating on the geographical hierarchy within each U.S.\ state. 
\end{enumerate}
Methods 2--5 are trained on the most recent $2d$ time points, and 3--5 are tuned
using 7-fold forward validation, where we allow them to choose a lag-specific
tuning parameter. Methods 1--2 are ``simple'' in the sense that for nowcasting
at a location $\ell$ they use sensors from $\ell$ only. Methods 3--5 are more
sophisticated in that they pool information across locations within the same
state. 

The KF-SF method requires a proper geographical hierarchy and thus we create
``rest-of-state'' jurisdictions by aggregating the remaining counties (outside
of the top 200 counties nationally) within each state, and to run KF-SF, we
create an AR(3) sensor at these rest-of-state locations (since one sensor at
each location is sufficient). It is worth noting that, as shown in
\citet{Jahja:2019}, KF-SF bears a close connection to ridge in Model 4: it 
is in fact equivalent to a modified ridge optimization problem that imposes
additional linear constraints. 

\section{Evaluation}
\label{sec:evaluation}

We now evaluate nowcasting performance over all locations and all but every 10th
nowcasting date in our evaluation period from October 1, 2020 to June 1,
2021. (We do this because it gives us a ``pure'' test set, since every 10th
nowcasting date was already used to choose the real-time deconvolution
methodology in Section~\ref{sec:deconv_realtime}.) As before, we compare to
finalized estimates of infection rates computed via retrospective deconvolution,
as in Section~\ref{sec:deconv_retro}. 

For the purposes of making fair comparisons, in every analysis (figure) that we
present, we only aggregate over the intersection of nowcasts dates and locations
at which the particular estimates under consideration---coming from real-time 
deconvolution, individual sensor models, or sensor fusion---are all
available. Abiding by this rule leads us to examine several different ways of
stratifying results, as the full intersection is fairly sparse (see the
second-to-last row in Figure~\ref{fig:sensor_avail}). In particular, we
consider the following two dimensions used to define strata:\footnote{To be
  explicit, when we say we do not ``include'' certain sensors, it means both 
  that we ignore results from their individual sensor models (in computing the
  common intersection of available nowcast dates and locations), and \emph{also}
  that we exclude them in running the sensor fusion methods.}  
\begin{itemize}
\item inclusion of Google-AA or not; 
\item inclusion of all claims-based sensors (CHNG-CLI, CHNG-COVID, and DV-CLI)
  or not. 
\end{itemize}
In what follows, we first examine the performance of individual sensor models
and a certain sensor fusion method (the simple average) compared to real-time
deconvolution, and then examine the relative performance of the different sensor
fusion methods.  

\subsection{Performance of Sensors and Sensor Fusion}

We begin by comparing the MAE of nowcasts from natural trend filtering (NTF) 
using tapered smoothing, as in \eqref{eq:tf_realtime4} (the real-time
deconvolution estimator chosen based on the analysis in
Section~\ref{sec:deconv_realtime}) to those from individual sensor models and
the simple average sensor fusion method. Despite its simplicity, the simple
average appears to be the best-performing sensor fusion method overall (details
in the next subsection), and so we stick with it as the de facto sensor fusion
method in this subsection. The results here do not include Google-AA; results
including Google-AA are shown in Appendix~\ref{app:eval_plots}. 

Figure~\ref{fig:mae_all_no_google_aa} displays the MAE from various methods as
a function of lag $k$. The top and bottom panels do not and do include the 
claims-based sensors, respectively. In either case, we see that up until
lag 6, all sensors outperform the real-time deconvolution estimate from NTF. The
simple average of all sensors improves accuracy even further, and achieves the
best MAE for all lags up through lag 6. We recall that NTF (with tapered
smoothing) itself already provides a huge increase in accuracy over the more
naive method for real-time deconvolution given by applying trend filtering
without extra boundary regularization (Figure~\ref{fig:extra_reg}). 
At lag 7, the NTF estimate catches up to about equal accuracy, and then
surpasses sensor fusion and all sensors in accuracy at lag 8 and onward. An
interpretation for this: right truncation ceases to be a significant problem
past lag 7, and thus we are better off performing deconvolution directly in
order to estimate infections more than a week into the past.   

\begin{figure}[tb]
\centering 
\includegraphics[width=0.85\linewidth]{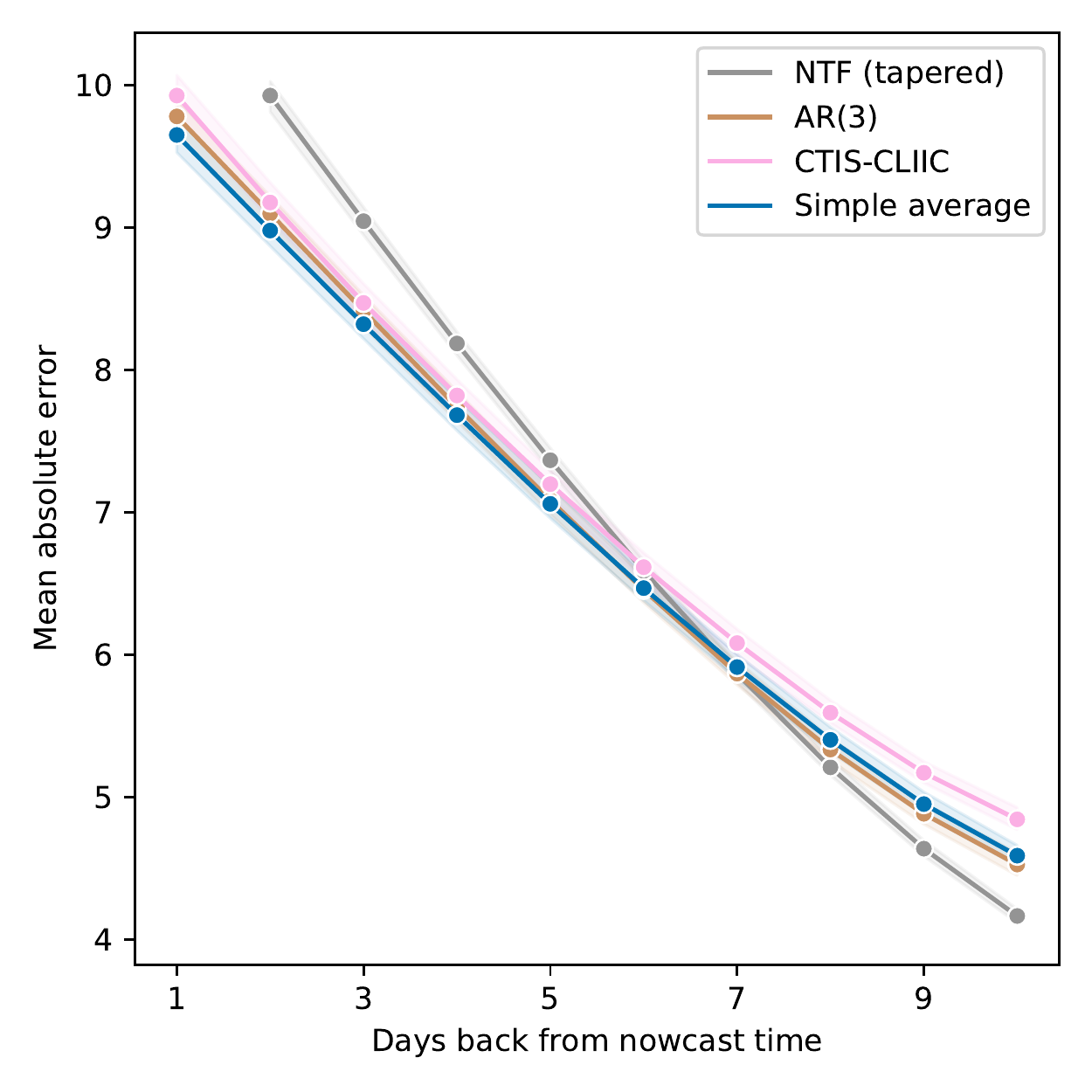} 
\includegraphics[width=0.85\linewidth]{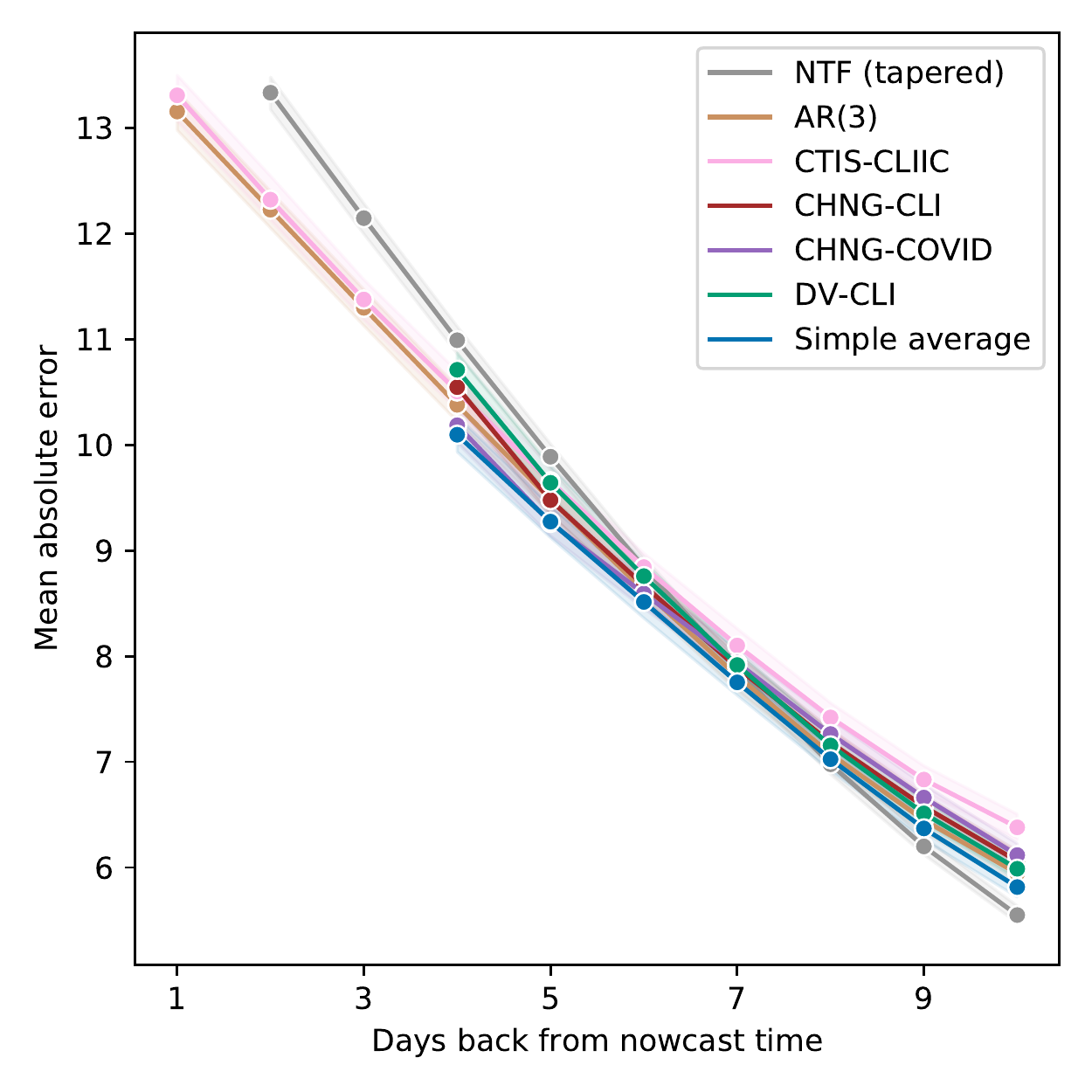}
\caption{Comparing NTF to individual sensor models and the simple average sensor
  fusion method by MAE for nowcasting. The top panel excludes the claims-based
  sensors, whereas the bottom includes them. For lags smaller than 7, all
  methods improve upon NTF (with tapered smoothing), with simple average being 
  the best among them.}    
\label{fig:mae_all_no_google_aa}
\end{figure}

\begin{figure}[tb]
\centering
\includegraphics[width=0.975\linewidth]{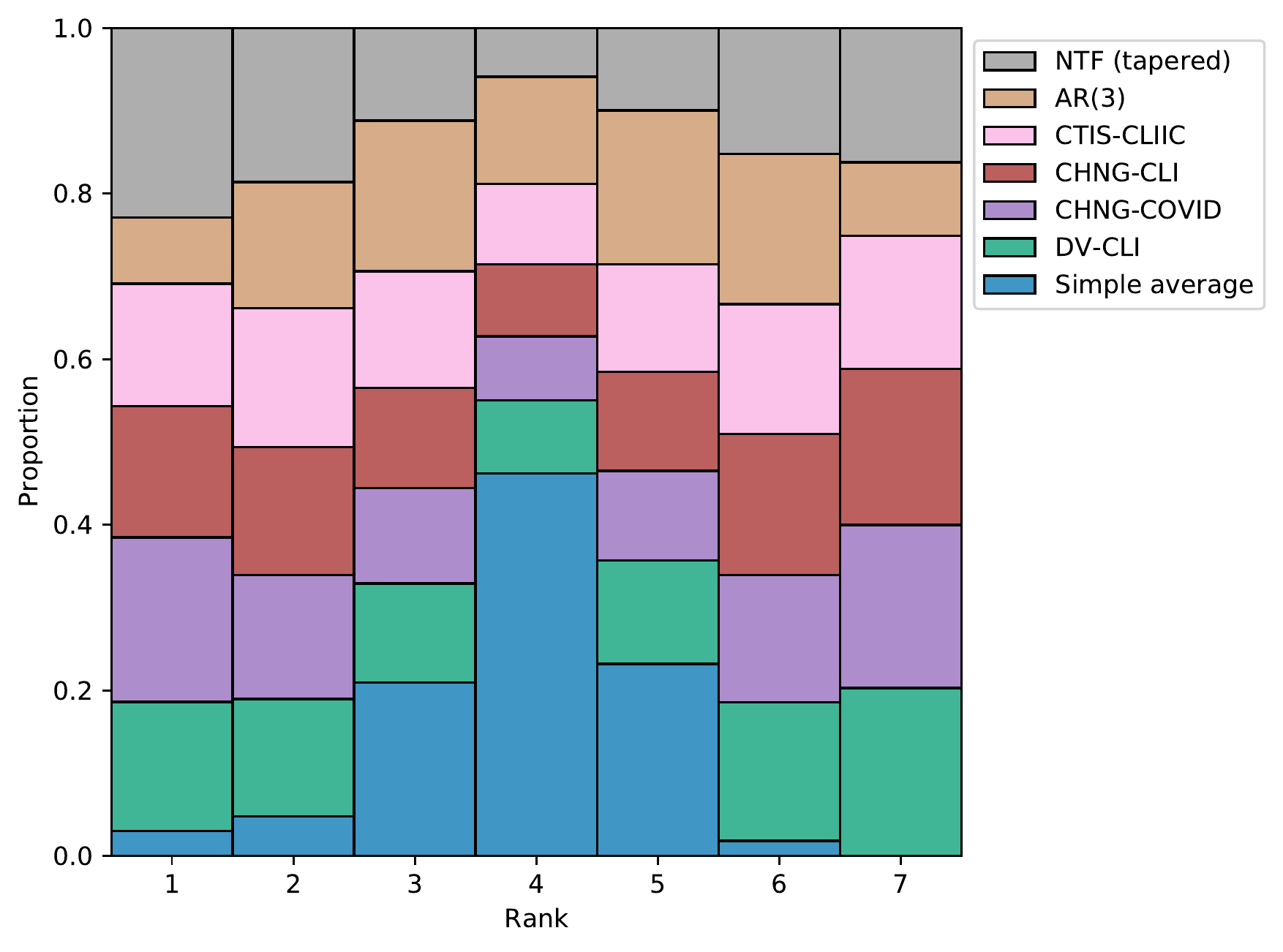}
\includegraphics[width=0.975\linewidth]{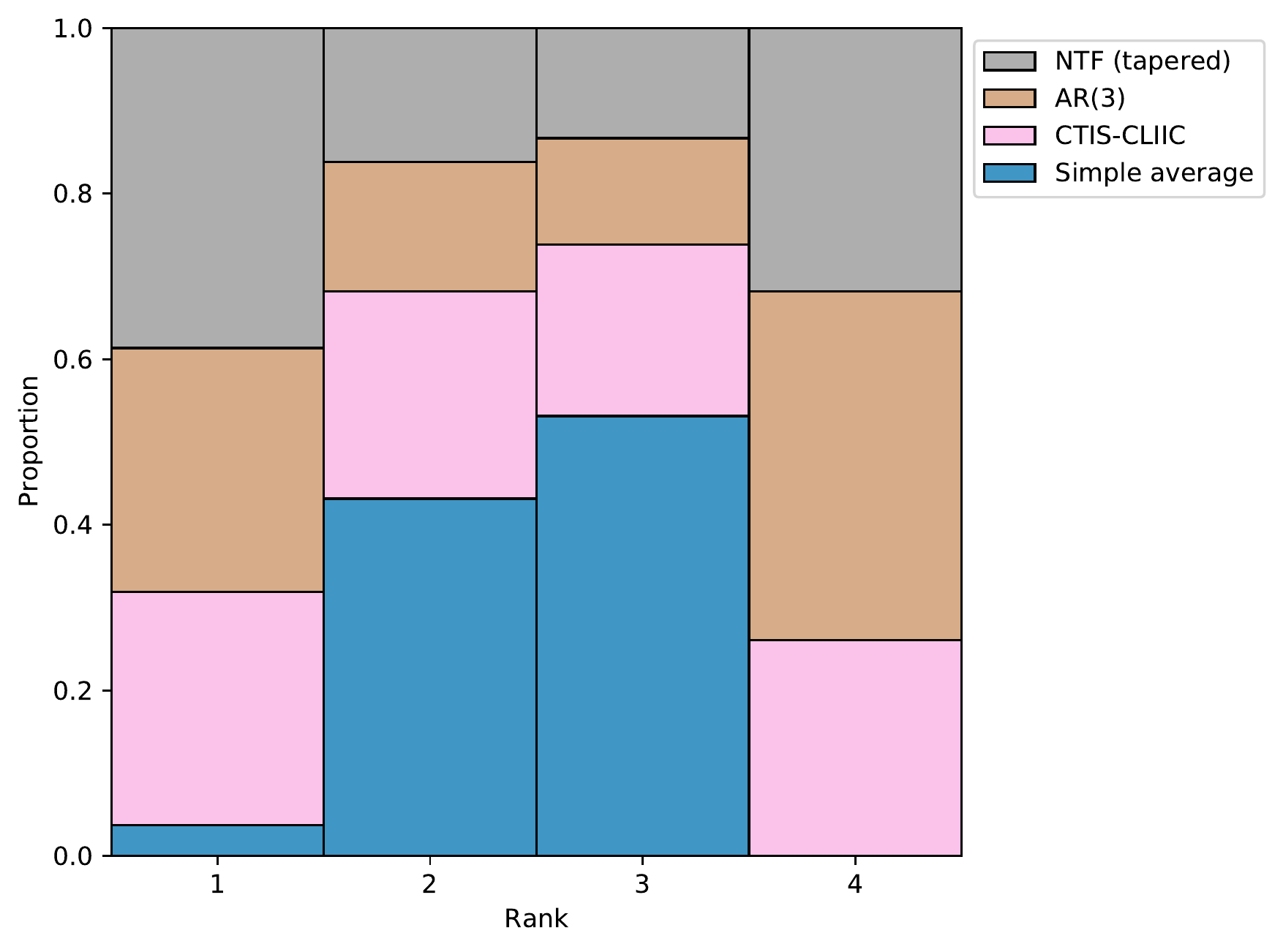}
\caption{Comparing NTF to individual sensor models and the simple average sensor
  fusion method by relative ranks over common nowcast tasks. The top panel
  excludes all claims-based sensors and considers lags 1--5, whereas the bottom
  panel includes them and considers lags 4--9 (the first 5 lags at which all
  methods are available, in either case). The simple average exhibits striking
  consistency: it is rarely the best, but also never the worst.} 
	\label{fig:sensor_rank_no_google_aa}
\end{figure}

Figure~\ref{fig:sensor_rank_no_google_aa} displays the empirical distributions
of ranks of nowcast errors coming from each method, computed with respect to
each other, over common nowcast tasks (defined by a location-date-lag
triplet). For example, in a particular nowcast task, we assign a rank of 1 to
the method with the smallest absolute error for that nowcast task. The top panel
again excludes claims-based signals, and the bottom panel includes them. The
striking feature in either panel, particularly the bottom panel, is that the
simple average has a highly distinctive distribution of ranks---it is rarely the
best method, but never the worst. While this is not particularly surprising
(averaging random variables tends to be variance-reducing, as long as the
variables are not too correlated), it also points to a key property of sensor
fusion---a certain kind of robustness, beyond accuracy.

\subsection{Relative Performance of Sensor Fusion Methods}
 
We now compare the various sensor fusion methods to each other. The results here
do not include claims-based signals; results including claims-based signals are
shown in Appendix~\ref{app:eval_plots}. Figure~\ref{fig:mae_sml} displays the
MAE of the various sensor fusion estimates, but divided up into three panels,
defined by averaging over small, medium, and large states (the figure caption
provides more details). Recall that for the lasso, ridge, and KF-SF approaches,
a model in a particular county is fit using the sensors from other counties in
the same state. Larger states have more pooling of information across locations
and present a greater potential for gains in accuracy. We see that the simple
average method is typically the best sensor fusion method at each lag, but for
medium and large states, KF-SF catches up with it and is just about as accurate.

\begin{figure*}[tb]
\centering
\includegraphics[width=0.95\linewidth]{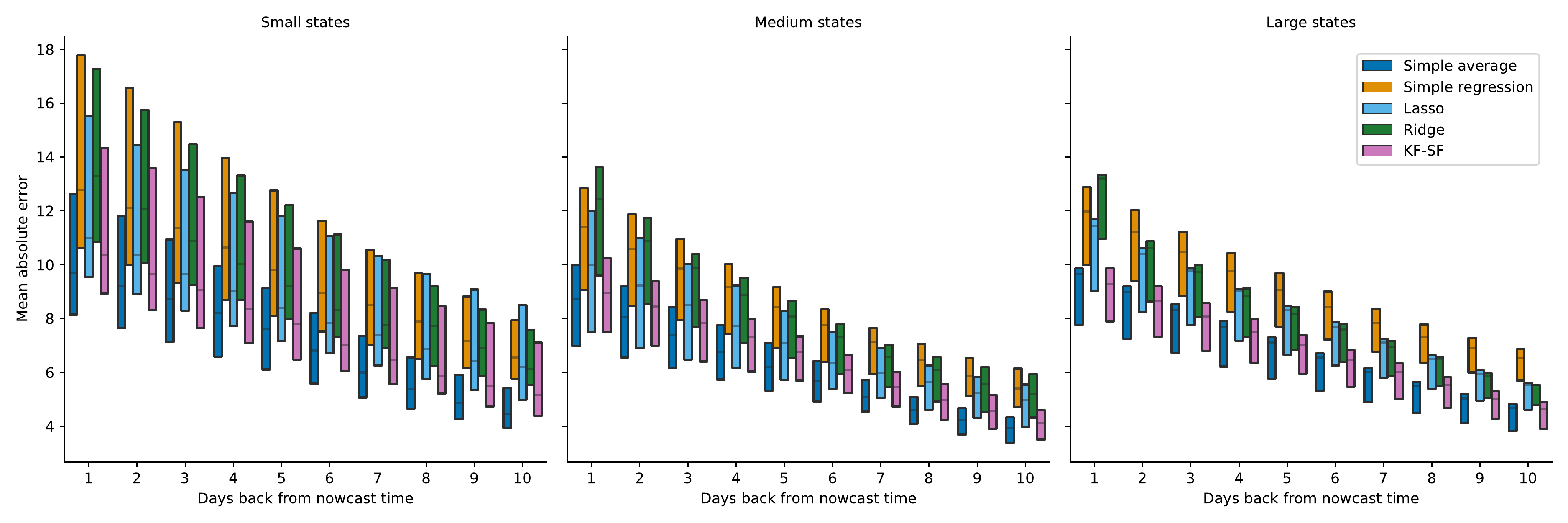}
\caption{Comparing sensor fusion methods by boxenplots of nowcasting
  errors (each box conveys the level 25\%, 50\%, and 75\% quantiles of the  
  absolute error distribution.) The three panels average over small (containing
  less than 5 locations), medium (between 5 and 14 locations), and large (more
  than 15 locations) states. Simple average performs generally the best
  throughout, but KF-SF catches up for medium and large states.}  
	\label{fig:mae_sml}
\end{figure*}

\begin{figure}[tb]
\centering
\includegraphics[width=0.975\linewidth]{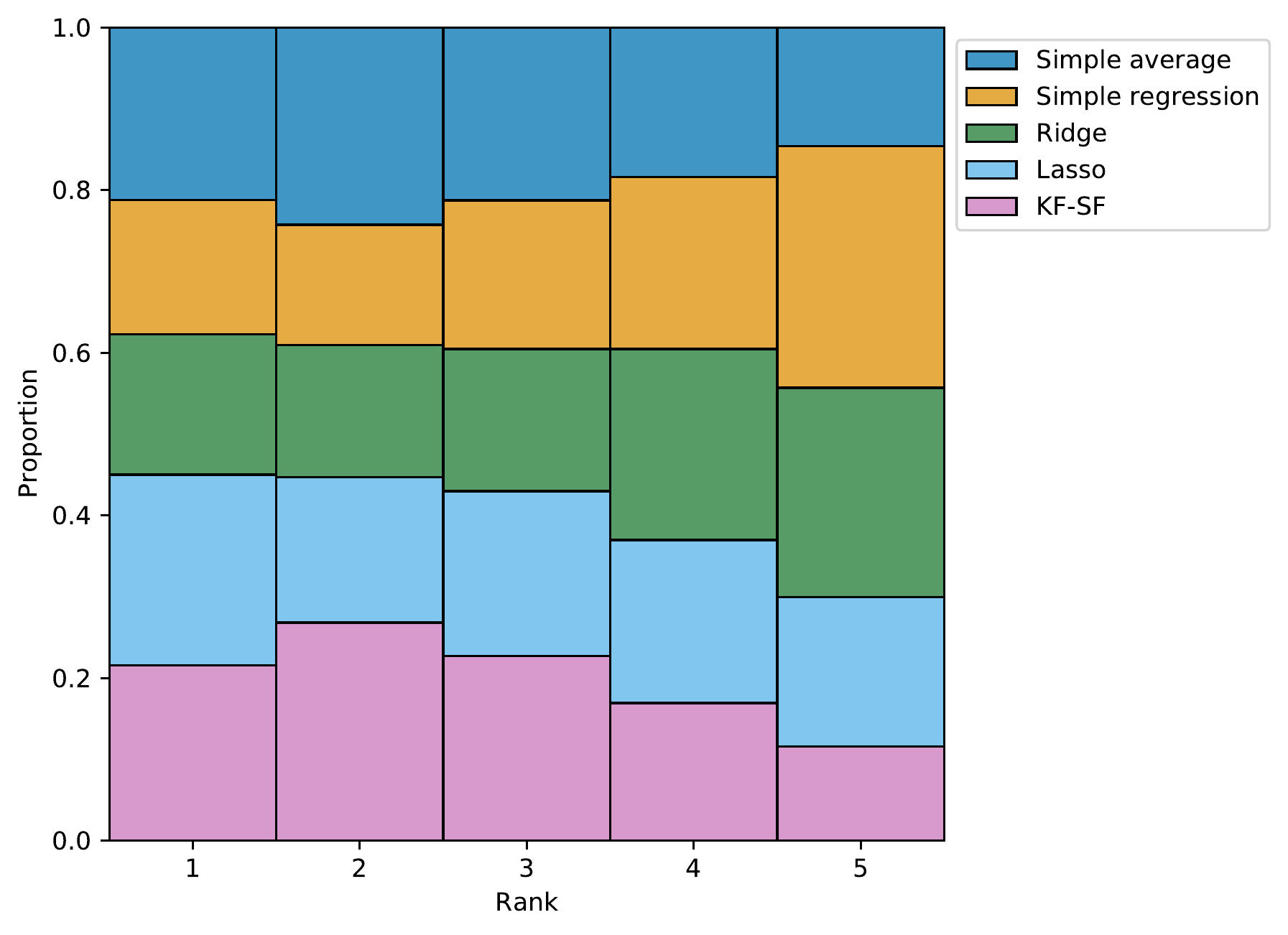}
\caption{Comparing sensor fusion methods by relative ranks over common nowcast 
  tasks, and considering only lags 1--5. The simple average and KF-SF methods
  consistently perform in the top half, while simple regression is most often
  the worst.}
	\label{fig:rank_ensemble}
\end{figure}

Figure~\ref{fig:rank_ensemble} displays the relative ranking of sensor fusion
methods. The simple average and KF-SF methods appear the most favorable
(often the best, and less so the worst), followed by lasso, then ridge, and 
lastly simple regression (most often the worst).   

\section{Discussion}
\label{sec:discussion}

In this work, we proposed, implemented, and evaluated a framework for real-time
estimation of new symptomatic COVID-19 infections from case reports. At time
$t$, in order to nowcast the infection rate at time $t-k$ (for small values of
$k$, such as $k=1,2,\ldots$), the main steps are to:  
\begin{enumerate}
\item estimate a symptom-onset-to-case-report delay distribution using the most 
  recent data available in a line list provided by the CDC;  
\item perform regularized deconvolution on the most recent case data available 
  from JHU CSSE; 
\item update models to track recent infection rates from various auxiliary
  signals (based on COVID-related data from medical insurance claims, online
  surveys, and Google searches), and fuse together the predictions from these
  models in order to stabilize recent estimates of infection rates. 
\end{enumerate}
In each step, we proposed methodological advances that improved the accuracy of
our nowcasts, when measured against finalized infection rate estimates obtained
by retrospective deconvolution (using data that would have only been available 
months later). While using auxiliary signals (step 3) did help in terms of
accuracy and robustness, the additional regularization devices that we
incorporated into real-time deconvolution (step 2) ended up providing the
biggest benefit to accuracy. 

To reiterate, we purposely defined our target of estimation to be symptomatic
infections that would eventually show up in public health reports, allowing us
to focus on developing and testing tools for real-time deconvolution and sensor
fusion, with minimal assumptions (e.g., without a mechanistic model for disease
spread). Estimating the number of true symptomatic infections at any point in
time---whether or not they will appear in case reports---is of course a much
harder problem. However, our methodology may be seen as a contribution toward
solving this larger problem in real-time; moreover, some simple post hoc
corrections could be applied to our real-time estimates in order to adjust for
confounding. For example, if $a_{\ell,t}$ is the fraction of untested
symptomatic infections in location $\ell$ at at time $t$, which (say) is
estimated from external data sources, then we could just multiply each element
\smash{$\hp^{(t)}_{\ell,s}$} of the delay distribution used in
\eqref{eq:tf_realtime4} by $b_{\ell,t}=1/(1-a_{\ell,t})$ in order to estimate
\emph{all} symptomatic infections from case reports. Due to the way we have set
up the deconvolution problem (cross-validating over optimal choices of tuning
parameters), this would be essentially equivalent to post-mulitplying the
nowcast \smash{$\hx^{(t)}_{\ell,s}$} we already produce by $b_{\ell,t}$.

We finish by describing a few directions for future work.

\smallskip
\paragraph*{Post Hoc Smoothing.}

As we saw in Section~\ref{sec:evaluation}, sensor fusion provides a real-time
improvement on pure deconvolution up until about a 10-day lag, and past that
point, the deconvolution estimates appear stable enough that sensor fusion
becomes unnecessary. While the quantative benefit of sensor fusion for small
lags is clear, sensor fusion is also lacking in the following qualitative
aspect: its estimates do not always appear visually smooth across time (this is
because the sensors themselves need not be smooth over time, and furthermore, 
sensor fusion may end up using a different subset of sensors at each lag,
creating additional jaggedness). Post smoothing techniques would be worth
investigating here, to aid visual consumption.

\smallskip
\paragraph*{$R_t$ Estimation.} 

The instantaneous reproductive number $R_t$, the average number of secondary
infections at time $t$ generated from a primary infection in the past, is a
useful and interpretable parameter that reflects the dynamics of epidemic growth
in a population. In the SIR model, the instantaneous reproductive number $R_t$
and growth rate $r_t$ at time $t$ obey the following relationship:
\[
R_t \approx 1 + \frac{r_t}{\gamma}, 
\] 
where $\gamma$ denotes the recovery rate in the SIR model. While this is
well-known in the literature on mathematical modeling of epidemics (and is exact
under local exponential growth; see, e.g., \citet{Wallinga:2007}), its use in
the presence of confounding seems to be underexplored and potentially
undervalued. If $I_t$ denotes the number of new infections at $t$, then using a
simple discrete difference approximation to $r_t$ leads to: 
\[
R_t \approx 1 + \frac{1}{\gamma}\bigg( \frac{I_{t+1}}{I_t} - 1 \bigg). 
\]
A similar though not identical approximation is given in
\citet{Bettencourt:2008}, where $I_{t+1}/I_t-1$ is replaced by
$\log(I_{t+1}/I_t)$. Critically, incident infections only enter right-hand side
above as a \emph{ratio} of values adjacent in time, and thus if we are only able
to estimate this up to an unknown multiplicative factor (due to confounding),
then this factor approximately cancels in the ratio as long as it is slowly
varying in time. In slightly more detail (and for simplicity, considering just a
single location), suppose as before that a fraction $a_t$ of infections go
untested at time $t$. Then $I_t = b_t x_t$ where $x_t$ is the number of new
infections at time $t$ that show up in case reports (i.e., the focus of this
paper) and $b_t = 1/(1-a_t)$. From the previous display,   
\[
R_t \approx 1 + \frac{1}{\gamma}\bigg( \frac{b_{t+1} x_{t+1}}{b_t x_t} - 1
\bigg) \approx 1 + \frac{1}{\gamma}\bigg( \frac{x_{t+1}}{x_t} - 1 \bigg),
\]
where the last approximation is motivated by an additional assumption the 
untested fraction varies slowly over time (so $b_{t+1}/b_t \approx
1$). This shows that estimates of $x_t$ can produce \emph{approximately
  unconfounded} estimates of $R_t$, even though $x_t$ is itself confounded
due to a lack of universal testing. This is true both in the retrospective and 
real-time sense, and will be the topic of future study.     


\smallskip
\paragraph*{Evaluation via Reconvolution.}

An important avenue for evaluating our methodology (beyond evaluating against
finalized infection rate estimates, as we do in this paper), would be to
reconvolve our real-time nowcasts of infection rates forward in time in order to
predict future case rates, and evaluate these predictions against finalized case
reporting data. Making and evaluating point predictions would be relatively
straightforward, however, distributional forecasts are currently the standard in
epidemiological forecasting (and also in COVID-19 forecasting), and adding a
distributional layer to our nowcasts (and propagating this through the
convolution operator) requires substantial new developments, and we leave it to
future work.

\begin{acks}[Acknowledgments]
The authors are grateful to Logan Brooks, Roni Rosenfeld, James Sharpnack, Sam 
Abbott, Joel Hellewell, and Sebastian Funk for several early insightful
conversations.   
 
MJ was supported by a fellowship from the Center for Machine Learning and 
Health at Carnegie Mellon. AC and RJT were supported by a gift from Google.org.   
\end{acks}



\begin{appendix}
	
\section{ADMM for Solving Deconvolution Problems}  
\label{app:tf_admm}

Here we give details on the ADMM approach used to solve the regularized least
squares deconvolution problems in Sections~\ref{sec:deconv_retro} and
\ref{sec:deconv_realtime}. We first focus on problem \eqref{eq:tf_retro}, and
then we discuss the modifications needed when incorporating extra regularization
for real-time deconvolution as in \eqref{eq:tf_realtime4}. To simplify notation,
we will henceforth drop the subscript dependnece of all quantities on the
location $\ell$, as well as the superscript dependence on the nowcast date $t$
for the real-time problems.

We also use \smash{$\hP$} to denote the (Toeplitz) convolution matrix with rows
determined by \smash{$\hp_s$}, $s<t$, i.e., such that for any vector $x$ (of
appropriate dimension) 
$$
(\hP x)_s = \sum_{k=1}^d \hp_k x_{s-k}.
$$
(We leave the dimensions of \smash{$\hP$} and $x$ here purposely ambiguous,
which should always be clear from the context anyway; this allows us to borrow
similar notation across problems with different underlying dimensions.) Thus we
can rewrite \eqref{eq:tf_retro} as 
$$
\minimize_x \; \| y - \hP x \|_2^2 + \lambda \|D^{(4)} x\|_1. 
$$
To apply ADMM, we must introduce auxiliary variables, and as in
\citet{ramdas2016fast}, we use the following ``specialized'' decomposition 
(which improves the convergence speed):
\begin{alignat*}{2}
&\minimize_x \; && \| y - \hP x \|_2^2 + \lambda \|D^{(1)} \alpha\|_1 \\ 
&\subjectto && \;\; \alpha = D^{(3)} x,
\end{alignat*}
where we used the recursive nature of the difference operators, writing the
4$\th$-order operator as a product of the 1$\st$- and 3$\rd$-order operators: 
\smash{$D^{(4)} = D^{(1)} D^{(3)}$}. The above problem gives rise to the
augmented Lagrangian: 
\begin{multline*}
\cL(x, \alpha, u) = \| y - \hP x \|_2^2 + \lambda \|D^{(1)} \alpha\|_1 +{} \\ 
\rho \|\alpha -  D^{(3)} x + u\|_2^2 -  \rho \|u\|_2^2, 
\end{multline*}
which corresponds to following ADMM updates, writing \smash{$D=D^{(3)}$} for 
brevity:  
\begin{align*}
x &\leftarrow (\hP^T \hP + \rho D^T D)^{-1} \big( \hP^T y + \rho D^T
    (\alpha + u)\big) \\   
\alpha &\leftarrow \argmin_z \; \| D x  - u - z \|_2^2 +
\frac{\lambda}{\rho} \|D^{(1)} \alpha\|_1 \\
u & \leftarrow u + \alpha - D x.
\end{align*}
The $\alpha$-update here requires solving a 1-dimensional fused lasso problem, 
which can be done in linear-time with the dynamic programming approach of 
\citet{johnson2013dynamic}. The $x$-update is more expensive than in pure
trend filtering (with no convolution operator) but owing to the bandedness of
\smash{$\hP$} (and $D$, though the bandwidth $d$ of \smash{$\hP$} dominates), it
can still be solved in $O(nd)$ operations. Further, in this and all applications  
of ADMM, we follow the recommendation of \citet{ramdas2016fast} and set the
Lagrangian parameter equal to the tuning parameter, $\rho = \lambda$. 

As for the two extensions presented in \eqref{eq:tf_realtime4}, the natural  
trend filtering constraints can be be enforced by introducing a linear
interpolant matrix as described in Section~11.2 of \citet{tibshirani2020divided}.
This effectively replaces the convolution matrix \smash{$\hP$} and the 3$\rd$
difference operator $D$, in the ADMM steps above, by \smash{$\tP$} and
\smash{$\tD$}, respectively, which are given by right multiplying $P$ and $D$ by
the interpolant matrix.  

Moreover, the additional tapered smoothing term can be pushed into the augmented
Lagrangian, and only alters the $x$-update, now becoming:       
\begin{multline*}
x \leftarrow (\tP^T \tP + \gamma M^T M + \rho \tD)^{-1} \cdot{} \\
 \big(\tP^T y + \rho \tD^T(\alpha + u)\big),
\end{multline*}
where $M$ is the matrix $W^{(t)} D^{(1)}$ in the tapered penalty in
\eqref{eq:tf_realtime4} times the linear interpolant matrix.

\section{Additional Evaluation Results}
\label{app:eval_plots}

\begin{figure}[tb]
\centering
\includegraphics[width=0.85\linewidth]{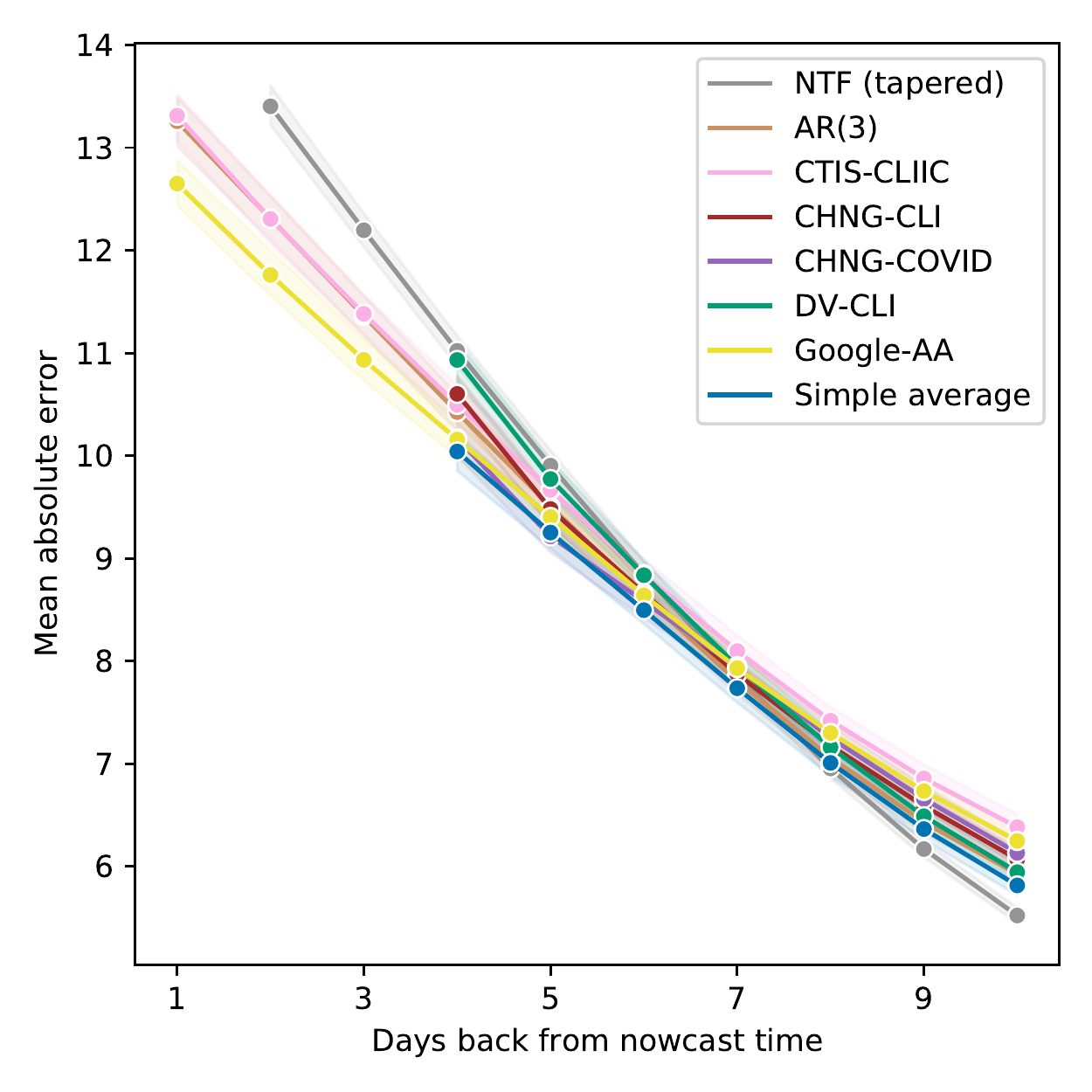}
\includegraphics[width=0.85\linewidth]{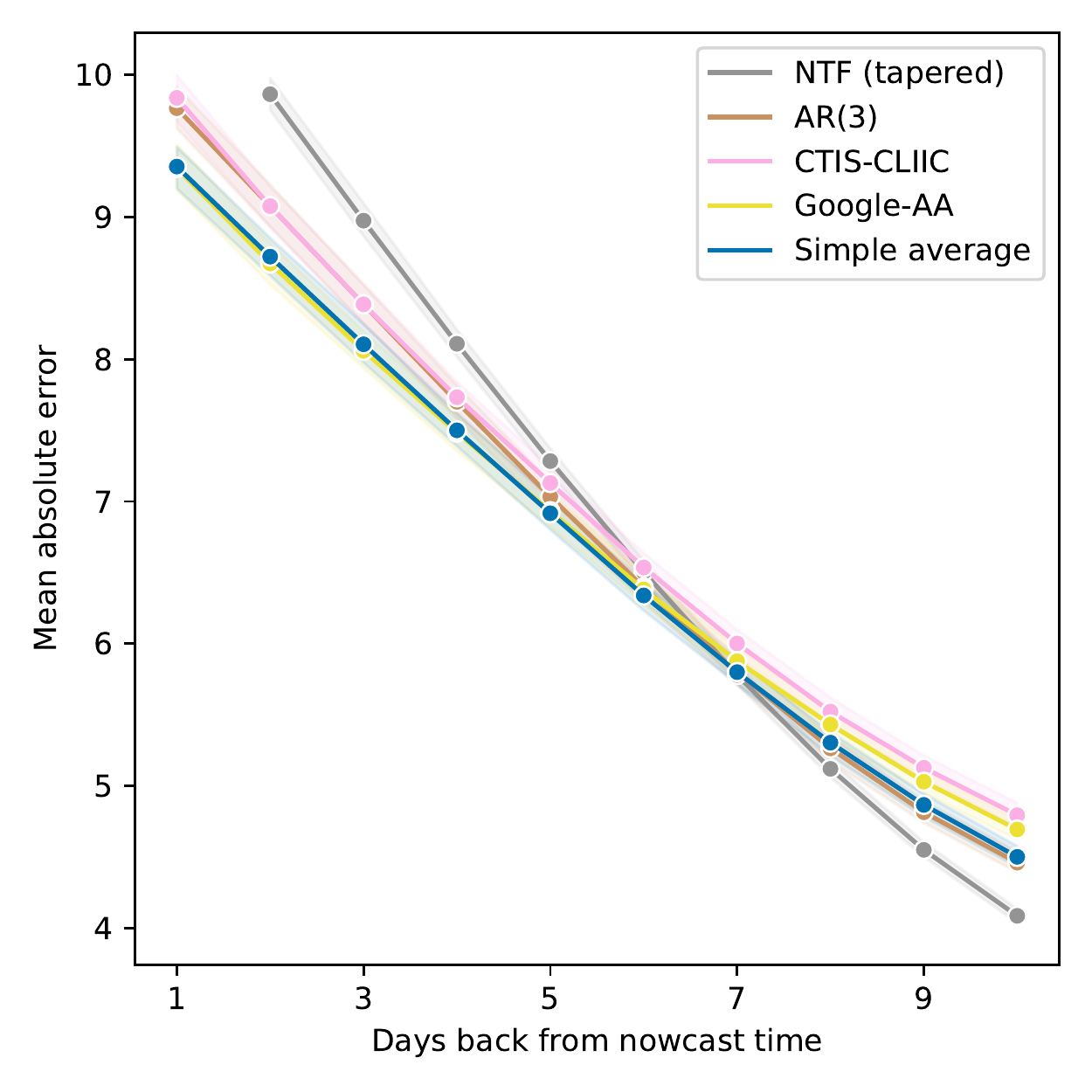}
\caption{As in Figure~\ref{fig:mae_all_no_google_aa}, but including Google-AA.}    
\label{fig:mae_all}
\end{figure}

\begin{figure}[tb]
\centering
\includegraphics[width=0.975\linewidth]{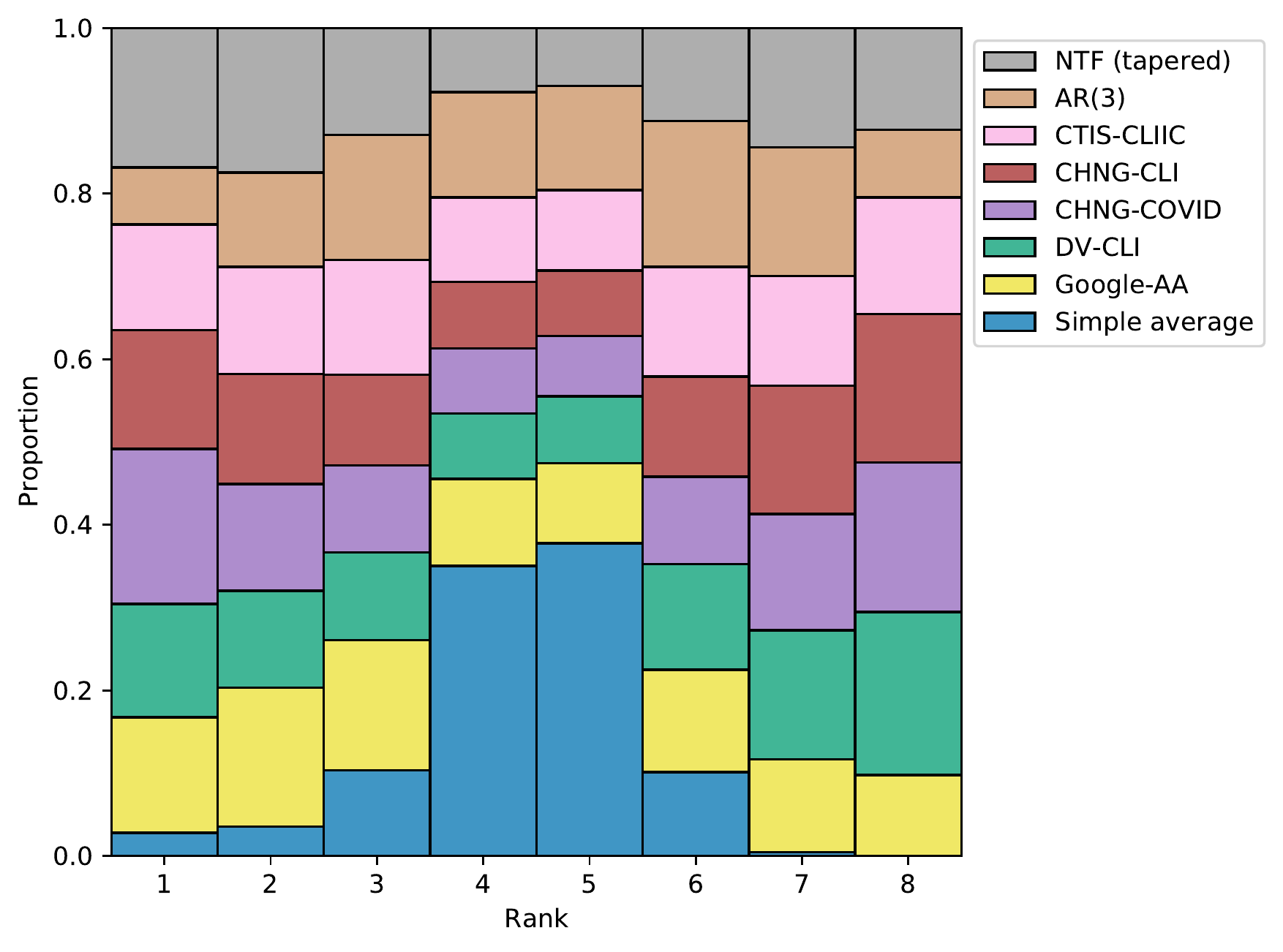}
\includegraphics[width=0.975\linewidth]{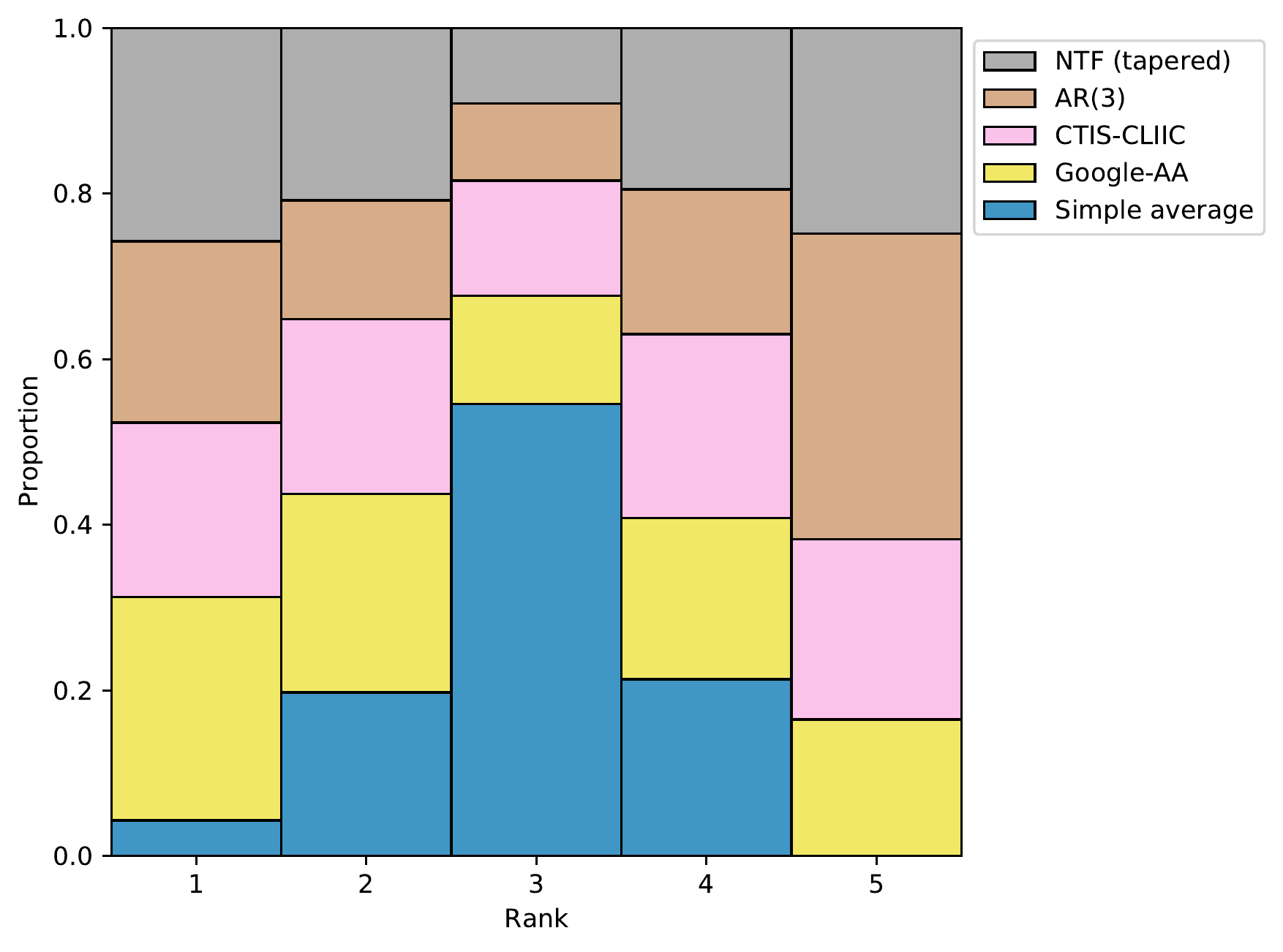}
\caption{As in Figure~\ref{fig:sensor_rank_no_google_aa}, but including Google-AA.}
\label{fig:sensor_rank}
\end{figure}

\begin{figure*}[tb]
\centering
\includegraphics[width=0.95\linewidth]{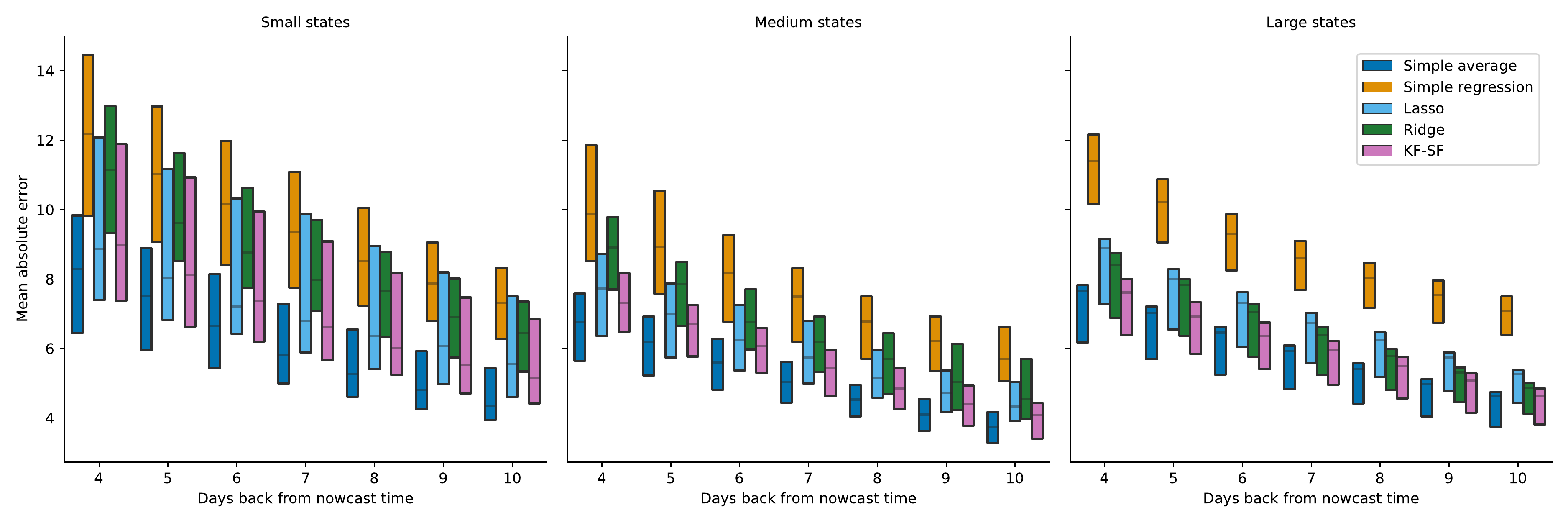}
\caption{As in Figure~\ref{fig:mae_sml}, but including claims-based signals.}
\label{fig:mae_sml_claims}
\end{figure*}

\begin{figure}[tb]
\centering
\includegraphics[width=0.975\linewidth]{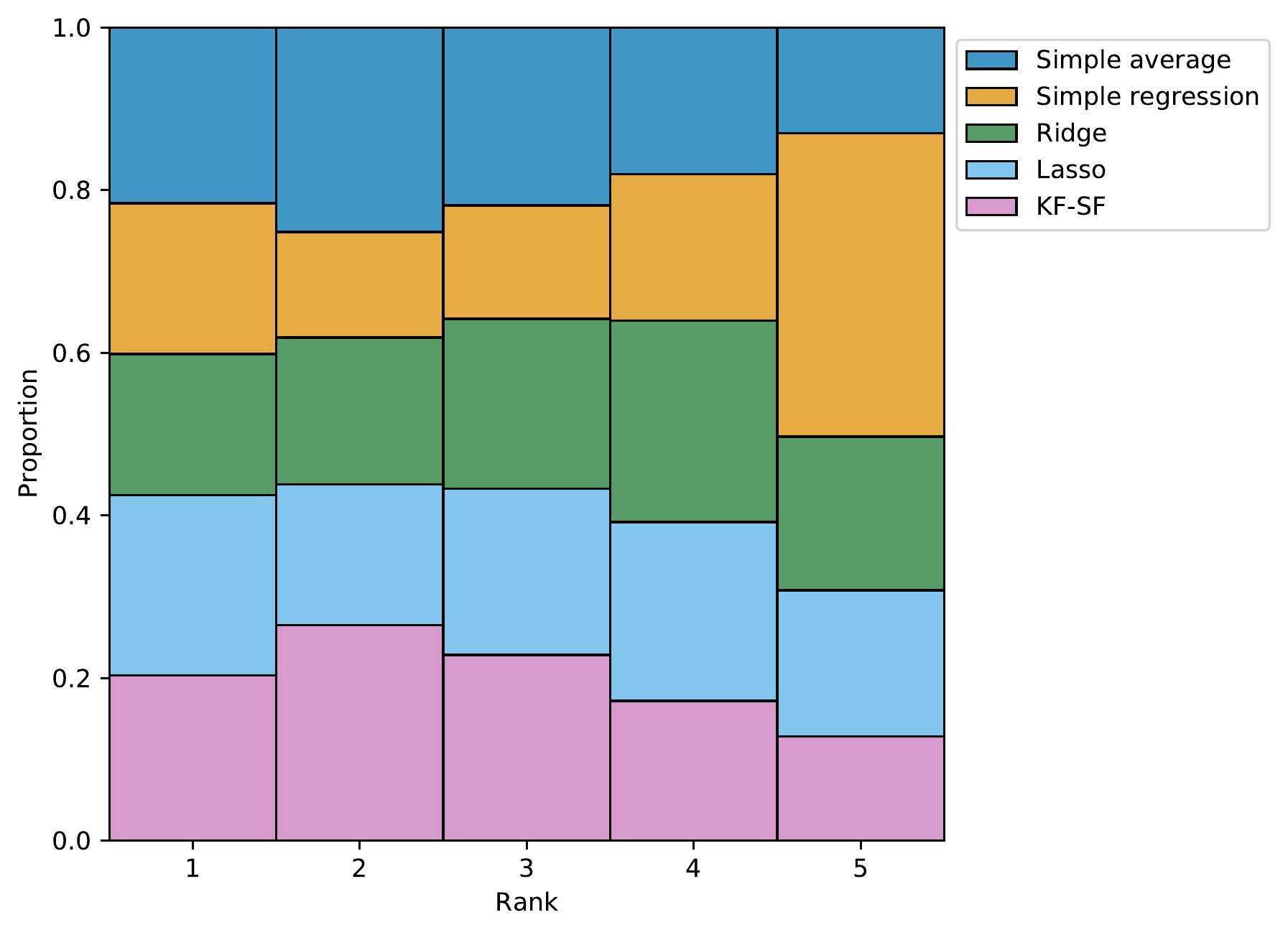}
\caption{As in Figure~\ref{fig:rank_ensemble}, but including claims-based
  signals.} 
\label{fig:rank_ensemble_claims}
\end{figure}

Figures~\ref{fig:mae_all} and \ref{fig:sensor_rank} are analogous to
Figures~\ref{fig:mae_all_no_google_aa} and \ref{fig:sensor_rank_no_google_aa},
but with the inclusion of the Google-AA sensor. Similarly,
Figures~\ref{fig:mae_sml_claims} and \ref{fig:rank_ensemble_claims} are the
counterparts to Figures~\ref{fig:mae_sml} and \ref{fig:rank_ensemble}, but with
the inclusion of claims-based sensors.

\end{appendix}

\bibliographystyle{imsart-nameyear}
\bibliography{ryantibs, nowcast}      

\begin{thebibliography}{51}

\bibitem[\protect\citeauthoryear{Abbott et~al.}{2020}]{Abbott:2020}
\begin{barticle}[author]
\bauthor{\bsnm{Abbott},~\bfnm{Sam}\binits{S.}},
  \bauthor{\bsnm{Hellewell},~\bfnm{Joel}\binits{J.}},
  \bauthor{\bsnm{Thompson},~\bfnm{Robin~N}\binits{R.~N.}},
  \bauthor{\bsnm{Sherratt},~\bfnm{Katharine}\binits{K.}},
  \bauthor{\bsnm{Gibbs},~\bfnm{Hamish~P}\binits{H.~P.}},
  \bauthor{\bsnm{Bosse},~\bfnm{Nikos~I}\binits{N.~I.}},
  \bauthor{\bsnm{Munday},~\bfnm{James~D}\binits{J.~D.}},
  \bauthor{\bsnm{Meakin},~\bfnm{Sophie}\binits{S.}},
  \bauthor{\bsnm{Doughty},~\bfnm{Emma~L}\binits{E.~L.}},
  \bauthor{\bsnm{Chun},~\bfnm{June~Young}\binits{J.~Y.}},
  \bauthor{\bsnm{Chan},~\bfnm{Yung-Wai~Desmond}\binits{Y.-W.~D.}},
  \bauthor{\bsnm{Finger},~\bfnm{Flavio}\binits{F.}},
  \bauthor{\bsnm{Campbell},~\bfnm{Paul}\binits{P.}},
  \bauthor{\bsnm{Endo},~\bfnm{Akira}\binits{A.}},
  \bauthor{\bsnm{Pearson},~\bfnm{Carl A~B}\binits{C.~A.~B.}},
  \bauthor{\bsnm{Gimma},~\bfnm{Amy}\binits{A.}},
  \bauthor{\bsnm{Russell},~\bfnm{Tim}\binits{T.}}, \bauthor{\bsnm{{{CMMID COVID
  modelling group}}}}, \bauthor{\bsnm{Flasche},~\bfnm{Stefan}\binits{S.}},
  \bauthor{\bsnm{Kucharski},~\bfnm{Adam~J}\binits{A.~J.}},
  \bauthor{\bsnm{Eggo},~\bfnm{Rosalind~M}\binits{R.~M.}} \AND
  \bauthor{\bsnm{Funk},~\bfnm{Sebastian}\binits{S.}}
(\byear{2020}).
\btitle{Estimating the time-varying reproduction number of {SARS-CoV-2} using
  national and subnational case counts}.
\bjournal{Wellcome Open Research}
\bvolume{5}.
\end{barticle}
\endbibitem

\bibitem[\protect\citeauthoryear{Ackley et~al.}{2020}]{Ackley:2020}
\begin{barticle}[author]
\bauthor{\bsnm{Ackley},~\bfnm{Aarah~F}\binits{A.~F.}},
  \bauthor{\bsnm{Pilewski},~\bfnm{Sarah}\binits{S.}},
  \bauthor{\bsnm{Petrovic},~\bfnm{Vladimir~S}\binits{V.~S.}},
  \bauthor{\bsnm{Worden},~\bfnm{Lee}\binits{L.}},
  \bauthor{\bsnm{Murray},~\bfnm{Erin}\binits{E.}} \AND
  \bauthor{\bsnm{Porco},~\bfnm{Travis~C}\binits{T.~C.}}
(\byear{2020}).
\btitle{assessing the utility of a smart thermometer and mobile application as
  a surveillance tool for influenza and influenza-like illness}.
\bjournal{Health Informatics Journal}
\bvolume{26}
\bpages{2148--2158}.
\end{barticle}
\endbibitem

\bibitem[\protect\citeauthoryear{Bavadekar et~al.}{2020}]{Bavadekar:2020}
\begin{bunpublished}[author]
\bauthor{\bsnm{Bavadekar},~\bfnm{Shailesh}\binits{S.}},
  \bauthor{\bsnm{Dai},~\bfnm{Andrew}\binits{A.}},
  \bauthor{\bsnm{Davis},~\bfnm{John}\binits{J.}},
  \bauthor{\bsnm{Desfontaines},~\bfnm{Damien}\binits{D.}},
  \bauthor{\bsnm{Eckstein},~\bfnm{Ilya}\binits{I.}},
  \bauthor{\bsnm{Everett},~\bfnm{Katie}\binits{K.}},
  \bauthor{\bsnm{Fabrikant},~\bfnm{Alex}\binits{A.}},
  \bauthor{\bsnm{Flores},~\bfnm{Gerardo}\binits{G.}},
  \bauthor{\bsnm{Gabrilovich},~\bfnm{Evgeniy}\binits{E.}},
  \bauthor{\bsnm{Gadepalli},~\bfnm{Krishna}\binits{K.}},
  \bauthor{\bsnm{Glass},~\bfnm{Shane}\binits{S.}},
  \bauthor{\bsnm{Huang},~\bfnm{Rayman}\binits{R.}},
  \bauthor{\bsnm{Kamath},~\bfnm{Chaitanya}\binits{C.}},
  \bauthor{\bsnm{Kraft},~\bfnm{Dennis}\binits{D.}},
  \bauthor{\bsnm{Kumok},~\bfnm{Akim}\binits{A.}},
  \bauthor{\bsnm{Marfatia},~\bfnm{Hinali}\binits{H.}},
  \bauthor{\bsnm{Mayer},~\bfnm{Yael}\binits{Y.}},
  \bauthor{\bsnm{Miller},~\bfnm{Benjamin}\binits{B.}},
  \bauthor{\bsnm{Pearce},~\bfnm{Adam}\binits{A.}},
  \bauthor{\bsnm{Perera},~\bfnm{Irippuge~Milinda}\binits{I.~M.}},
  \bauthor{\bsnm{Ramachandran},~\bfnm{Venky}\binits{V.}},
  \bauthor{\bsnm{Raman},~\bfnm{Karthik}\binits{K.}},
  \bauthor{\bsnm{Roessler},~\bfnm{Thomas}\binits{T.}},
  \bauthor{\bsnm{Shafran},~\bfnm{Izhak}\binits{I.}},
  \bauthor{\bsnm{Shekel},~\bfnm{Tomer}\binits{T.}},
  \bauthor{\bsnm{Stanton},~\bfnm{Charlotte}\binits{C.}},
  \bauthor{\bsnm{Stimes},~\bfnm{Jacob}\binits{J.}},
  \bauthor{\bsnm{Sun},~\bfnm{Mimi}\binits{M.}},
  \bauthor{\bsnm{Wellenius},~\bfnm{Gregory}\binits{G.}}, \bauthor{} \AND
  \bauthor{\bsnm{Zoghi},~\bfnm{Masrour}\binits{M.}}
(\byear{2020}).
\btitle{Google {COVID}-19 search trends symptoms dataset: {A}nonymization
  process description}.
\bnote{arXiv: 2009.01265}.
\end{bunpublished}
\endbibitem

\bibitem[\protect\citeauthoryear{Bettencourt and
  Ribeiro}{2008}]{Bettencourt:2008}
\begin{barticle}[author]
\bauthor{\bsnm{Bettencourt},~\bfnm{Luis~MA}\binits{L.~M.}} \AND
  \bauthor{\bsnm{Ribeiro},~\bfnm{Ruy~M}\binits{R.~M.}}
(\byear{2008}).
\btitle{Real time {Bayesian} estimation of the epidemic potential of emerging
  infectious diseases}.
\bjournal{PLOS ONE}
\bvolume{3}
\bpages{e2185}.
\end{barticle}
\endbibitem

\bibitem[\protect\citeauthoryear{Brooks}{2020}]{Brooks:2020}
\begin{bphdthesis}[author]
\bauthor{\bsnm{Brooks},~\bfnm{Logan~C}\binits{L.~C.}}
(\byear{2020}).
\btitle{Pancasting: {F}orecasting epidemics from provisional data},
\btype{PhD thesis},
\bpublisher{Carnegie Mellon University}.
\end{bphdthesis}
\endbibitem

\bibitem[\protect\citeauthoryear{Brownstein, Freifeld and
  Madoff}{2009}]{Brownstein:2009}
\begin{barticle}[author]
\bauthor{\bsnm{Brownstein},~\bfnm{John~S}\binits{J.~S.}},
  \bauthor{\bsnm{Freifeld},~\bfnm{Clark~C}\binits{C.~C.}} \AND
  \bauthor{\bsnm{Madoff},~\bfnm{Lawrence~C}\binits{L.~C.}}
(\byear{2009}).
\btitle{Digital disease detection --- harnessing the web for public health
  surveillance}.
\bjournal{New England Journal of Medicine}
\bvolume{360}
\bpages{2153--2157}.
\end{barticle}
\endbibitem

\bibitem[\protect\citeauthoryear{Carlson et~al.}{2013}]{Carlson:2013}
\begin{barticle}[author]
\bauthor{\bsnm{Carlson},~\bfnm{Sandra~J}\binits{S.~J.}},
  \bauthor{\bsnm{Dalton},~\bfnm{Craig~B}\binits{C.~B.}},
  \bauthor{\bsnm{Butler},~\bfnm{Michelle~T}\binits{M.~T.}},
  \bauthor{\bsnm{Fejsa},~\bfnm{John}\binits{J.}},
  \bauthor{\bsnm{Elvidge},~\bfnm{Elissa}\binits{E.}} \AND
  \bauthor{\bsnm{Durrheim},~\bfnm{David~N}\binits{D.~N.}}
(\byear{2013}).
\btitle{Flutracking weekly online community survey of influenza-like illness
  annual report 2011 and 2012}.
\bjournal{Communicable diseases intelligence quarterly report}
\bvolume{37}
\bpages{E398--406}.
\end{barticle}
\endbibitem

\bibitem[\protect\citeauthoryear{Charu et~al.}{2017}]{Charu:2017}
\begin{barticle}[author]
\bauthor{\bsnm{Charu},~\bfnm{Vivek}\binits{V.}},
  \bauthor{\bsnm{Zeger},~\bfnm{Scott}\binits{S.}},
  \bauthor{\bsnm{Gog},~\bfnm{Julia}\binits{J.}},
  \bauthor{\bsnm{Bjørnstad},~\bfnm{Ottar~N.}\binits{O.~N.}},
  \bauthor{\bsnm{Kissler},~\bfnm{Stephen}\binits{S.}},
  \bauthor{\bsnm{Simonsen},~\bfnm{Lone}\binits{L.}},
  \bauthor{\bsnm{Grenfell},~\bfnm{Bryan~T.}\binits{B.~T.}} \AND
  \bauthor{\bsnm{Viboud},~\bfnm{C{\'e}cile}\binits{C.}}
(\byear{2017}).
\btitle{Human mobility and the spatial transmission of influenza in the {United
  States}}.
\bjournal{PLOS Computational Biology}
\bvolume{13}
\bpages{1--23}.
\end{barticle}
\endbibitem

\bibitem[\protect\citeauthoryear{Chitwood et~al.}{2021}]{Chitwood:2021}
\begin{bunpublished}[author]
\bauthor{\bsnm{Chitwood},~\bfnm{Melanie~H}\binits{M.~H.}},
  \bauthor{\bsnm{Russi},~\bfnm{Marcus}\binits{M.}},
  \bauthor{\bsnm{Gunasekera},~\bfnm{Kenneth}\binits{K.}},
  \bauthor{\bsnm{Havumaki},~\bfnm{Joshua}\binits{J.}},
  \bauthor{\bsnm{Pitzer},~\bfnm{Virginia~E}\binits{V.~E.}},
  \bauthor{\bsnm{Salomon},~\bfnm{Joshua~A}\binits{J.~A.}},
  \bauthor{\bsnm{Swartwood},~\bfnm{Nicole}\binits{N.}},
  \bauthor{\bsnm{Warren},~\bfnm{Joshua~L}\binits{J.~L.}},
  \bauthor{\bsnm{Weinberger},~\bfnm{Daniel~M}\binits{D.~M.}} \AND
  \bauthor{\bsnm{Cohen},~\bfnm{Ted}\binits{T.}}
(\byear{2021}).
\btitle{Reconstructing the course of the {COVID}-19 epidemic over 2020 for {US}
  states and counties: results of a {Bayesian} evidence synthesis model}.
\bdoi{10.1101/2020.06.17.20133983}
\end{bunpublished}
\endbibitem

\bibitem[\protect\citeauthoryear{Cori et~al.}{2013}]{Cori:2013}
\begin{barticle}[author]
\bauthor{\bsnm{Cori},~\bfnm{Anne}\binits{A.}},
  \bauthor{\bsnm{Ferguson},~\bfnm{Neil~M.}\binits{N.~M.}},
  \bauthor{\bsnm{Fraser},~\bfnm{Christophe}\binits{C.}} \AND
  \bauthor{\bsnm{Cauchemez},~\bfnm{Simon}\binits{S.}}
(\byear{2013}).
\btitle{A new framework and software to estimate time-varying reproduction
  numbers during epidemics}.
\bjournal{American Journal of Epidemiology}
\bvolume{178}
\bpages{1505--1512}.
\end{barticle}
\endbibitem

\bibitem[\protect\citeauthoryear{Debeye and Van~Riel}{1990}]{Debeye:1990}
\begin{barticle}[author]
\bauthor{\bsnm{Debeye},~\bfnm{H.~W.~J.}\binits{H.~W.~J.}} \AND
  \bauthor{\bsnm{Van~Riel},~\bfnm{P.}\binits{P.}}
(\byear{1990}).
\btitle{$\mathnormal{L}_p$-norm deconvolution}.
\bjournal{Geophysical Prospecting}
\bvolume{38}
\bpages{381--403}.
\end{barticle}
\endbibitem

\bibitem[\protect\citeauthoryear{Dong, Du and Gardner}{2020}]{Dong:2020}
\begin{barticle}[author]
\bauthor{\bsnm{Dong},~\bfnm{Ensheng}\binits{E.}},
  \bauthor{\bsnm{Du},~\bfnm{Hongru}\binits{H.}} \AND
  \bauthor{\bsnm{Gardner},~\bfnm{Lauren}\binits{L.}}
(\byear{2020}).
\btitle{An interactive web-based dashboard to track {COVID}-19 in real time}.
\bjournal{The Lancet Infectious Diseases}
\bvolume{20}
\bpages{533--544}.
\end{barticle}
\endbibitem

\bibitem[\protect\citeauthoryear{Farrow}{2016}]{Farrow:2016}
\begin{bphdthesis}[author]
\bauthor{\bsnm{Farrow},~\bfnm{David~C}\binits{D.~C.}}
(\byear{2016}).
\btitle{Modeling the past, present, and future of influenza},
\btype{PhD thesis},
\bpublisher{Carnegie Mellon University}.
\end{bphdthesis}
\endbibitem

\bibitem[\protect\citeauthoryear{{Centers for Disease Control and Prevention,
  COVID-19 Response}}{2020a}]{cdc_public}
\begin{bmisc}[author]
\bauthor{\bsnm{{Centers for Disease Control and Prevention, COVID-19
  Response}}}
(\byear{2020}a).
\btitle{{COVID-19 Case Surveillance Public Use Data}}.
\bhowpublished{\url{https://data.cdc.gov/Case-Surveillance/COVID-19-Case-Surveillance-Public-Use-Data/vbim-akqf}}.
\bnote{Data accessed on November 3, 2021}.
\end{bmisc}
\endbibitem

\bibitem[\protect\citeauthoryear{{Centers for Disease Control and Prevention,
  COVID-19 Response}}{2020b}]{cdc_restricted}
\begin{bmisc}[author]
\bauthor{\bsnm{{Centers for Disease Control and Prevention, COVID-19
  Response}}}
(\byear{2020}b).
\btitle{{COVID-19 Case Surveillance Restricted Access Detailed Data}}.
\bhowpublished{\url{https://data.cdc.gov/Case-Surveillance/COVID-19-Case-Surveillance-Restricted-Access-Detai/mbd7-r32t}}.
\bnote{Data accessed on November 3, 2021}.
\end{bmisc}
\endbibitem

\bibitem[\protect\citeauthoryear{Ginsberg et~al.}{2009}]{Ginsberg:2009}
\begin{barticle}[author]
\bauthor{\bsnm{Ginsberg},~\bfnm{Jeremy}\binits{J.}},
  \bauthor{\bsnm{Mohebbi},~\bfnm{Matthew~H}\binits{M.~H.}},
  \bauthor{\bsnm{Patel},~\bfnm{Rajan~S}\binits{R.~S.}},
  \bauthor{\bsnm{Brammer},~\bfnm{Lynnette}\binits{L.}},
  \bauthor{\bsnm{Smolinski},~\bfnm{Mark~S}\binits{M.~S.}} \AND
  \bauthor{\bsnm{Brilliant},~\bfnm{Larry}\binits{L.}}
(\byear{2009}).
\btitle{Detecting influenza epidemics using search engine query data}.
\bjournal{Nature}
\bvolume{457}
\bpages{1012--1014}.
\end{barticle}
\endbibitem

\bibitem[\protect\citeauthoryear{Goldstein et~al.}{2009}]{Goldstein:2009}
\begin{barticle}[author]
\bauthor{\bsnm{Goldstein},~\bfnm{Edward}\binits{E.}},
  \bauthor{\bsnm{Dushoff},~\bfnm{Jonathan}\binits{J.}},
  \bauthor{\bsnm{Ma},~\bfnm{Junling}\binits{J.}},
  \bauthor{\bsnm{Plotkin},~\bfnm{Joshua~B}\binits{J.~B.}},
  \bauthor{\bsnm{Earn},~\bfnm{David~JD}\binits{D.~J.}} \AND
  \bauthor{\bsnm{Lipsitch},~\bfnm{Marc}\binits{M.}}
(\byear{2009}).
\btitle{Reconstructing influenza incidence by deconvolution of daily mortality
  time series}.
\bjournal{Proceedings of the National Academy of Sciences}
\bvolume{106}
\bpages{21825--21829}.
\end{barticle}
\endbibitem

\bibitem[\protect\citeauthoryear{Gostic et~al.}{2020}]{Gostic:2020}
\begin{barticle}[author]
\bauthor{\bsnm{Gostic},~\bfnm{Katelyn~M.}\binits{K.~M.}},
  \bauthor{\bsnm{McGough},~\bfnm{Lauren}\binits{L.}},
  \bauthor{\bsnm{Baskerville},~\bfnm{Edward~B.}\binits{E.~B.}},
  \bauthor{\bsnm{Abbott},~\bfnm{Sam}\binits{S.}},
  \bauthor{\bsnm{Joshi},~\bfnm{Keya}\binits{K.}},
  \bauthor{\bsnm{Tedijanto},~\bfnm{Christine}\binits{C.}},
  \bauthor{\bsnm{Kahn},~\bfnm{Rebecca}\binits{R.}},
  \bauthor{\bsnm{Niehus},~\bfnm{Rene}\binits{R.}},
  \bauthor{\bsnm{Hay},~\bfnm{James~A.}\binits{J.~A.}},
  \bauthor{\bsnm{De~Salazar},~\bfnm{Pablo~M.}\binits{P.~M.}},
  \bauthor{\bsnm{Hellewell},~\bfnm{Joel}\binits{J.}},
  \bauthor{\bsnm{Meakin},~\bfnm{Sophie}\binits{S.}},
  \bauthor{\bsnm{Munday},~\bfnm{James~D.}\binits{J.~D.}},
  \bauthor{\bsnm{Bosse},~\bfnm{Nikos~I.}\binits{N.~I.}},
  \bauthor{\bsnm{Sherrat},~\bfnm{Katharine}\binits{K.}},
  \bauthor{\bsnm{Thompson},~\bfnm{Robin~N.}\binits{R.~N.}},
  \bauthor{\bsnm{White},~\bfnm{Laura~F.}\binits{L.~F.}},
  \bauthor{\bsnm{Huisman},~\bfnm{Jana~S.}\binits{J.~S.}},
  \bauthor{\bsnm{Scire},~\bfnm{Jérémie}\binits{J.}},
  \bauthor{\bsnm{Bonhoeffer},~\bfnm{Sebastian}\binits{S.}},
  \bauthor{\bsnm{Stadler},~\bfnm{Tanja}\binits{T.}},
  \bauthor{\bsnm{Wallinga},~\bfnm{Jacco}\binits{J.}},
  \bauthor{\bsnm{Funk},~\bfnm{Sebastian}\binits{S.}},
  \bauthor{\bsnm{Lipsitch},~\bfnm{Marc}\binits{M.}} \AND
  \bauthor{\bsnm{Cobey},~\bfnm{Sarah}\binits{S.}}
(\byear{2020}).
\btitle{Practical considerations for measuring the effective reproductive
  number, $R_t$}.
\bjournal{PLOS Computational Biology}
\bvolume{16}
\bpages{1--21}.
\end{barticle}
\endbibitem

\bibitem[\protect\citeauthoryear{Hawryluk et~al.}{2021}]{Hawryluk:2021}
\begin{binproceedings}[author]
\bauthor{\bsnm{Hawryluk},~\bfnm{Iwona}\binits{I.}},
  \bauthor{\bsnm{Hoeltgebaum},~\bfnm{Henrique}\binits{H.}},
  \bauthor{\bsnm{Mishra},~\bfnm{Swapnil}\binits{S.}},
  \bauthor{\bsnm{Miscouridou},~\bfnm{Xenia}\binits{X.}},
  \bauthor{\bsnm{Schnekenberg},~\bfnm{Ricardo~P}\binits{R.~P.}},
  \bauthor{\bsnm{Whittaker},~\bfnm{Charles}\binits{C.}},
  \bauthor{\bsnm{Vollmer},~\bfnm{Michaela}\binits{M.}},
  \bauthor{\bsnm{Flaxman},~\bfnm{Seth}\binits{S.}},
  \bauthor{\bsnm{Bhatt},~\bfnm{Samir}\binits{S.}} \AND
  \bauthor{\bsnm{Mellan},~\bfnm{Thomas~A}\binits{T.~A.}}
(\byear{2021}).
\btitle{Gaussian process nowcasting: {Application} to {COVID}-19 mortality
  reporting}.
In \bbooktitle{Conference on Uncertainty in Artificial Intelligence}.
\end{binproceedings}
\endbibitem

\bibitem[\protect\citeauthoryear{Jahja et~al.}{2019}]{Jahja:2019}
\begin{binproceedings}[author]
\bauthor{\bsnm{Jahja},~\bfnm{Maria}\binits{M.}},
  \bauthor{\bsnm{Farrow},~\bfnm{David}\binits{D.}},
  \bauthor{\bsnm{Rosenfeld},~\bfnm{Roni}\binits{R.}} \AND
  \bauthor{\bsnm{Tibshirani},~\bfnm{Ryan~J.}\binits{R.~J.}}
(\byear{2019}).
\btitle{{Kalman Filter}, sensor fusion, and constrained regression:
  equivalences and insights}.
In \bbooktitle{Advances in Neural Information Processing Systems}.
\end{binproceedings}
\endbibitem

\bibitem[\protect\citeauthoryear{Johnson}{2013}]{johnson2013dynamic}
\begin{barticle}[author]
\bauthor{\bsnm{Johnson},~\bfnm{Nicholas}\binits{N.}}
(\byear{2013}).
\btitle{A dynamic programming algorithm for the fused lasso and
  $L_0$-segmentation}.
\bjournal{Journal of Computational and Graphical Statistics}
\bvolume{22}
\bpages{246--260}.
\end{barticle}
\endbibitem

\bibitem[\protect\citeauthoryear{Kaplan and
  Meier}{1958}]{kaplan1958nonparametric}
\begin{barticle}[author]
\bauthor{\bsnm{Kaplan},~\bfnm{Edward~L.}\binits{E.~L.}} \AND
  \bauthor{\bsnm{Meier},~\bfnm{Paul}\binits{P.}}
(\byear{1958}).
\btitle{Nonparametric Estimation from Incomplete Observations}.
\bjournal{Journal of the American Statistical Association}
\bvolume{53}
\bpages{457--481}.
\end{barticle}
\endbibitem

\bibitem[\protect\citeauthoryear{Kass-Hout and
  Alhinnawi}{2013}]{Kass-Hout:2013}
\begin{barticle}[author]
\bauthor{\bsnm{Kass-Hout},~\bfnm{Taha~A}\binits{T.~A.}} \AND
  \bauthor{\bsnm{Alhinnawi},~\bfnm{Hend}\binits{H.}}
(\byear{2013}).
\btitle{Social media in public health}.
\bjournal{British Medical Bulletin}
\bvolume{108}
\bpages{5--24}.
\end{barticle}
\endbibitem

\bibitem[\protect\citeauthoryear{Kass-Hout and Zhang}{2011}]{Kass-Hout:2011}
\begin{bbook}[author]
\bauthor{\bsnm{Kass-Hout},~\bfnm{Taha~A}\binits{T.~A.}} \AND
  \bauthor{\bsnm{Zhang},~\bfnm{Xiaohui}\binits{X.}}
(\byear{2011}).
\btitle{Biosurveillance: {M}ethods and Case Studies}.
\bpublisher{CRC Press}.
\end{bbook}
\endbibitem

\bibitem[\protect\citeauthoryear{{Reich Lab}}{2020}]{ForecastHub}
\begin{bmisc}[author]
\bauthor{\bsnm{{Reich Lab}}}
(\byear{2020}).
\btitle{{The COVID-19 Forecast Hub}}.
\bhowpublished{\url{https://covid19forecasthub.org}}.
\end{bmisc}
\endbibitem

\bibitem[\protect\citeauthoryear{Leuba et~al.}{2020}]{Leuba:2020}
\begin{barticle}[author]
\bauthor{\bsnm{Leuba},~\bfnm{Sequoia~I.}\binits{S.~I.}},
  \bauthor{\bsnm{Yaesoubi},~\bfnm{Reza}\binits{R.}},
  \bauthor{\bsnm{Antillon},~\bfnm{Marina}\binits{M.}},
  \bauthor{\bsnm{Cohen},~\bfnm{Ted}\binits{T.}} \AND
  \bauthor{\bsnm{Zimmer},~\bfnm{Christoph}\binits{C.}}
(\byear{2020}).
\btitle{Tracking and predicting {U.S.} influenza activity with a real-time
  surveillance network}.
\bjournal{PLOS Computational Biology}
\bvolume{16}
\bpages{1--14}.
\end{barticle}
\endbibitem

\bibitem[\protect\citeauthoryear{McDonald et~al.}{2021}]{McDonald:2021}
\begin{bunpublished}[author]
\bauthor{\bsnm{McDonald},~\bfnm{Daniel~J.}\binits{D.~J.}},
  \bauthor{\bsnm{Bien},~\bfnm{Jacob}\binits{J.}},
  \bauthor{\bsnm{Green},~\bfnm{Alden}\binits{A.}},
  \bauthor{\bsnm{Hu},~\bfnm{Addison~J.}\binits{A.~J.}},
  \bauthor{\bsnm{DeFries},~\bfnm{Nat}\binits{N.}},
  \bauthor{\bsnm{Hyun},~\bfnm{Sangwon}\binits{S.}},
  \bauthor{\bsnm{Oliveira},~\bfnm{Natalia~L.}\binits{N.~L.}},
  \bauthor{\bsnm{Sharpnack},~\bfnm{James}\binits{J.}},
  \bauthor{\bsnm{Tang},~\bfnm{Jingjing}\binits{J.}},
  \bauthor{\bsnm{Tibshirani},~\bfnm{Robert}\binits{R.}},
  \bauthor{\bsnm{Ventura},~\bfnm{Val{\'e}rie}\binits{V.}},
  \bauthor{\bsnm{Wasserman},~\bfnm{Larry}\binits{L.}} \AND
  \bauthor{\bsnm{Tibshirani},~\bfnm{Ryan~J.}\binits{R.~J.}}
(\byear{2021}).
\btitle{Can auxiliary indicators improve {COVID}-19 forecasting and hotspot
  prediction?}
\bnote{To appear, PNAS}.
\end{bunpublished}
\endbibitem

\bibitem[\protect\citeauthoryear{McGough et~al.}{2020}]{McGough:2020}
\begin{barticle}[author]
\bauthor{\bsnm{McGough},~\bfnm{Sarah~F}\binits{S.~F.}},
  \bauthor{\bsnm{Johansson},~\bfnm{Michael~A}\binits{M.~A.}},
  \bauthor{\bsnm{Lipsitch},~\bfnm{Marc}\binits{M.}} \AND
  \bauthor{\bsnm{Menzies},~\bfnm{Nicolas~A}\binits{N.~A.}}
(\byear{2020}).
\btitle{Nowcasting by {Bayesian} smoothing: {A} flexible, generalizable model
  for real-time epidemic tracking}.
\bjournal{PLOS Computational Biology}
\bvolume{16}
\bpages{e1007735}.
\end{barticle}
\endbibitem

\bibitem[\protect\citeauthoryear{McIver and Brownstein}{2014}]{McIver:2014}
\begin{barticle}[author]
\bauthor{\bsnm{McIver},~\bfnm{David~J.}\binits{D.~J.}} \AND
  \bauthor{\bsnm{Brownstein},~\bfnm{John~S.}\binits{J.~S.}}
(\byear{2014}).
\btitle{{Wikipedia} usage estimates prevalence of influenza-like illness in the
  {United States} in near real-time}.
\bjournal{PLOS Computational Biology}
\bvolume{10}
\bpages{e1003581}.
\end{barticle}
\endbibitem

\bibitem[\protect\citeauthoryear{Oppenheim and Verghese}{2017}]{Oppenheim:2017}
\begin{bbook}[author]
\bauthor{\bsnm{Oppenheim},~\bfnm{Alan~V.}\binits{A.~V.}} \AND
  \bauthor{\bsnm{Verghese},~\bfnm{George~C.}\binits{G.~C.}}
(\byear{2017}).
\btitle{Signals, Systems and Inference}.
\bpublisher{Pearson}.
\end{bbook}
\endbibitem

\bibitem[\protect\citeauthoryear{Paul and Dredze}{2017}]{Paul:2017}
\begin{barticle}[author]
\bauthor{\bsnm{Paul},~\bfnm{Michael~J}\binits{M.~J.}} \AND
  \bauthor{\bsnm{Dredze},~\bfnm{Mark}\binits{M.}}
(\byear{2017}).
\btitle{Social monitoring for public health}.
\bjournal{Synthesis Lectures on Information Concepts, Retrieval, and Services}
\bvolume{9}
\bpages{1--183}.
\end{barticle}
\endbibitem

\bibitem[\protect\citeauthoryear{Radin et~al.}{2020}]{Radin:2020}
\begin{barticle}[author]
\bauthor{\bsnm{Radin},~\bfnm{Jennifer~M}\binits{J.~M.}},
  \bauthor{\bsnm{Wineinger},~\bfnm{Nathan~E}\binits{N.~E.}},
  \bauthor{\bsnm{Topol},~\bfnm{Eric~J}\binits{E.~J.}} \AND
  \bauthor{\bsnm{Steinhubl},~\bfnm{Steven~R}\binits{S.~R.}}
(\byear{2020}).
\btitle{Harnessing wearable device data to improve state-level real-time
  surveillance of influenza-like illness in the {USA}: {A} population-based
  study}.
\bjournal{The Lancet Digital Health}
\bvolume{2}
\bpages{e85--e93}.
\end{barticle}
\endbibitem

\bibitem[\protect\citeauthoryear{Ramdas and Tibshirani}{2016}]{ramdas2016fast}
\begin{barticle}[author]
\bauthor{\bsnm{Ramdas},~\bfnm{Aaditya}\binits{A.}} \AND
  \bauthor{\bsnm{Tibshirani},~\bfnm{Ryan~J.}\binits{R.~J.}}
(\byear{2016}).
\btitle{Fast and Flexible {ADMM} Algorithms for Trend Filtering}.
\bjournal{Journal of Computational and Graphical Statistics}
\bvolume{25}
\bpages{839--858}.
\end{barticle}
\endbibitem

\bibitem[\protect\citeauthoryear{Reinhart et~al.}{2021}]{Reinhart:2021}
\begin{bunpublished}[author]
\bauthor{\bsnm{Reinhart},~\bfnm{Alex}\binits{A.}},
  \bauthor{\bsnm{Brooks},~\bfnm{Logan}\binits{L.}},
  \bauthor{\bsnm{Jahja},~\bfnm{Maria}\binits{M.}},
  \bauthor{\bsnm{Rumack},~\bfnm{Aaron}\binits{A.}},
  \bauthor{\bsnm{Tang},~\bfnm{Jingjing}\binits{J.}},
  \bauthor{\bsnm{Agrawal},~\bfnm{Sumit}\binits{S.}},
  \bauthor{\bsnm{Saeed},~\bfnm{Wael~Al}\binits{W.~A.}},
  \bauthor{\bsnm{Arnold},~\bfnm{Taylor}\binits{T.}},
  \bauthor{\bsnm{Basu},~\bfnm{Amartya}\binits{A.}},
  \bauthor{\bsnm{Bien},~\bfnm{Jacob}\binits{J.}},
  \bauthor{\bsnm{Cabrera},~\bfnm{{\'A}ngel~A.}\binits{{\'A}.~A.}},
  \bauthor{\bsnm{Chin},~\bfnm{Andrew}\binits{A.}},
  \bauthor{\bsnm{Chua},~\bfnm{Eu~Jing}\binits{E.~J.}},
  \bauthor{\bsnm{Clark},~\bfnm{Brian}\binits{B.}},
  \bauthor{\bsnm{Colquhoun},~\bfnm{Sarah}\binits{S.}},
  \bauthor{\bsnm{DeFries},~\bfnm{Nat}\binits{N.}},
  \bauthor{\bsnm{Farrow},~\bfnm{David~C.}\binits{D.~C.}},
  \bauthor{\bsnm{Forlizzi},~\bfnm{Jodi}\binits{J.}},
  \bauthor{\bsnm{Grabman},~\bfnm{Jed}\binits{J.}},
  \bauthor{\bsnm{Gratzl},~\bfnm{Samuel}\binits{S.}},
  \bauthor{\bsnm{Green},~\bfnm{Alden}\binits{A.}},
  \bauthor{\bsnm{Haff},~\bfnm{George}\binits{G.}},
  \bauthor{\bsnm{Han},~\bfnm{Robin}\binits{R.}},
  \bauthor{\bsnm{Harwood},~\bfnm{Kate}\binits{K.}},
  \bauthor{\bsnm{Hu},~\bfnm{Addison~J.}\binits{A.~J.}},
  \bauthor{\bsnm{Hyde},~\bfnm{Raphael}\binits{R.}},
  \bauthor{\bsnm{Hyun},~\bfnm{Sangwon}\binits{S.}},
  \bauthor{\bsnm{Joshi},~\bfnm{Ananya}\binits{A.}},
  \bauthor{\bsnm{Kim},~\bfnm{Jimi}\binits{J.}},
  \bauthor{\bsnm{Kuznetsov},~\bfnm{Andrew}\binits{A.}},
  \bauthor{\bsnm{Motte-Kerr},~\bfnm{Wichada~La}\binits{W.~L.}},
  \bauthor{\bsnm{Lee},~\bfnm{Yeon~Jin}\binits{Y.~J.}},
  \bauthor{\bsnm{Lee},~\bfnm{Kenneth}\binits{K.}},
  \bauthor{\bsnm{Lipton},~\bfnm{Zachary~C.}\binits{Z.~C.}},
  \bauthor{\bsnm{Liu},~\bfnm{Michael~X.}\binits{M.~X.}},
  \bauthor{\bsnm{Mackey},~\bfnm{Lester}\binits{L.}},
  \bauthor{\bsnm{Mazaitis},~\bfnm{Kathryn}\binits{K.}},
  \bauthor{\bsnm{McDonald},~\bfnm{Daniel~J.}\binits{D.~J.}},
  \bauthor{\bsnm{McGuinness},~\bfnm{Phillip}\binits{P.}},
  \bauthor{\bsnm{Narasimhan},~\bfnm{Balasubramanian}\binits{B.}},
  \bauthor{\bsnm{O'Brien},~\bfnm{Michael~P.}\binits{M.~P.}},
  \bauthor{\bsnm{Oliveira},~\bfnm{Natalia~L.}\binits{N.~L.}},
  \bauthor{\bsnm{Patil},~\bfnm{Pratik}\binits{P.}},
  \bauthor{\bsnm{Perer},~\bfnm{Adam}\binits{A.}},
  \bauthor{\bsnm{Politsch},~\bfnm{Collin~A.}\binits{C.~A.}},
  \bauthor{\bsnm{Rajanala},~\bfnm{Samyak}\binits{S.}},
  \bauthor{\bsnm{Rucker},~\bfnm{Dawn}\binits{D.}},
  \bauthor{\bsnm{Scott},~\bfnm{Chris}\binits{C.}},
  \bauthor{\bsnm{Shah},~\bfnm{Nigam~H.}\binits{N.~H.}},
  \bauthor{\bsnm{Shankar},~\bfnm{Vishnu}\binits{V.}},
  \bauthor{\bsnm{Sharpnack},~\bfnm{James}\binits{J.}},
  \bauthor{\bsnm{Shemetov},~\bfnm{Dmitry}\binits{D.}},
  \bauthor{\bsnm{Simon},~\bfnm{Noah}\binits{N.}},
  \bauthor{\bsnm{Smith},~\bfnm{Benjamin~Y.}\binits{B.~Y.}},
  \bauthor{\bsnm{Srivastava},~\bfnm{Vishakha}\binits{V.}},
  \bauthor{\bsnm{Tan},~\bfnm{Shuyi}\binits{S.}},
  \bauthor{\bsnm{Tibshirani},~\bfnm{Robert}\binits{R.}},
  \bauthor{\bsnm{Tuzhilina},~\bfnm{Elena}\binits{E.}},
  \bauthor{\bsnm{Nortwick},~\bfnm{Ana Karina~Van}\binits{A.~K.~V.}},
  \bauthor{\bsnm{Ventura},~\bfnm{Val{\'e}rie}\binits{V.}},
  \bauthor{\bsnm{Wasserman},~\bfnm{Larry}\binits{L.}},
  \bauthor{\bsnm{Weaver},~\bfnm{Benjamin}\binits{B.}},
  \bauthor{\bsnm{Weiss},~\bfnm{Jeremy~C.}\binits{J.~C.}},
  \bauthor{\bsnm{Whitman},~\bfnm{Spencer}\binits{S.}},
  \bauthor{\bsnm{Williams},~\bfnm{Kristin}\binits{K.}},
  \bauthor{\bsnm{Rosenfeld},~\bfnm{Roni}\binits{R.}} \AND
  \bauthor{\bsnm{Tibshirani},~\bfnm{Ryan~J.}\binits{R.~J.}}
(\byear{2021}).
\btitle{An open repository of real-time {COVID}-19 indicators}.
\bnote{To appear, PNAS}.
\end{bunpublished}
\endbibitem

\bibitem[\protect\citeauthoryear{Rosenfeld and
  Tibshirani}{2021}]{Rosenfeld:2021}
\begin{bunpublished}[author]
\bauthor{\bsnm{Rosenfeld},~\bfnm{Roni}\binits{R.}} \AND
  \bauthor{\bsnm{Tibshirani},~\bfnm{Ryan~J.}\binits{R.~J.}}
(\byear{2021}).
\btitle{Epidemic tracking and forecasting: {L}essons learned from a tumultuous
  year}.
\bnote{To appear, PNAS}.
\end{bunpublished}
\endbibitem

\bibitem[\protect\citeauthoryear{Rudin and Osher}{1994}]{Rudin:1994}
\begin{binproceedings}[author]
\bauthor{\bsnm{Rudin},~\bfnm{L.~I.}\binits{L.~I.}} \AND
  \bauthor{\bsnm{Osher},~\bfnm{S.}\binits{S.}}
(\byear{1994}).
\btitle{Total variation based image restoration with free local constraints}.
In \bbooktitle{International Conference on Image Processing}
\bvolume{1}
\bpages{31--35}.
\end{binproceedings}
\endbibitem

\bibitem[\protect\citeauthoryear{Salath{\'e} et~al.}{2012}]{Salathe:2012}
\begin{barticle}[author]
\bauthor{\bsnm{Salath{\'e}},~\bfnm{Marcel}\binits{M.}},
  \bauthor{\bsnm{Bengtsson},~\bfnm{Linus}\binits{L.}},
  \bauthor{\bsnm{Bodnar},~\bfnm{Todd~J}\binits{T.~J.}},
  \bauthor{\bsnm{Brewer},~\bfnm{Devon~D}\binits{D.~D.}},
  \bauthor{\bsnm{Brownstein},~\bfnm{John~S}\binits{J.~S.}},
  \bauthor{\bsnm{Buckee},~\bfnm{Caroline}\binits{C.}},
  \bauthor{\bsnm{Campbell},~\bfnm{Ellsworth~M}\binits{E.~M.}},
  \bauthor{\bsnm{Cattuto},~\bfnm{Ciro}\binits{C.}},
  \bauthor{\bsnm{Khandelwal},~\bfnm{Shashank}\binits{S.}},
  \bauthor{\bsnm{Mabry},~\bfnm{Patricia~L}\binits{P.~L.}} \AND
  \bauthor{\bsnm{Vespignani},~\bfnm{Alessandro}\binits{A.}}
(\byear{2012}).
\btitle{Digital epidemiology}.
\bjournal{PLOS Computational Biology}
\bvolume{8}
\bpages{1--3}.
\end{barticle}
\endbibitem

\bibitem[\protect\citeauthoryear{Salomon et~al.}{2021}]{Salomon:2021}
\begin{bunpublished}[author]
\bauthor{\bsnm{Salomon},~\bfnm{Joshua~A.}\binits{J.~A.}},
  \bauthor{\bsnm{Reinhart},~\bfnm{Alex}\binits{A.}},
  \bauthor{\bsnm{Bilinski},~\bfnm{Alyssa}\binits{A.}},
  \bauthor{\bsnm{Chua},~\bfnm{Eu~Jing}\binits{E.~J.}},
  \bauthor{\bsnm{La~Motte-Kerr},~\bfnm{Wichada}\binits{W.}},
  \bauthor{\bsnm{R{\"o}nn},~\bfnm{Minttu~M.}\binits{M.~M.}},
  \bauthor{\bsnm{Reitsma},~\bfnm{Marissa}\binits{M.}},
  \bauthor{\bsnm{Morris},~\bfnm{Katherine~Ann}\binits{K.~A.}},
  \bauthor{\bsnm{LaRocca},~\bfnm{Sarah}\binits{S.}},
  \bauthor{\bsnm{Farag},~\bfnm{Tamer}\binits{T.}},
  \bauthor{\bsnm{Kreuter},~\bfnm{Frauke}\binits{F.}},
  \bauthor{\bsnm{Rosenfeld},~\bfnm{Roni}\binits{R.}} \AND
  \bauthor{\bsnm{Tibshirani},~\bfnm{Ryan~J.}\binits{R.~J.}}
(\byear{2021}).
\btitle{The {COVID}-19 trends and impact survey: {C}ontinuous real-time
  measurement of {COVID}-19 symptoms, risks, protective behaviors, testing and
  vaccination}.
\bnote{To appear, PNAS}.
\end{bunpublished}
\endbibitem

\bibitem[\protect\citeauthoryear{Santillana et~al.}{2015}]{Santillana:2015}
\begin{barticle}[author]
\bauthor{\bsnm{Santillana},~\bfnm{Mauricio}\binits{M.}},
  \bauthor{\bsnm{Nguyen},~\bfnm{Andr{\'e}~T}\binits{A.~T.}},
  \bauthor{\bsnm{Dredze},~\bfnm{Mark}\binits{M.}},
  \bauthor{\bsnm{Paul},~\bfnm{Michael~J}\binits{M.~J.}},
  \bauthor{\bsnm{Nsoesie},~\bfnm{Elaine~O}\binits{E.~O.}} \AND
  \bauthor{\bsnm{Brownstein},~\bfnm{John~S}\binits{J.~S.}}
(\byear{2015}).
\btitle{Combining search, social media, and traditional data sources to improve
  influenza surveillance}.
\bjournal{PLOS Computational Biology}
\bvolume{11}
\bpages{e1004513}.
\end{barticle}
\endbibitem

\bibitem[\protect\citeauthoryear{Santillana et~al.}{2016}]{Santillana:2016}
\begin{barticle}[author]
\bauthor{\bsnm{Santillana},~\bfnm{Mauricio}\binits{M.}},
  \bauthor{\bsnm{Nguyen},~\bfnm{Andre~T}\binits{A.~T.}},
  \bauthor{\bsnm{Louie},~\bfnm{Tamara}\binits{T.}},
  \bauthor{\bsnm{Zink},~\bfnm{Anna}\binits{A.}},
  \bauthor{\bsnm{Gray},~\bfnm{Josh}\binits{J.}},
  \bauthor{\bsnm{Sung},~\bfnm{Iyue}\binits{I.}} \AND
  \bauthor{\bsnm{Brownstein},~\bfnm{John~S}\binits{J.~S.}}
(\byear{2016}).
\btitle{Cloud-based electronic health records for real-time, region-specific
  influenza surveillance}.
\bjournal{Scientific Reports}
\bvolume{6}
\bpages{1--8}.
\end{barticle}
\endbibitem

\bibitem[\protect\citeauthoryear{Smolinski et~al.}{2015}]{Smolinski:2015}
\begin{barticle}[author]
\bauthor{\bsnm{Smolinski},~\bfnm{Mark~S.}\binits{M.~S.}},
  \bauthor{\bsnm{Crawley},~\bfnm{Adam~W.}\binits{A.~W.}},
  \bauthor{\bsnm{Baltrusaitis},~\bfnm{Kristin}\binits{K.}},
  \bauthor{\bsnm{Chunara},~\bfnm{Rumi}\binits{R.}},
  \bauthor{\bsnm{Olsen},~\bfnm{Jennifer~M.}\binits{J.~M.}},
  \bauthor{\bsnm{Wójcik},~\bfnm{Oktawia}\binits{O.}},
  \bauthor{\bsnm{Santillana},~\bfnm{Mauricio}\binits{M.}},
  \bauthor{\bsnm{Nguyen},~\bfnm{Andre}\binits{A.}} \AND
  \bauthor{\bsnm{Brownstein},~\bfnm{John~S.}\binits{J.~S.}}
(\byear{2015}).
\btitle{{Flu Near You}: {C}rowdsourced symptom reporting spanning 2 influenza
  seasons}.
\bjournal{American Journal of Public Health}
\bvolume{105}
\bpages{2124--2130}.
\end{barticle}
\endbibitem

\bibitem[\protect\citeauthoryear{Systrom, Vladek and Krieger}{2020}]{rtlive}
\begin{bmisc}[author]
\bauthor{\bsnm{Systrom},~\bfnm{Kevin}\binits{K.}},
  \bauthor{\bsnm{Vladek},~\bfnm{Thomas}\binits{T.}} \AND
  \bauthor{\bsnm{Krieger},~\bfnm{Mike}\binits{M.}}
(\byear{2020}).
\btitle{Rt.live}.
\bhowpublished{\url{https://github.com/rtcovidlive/covid-model}}.
\end{bmisc}
\endbibitem

\bibitem[\protect\citeauthoryear{Taylor, Banks and McCoy}{1979}]{Taylor:1979}
\begin{barticle}[author]
\bauthor{\bsnm{Taylor},~\bfnm{Howard~L.}\binits{H.~L.}},
  \bauthor{\bsnm{Banks},~\bfnm{Stephen~C.}\binits{S.~C.}} \AND
  \bauthor{\bsnm{McCoy},~\bfnm{John~F.}\binits{J.~F.}}
(\byear{1979}).
\btitle{Deconvolution with the $\ell_1$ norm}.
\bjournal{Geophysics}
\bvolume{44}
\bpages{39--52}.
\end{barticle}
\endbibitem

\bibitem[\protect\citeauthoryear{Thompson et~al.}{2019}]{Thompson:2019}
\begin{barticle}[author]
\bauthor{\bsnm{Thompson},~\bfnm{R.~N.}\binits{R.~N.}},
  \bauthor{\bsnm{Stockwin},~\bfnm{J.~E.}\binits{J.~E.}}, \bauthor{\bsnm{{van
  Gaalen}},~\bfnm{R.~D.}\binits{R.~D.}},
  \bauthor{\bsnm{Polonsky},~\bfnm{J.~A.}\binits{J.~A.}},
  \bauthor{\bsnm{Kamvar},~\bfnm{Z.~N.}\binits{Z.~N.}},
  \bauthor{\bsnm{Demarsh},~\bfnm{P.~A.}\binits{P.~A.}},
  \bauthor{\bsnm{Dahlqwist},~\bfnm{E.}\binits{E.}},
  \bauthor{\bsnm{Li},~\bfnm{S.}\binits{S.}},
  \bauthor{\bsnm{Miguel},~\bfnm{E.}\binits{E.}},
  \bauthor{\bsnm{Jombart},~\bfnm{T.}\binits{T.}},
  \bauthor{\bsnm{Lessler},~\bfnm{J.}\binits{J.}},
  \bauthor{\bsnm{Cauchemez},~\bfnm{S.}\binits{S.}} \AND
  \bauthor{\bsnm{Cori},~\bfnm{A.}\binits{A.}}
(\byear{2019}).
\btitle{Improved inference of time-varying reproduction numbers during
  infectious disease outbreaks}.
\bjournal{Epidemics}
\bvolume{29}
\bpages{100356}.
\end{barticle}
\endbibitem

\bibitem[\protect\citeauthoryear{Tibshirani}{2014}]{tibshirani2014adaptive}
\begin{barticle}[author]
\bauthor{\bsnm{Tibshirani},~\bfnm{Ryan~J.}\binits{R.~J.}}
(\byear{2014}).
\btitle{Adaptive piecewise polynomial estimation via trend filtering}.
\bjournal{Annals of Statistics}
\bvolume{42}
\bpages{285--323}.
\end{barticle}
\endbibitem

\bibitem[\protect\citeauthoryear{Tibshirani}{2020}]{tibshirani2020divided}
\begin{bunpublished}[author]
\bauthor{\bsnm{Tibshirani},~\bfnm{Ryan~J.}\binits{R.~J.}}
(\byear{2020}).
\btitle{Divided differences, falling factorials, and discrete splines:
  {A}nother look at trend filtering and related problems}.
\bnote{arXiv: 2003.03886}.
\end{bunpublished}
\endbibitem

\bibitem[\protect\citeauthoryear{Viboud et~al.}{2014}]{Viboud:2014}
\begin{barticle}[author]
\bauthor{\bsnm{Viboud},~\bfnm{C{\'e}cile}\binits{C.}},
  \bauthor{\bsnm{Charu},~\bfnm{Vivek}\binits{V.}},
  \bauthor{\bsnm{Olson},~\bfnm{Donald}\binits{D.}},
  \bauthor{\bsnm{Ballesteros},~\bfnm{Sébastien}\binits{S.}},
  \bauthor{\bsnm{Gog},~\bfnm{Julia}\binits{J.}},
  \bauthor{\bsnm{Khan},~\bfnm{Farid}\binits{F.}},
  \bauthor{\bsnm{Grenfell},~\bfnm{Bryan}\binits{B.}} \AND
  \bauthor{\bsnm{Simonsen},~\bfnm{Lone}\binits{L.}}
(\byear{2014}).
\btitle{Demonstrating the use of high-volume electronic medical claims data to
  monitor local and regional influenza activity in the {US}}.
\bjournal{PLOS ONE}
\bvolume{9}
\bpages{1--12}.
\end{barticle}
\endbibitem

\bibitem[\protect\citeauthoryear{Wallinga and Lipsitch}{2007}]{Wallinga:2007}
\begin{barticle}[author]
\bauthor{\bsnm{Wallinga},~\bfnm{Jacco}\binits{J.}} \AND
  \bauthor{\bsnm{Lipsitch},~\bfnm{Marc}\binits{M.}}
(\byear{2007}).
\btitle{How generation intervals shape the relationship between growth rates
  and reproductive numbers}.
\bjournal{Proceedings of the Royal Society B: Biological Sciences}
\bvolume{274}
\bpages{599--604}.
\end{barticle}
\endbibitem

\bibitem[\protect\citeauthoryear{Wiener}{1964}]{Wiener:1964}
\begin{bbook}[author]
\bauthor{\bsnm{Wiener},~\bfnm{Norbert}\binits{N.}}
(\byear{1964}).
\btitle{Extrapolation, Interpolation, and Smoothing of Stationary Time Series}.
\bpublisher{The MIT Press}.
\end{bbook}
\endbibitem

\bibitem[\protect\citeauthoryear{Yang, Santillana and Kou}{2015}]{Yang:2015}
\begin{barticle}[author]
\bauthor{\bsnm{Yang},~\bfnm{Shihao}\binits{S.}},
  \bauthor{\bsnm{Santillana},~\bfnm{Mauricio}\binits{M.}} \AND
  \bauthor{\bsnm{Kou},~\bfnm{S.~C.}\binits{S.~C.}}
(\byear{2015}).
\btitle{Accurate estimation of influenza epidemics using {Google} search data
  via {ARGO}}.
\bjournal{Proceedings of the National Academy of Sciences}
\bvolume{112}
\bpages{14473--14478}.
\end{barticle}
\endbibitem

\bibitem[\protect\citeauthoryear{Yang et~al.}{2019}]{Yang:2019}
\begin{barticle}[author]
\bauthor{\bsnm{Yang},~\bfnm{Cheng-Yi}\binits{C.-Y.}},
  \bauthor{\bsnm{Chen},~\bfnm{Ray-Jade}\binits{R.-J.}},
  \bauthor{\bsnm{Chou},~\bfnm{Wan-Lin}\binits{W.-L.}},
  \bauthor{\bsnm{Lee},~\bfnm{Yuarn-Jang}\binits{Y.-J.}} \AND
  \bauthor{\bsnm{Lo},~\bfnm{Yu-Sheng}\binits{Y.-S.}}
(\byear{2019}).
\btitle{An integrated influenza surveillance framework based on national
  influenza-like illness incidence and multiple hospital electronic medical
  records for early prediction of influenza epidemics: {D}esign and
  evaluation}.
\bjournal{Journal of Medical Internet Research}
\bvolume{21}
\bpages{e12341}.
\end{barticle}
\endbibitem

\end{thebibliography}
\end{document}